\documentclass[fleqn]{report}

\usepackage{amssymb}
\usepackage{amsfonts}
\usepackage{amsmath}

\begin{document}

\title{PRICING\ OF\ HIGH-DIMENSIONAL\ OPTIONS}
\author{Alexander Kushpel}
\date{25 September 2015}
\maketitle


\tableofcontents

\chapter{Preface}


Pricing of high dimensional or spread options is one of the oldest and
important problems in Mathematical Finance. Such options are important in
equity, foreign exchange and commodity markets. Electricity spark spread
options are traded over a wide range of markets for exchanging a specific
fuel for electricity. A class of spreads which exchanges raw soybeans with a
combination of soybean oil and soybean meal is popular in agricultural
markets \cite{carmona}. The spread is defined as the instrument $S_{t}$, $%
t\geq 0$ whose value at time $t$ is given by the difference $%
S_{t}=S_{1,t}-S_{2,t},$ $t\geq 0$. Buying such a spread is buying $S_{1,t}$
and selling $S_{2,t}$. We should not limit ourselves to the case of the
spread defined by $S_{t}$, and instead we think of $S_{t}$ as a price of
traded financial instrument.

It is known that pricing of spread options requires models with jumps very
different from geometric Brownian motion, and pricing of such options can be
challenging \cite{hurd}. We use L\'{e}vy processes to model returns.

Known methods for pricing spread options can be divided into two big groups:
analytical approximations (approximating formulas) and numerical methods. We
shall concentrate on analytical methods which are aimed to develop
closed-form formulas to approximate the spread option price. There are two
main approaches here: PDE's and martingales. Experience shows that PDE's
methods are suitable if the dimension is low \cite{Duffy, Tavella}. We shall
adapt martingale pricing approach. In this case the price $V$ of the common
spread option at time $0$ is given by $V=\exp \left( -rT\right) \mathbb{E}^{%
\mathbb{Q}}\left[ H\right] ,$ where $H:\mathbb{R}^{n}\rightarrow \left[
0,\infty \right) $ is the reward (payoff) function, $T>0$ is maturity time
and the expectation is taken with respect to the equivalent martingale
measure $\mathbb{Q}$ which corresponds to the chosen model (see Appendixes I
and III for more details). In many cases of practical interest $\mathbb{Q}$
admits a density function $p_{T}^{\mathbb{Q}}$. Hence, in this case,%
\begin{equation}
V=\exp \left( -rT\right) \int_{\mathbb{R}^{n}}Hp_{T}^{\mathbb{Q}}d\mathbf{x}.
\label{price11}
\end{equation}%
It is important in applications to construct a pricing theory which includes
a wide range of reward functions $H$. In many practical cases the reward
function $H$ grows exponentially. For example, consider a frictionless
market with no arbitrage opportunities and with a constant riskless interest
rate $r>0$. Let $\mathbf{S}_{t}=\left\{ S_{j,t},1\leq j\leq n,t\geq
0\right\} $, be $n$ asset prices which are modeled by an exponential L\'{e}%
vy processes $S_{j,t}=S_{j,0}\exp \left( X_{j,t}\right) $. A European call
option is defined by date $T$, called the date of maturity, and a number $%
K>0 $, called the strike of exercise price, and it gives the right to its
owner to acquire at time $T$ one unit of the underlying instrument at the
unit price $K$. Assuming that this instrument can be resold for $S_{T}$,
this means that the owner of the option will receive the payout $\max
\left\{ S_{T}-K,0\right\} $ at maturity $T$. Consider an option on the price
spread $S_{T}=S_{1,T}-\sum_{j=1}^{n}S_{j,T}$. The common \ \textit{spread}
with maturity $T>0$ and strike $K\geq 0$ is the contract that pays%
\[
H\left( X_{1,T},\cdot \cdot \cdot ,X_{n,T}\right) =\left( S_{T}-K\right)
\chi _{\left\{ S_{T}>K\right\} }
\]%
\[
=\max \left\{ S_{1,0}\exp \left( X_{1,T}\right) -\sum_{j=2}^{n}S_{j,0}\exp
\left( X_{j,T}\right) -K,\text{ }0\right\}
\]%
at time $T>0$. Clearly $H\left( x_{1},\cdot \cdot \cdot ,x_{n}\right) \sim
\exp \left( x_{1}\right) $, $x_{1}\rightarrow \infty $. Hence the
characteristic function $\Phi ^{\mathbb{Q}}\left( \mathbf{x,}T\right) $ of
our model process, which is the Fourier transform of $p_{T}^{\mathbb{Q}%
}\left( \mathbf{x}\right) $ must admit an analytic extension into\
sufficiently wide strip to guarantee convergence of the pricing integral (%
\ref{price11}). Thus, we say that the model process is \textit{adapted} to
the payoff $H$ if $\mathbb{E}^{\mathbb{Q}}\left[ H\right] <\infty $. This is
a very restrictive condition on the model.

Let us discuss now customary used models in the one-dimensional situation.
Consider a common frictionless market consisting of a riskless bond and
stock which is modeled by an exponential L\'{e}vy process $S_{t}=S_{0}\exp
\left( X_{t}\right) $ under a fixed equivalent martingale measure $\mathbb{Q}
$ with a given constant riskless rate $r>0$. Observe that the idea of
modeling the option price via a log-normal distribution is due to Samuelson
\cite{sam1}. Since in our model the stock does not pay dividends then the
discounted stock price $\exp \left( -rt\right) S_{t}$ must be a martingale
under $\mathbb{Q}$. Consider a contract (European call option) which gives
to its owner the right but not the obligation to buy the underlying asset
for the fixed strike price $K$ at the specified expiry date $T$. We need to
evaluate its price $V$. In this case the payoff has the form
\begin{equation}
H(x)=(S_{0}\exp \left( x\right) -K)_{+},  \label{payoff}
\end{equation}%
where for any $a\in \mathbb{R}$, $(a)_{+}=\max \{a,0\}$, $K$ is the strike
price.

In the classical Black-Scholes model \cite{black1} the price of a stock
follows the \textit{Geometric Brownian motion} defined as $S_{t}=S_{0}\exp
(X_{t})$, where $X_{t}$, $t\geq 0$ is the Brownian motion with the \textit{%
probability density function}
\[
p_{\Delta t}(x)=\left( 2\pi \sigma ^{2}\Delta t\right) ^{-1/2}\exp \left( -%
\frac{(x-\mu \Delta t)^{2}}{2\sigma ^{2}\Delta t}\right)
\]%
for the increments $X_{t+\Delta t}-X_{t}$ and parameters $\mu $ and $\sigma $
are known as \textit{drift} and \textit{volatility} respectively \cite{bl1},
p. 2. The dynamics for stock prices are given by
\[
dS_{t}=\mu S_{t}dt+\sigma S_{t}dW_{t},
\]%
where $W_{t}$ is a standard Brownian motion. This stochastic differential
equation can be solved,
\[
S_{t}=S_{0}\exp \left( \left( \mu -\frac{\sigma ^{2}}{2}\right) t+\sigma
W_{t}\right)
\]%
and the arbitrage free price $V$ at time $t=0$ of a call option with
maturity $T$ and strike price $K$ can be expressed as%
\begin{equation}
V=S_{0}\Phi \left( b_{1}\right) -K\exp \left( -rT\right) \Phi \left(
b_{2}\right) ,  \label{bsm1}
\end{equation}%
where%
\[
b_{1}=\frac{\ln \left( S_{0}/K\right) +\left( r+\sigma ^{2}/2\right) T}{%
\sigma T^{1/2}},
\]%
\[
b_{2}=\frac{\ln \left( S_{0}/K\right) +\left( r-\sigma ^{2}/2\right) T}{%
\sigma T^{1/2}}
\]%
and $\Phi $ is the standard Normal cumulative distribution function \cite%
{black1}. In this model there exists \ a unique martingale measure $\mathbb{Q%
}$ which is given by Girsanov theorem presented in Appendix III. See \cite%
{Lamberton1}, \cite{Irle1} and \cite{Elliot1} for more information. As we
can see, only the volatility parameter $\sigma $ appears in (\ref{bsm1}) and
the drift $\mu $ term vanishes. There are two common approaches to estimate $%
\sigma $. The first is based on empirical estimation from historical data.
The stock price is observed at fixed time intervals (e.g. every day). Then
we calculate the log-returns and estimate $\sigma $ by $sa^{1/2}$, where $s$
is the standard deviation and $a$ is the number of trading days. The second
approach is connected with the so-called implied volatility, which is the
volatility of the underlying which, when substituted \ into (\ref{bsm1})
gives a theoretical price equal to the market price. This equation can be
solved numerically. If we calculate the implied volatility for different
strikes $K$ and expiration times $T$ then we find that the volatility is not
constant. The shape of the implied volatility versus $S_{T}/K$ for a fixed $%
T $ is called \textit{volatility smile}. This phenomenon is a consequence of
the fact that the Normal distribution is a poor model for the log-returns
\cite{Chen1}. Observe that if Black-Scholes's formula (\ref{bsm1}) were
correct the implied volatility would be independent on $T$ and $K$ and equal
to the historic volatility $sa^{1/2}$, which is not true in reality. During
the past decades the Black-Scholes model was increasingly criticized.
Mandelbrot was the first who presented evidence against the log-normal
distribution hypothesis \cite{man1}. Namely, he found that the empirical
distribution is more concentrated in the tails and around the origin when
compared with the Normal distribution. On the basis of these observations
Mandelbrot proposed to consider a class of a pure jump processes instead of
the continuous Brownian motion.

It is well-known that the Black-Scholes theory becomes much more efficient
if additional stochastic factors are introduced. Consequently, it is
important to consider a wider family of L\'{e}vy processes. Stable L\'{e}vy
processes have been used first in this context by Mandelbrot \cite{man1} and
Fama \cite{f11}. From the 90$^{s}$ L\'{e}vy processes became more popular
(see e.g. \cite{ms1, ms2, bp1, bl1, koponen, bn2, bn3, bn5, bn6, 31-2015}
and references therein).

There are several different ways to construct high dimensional L\'{e}vy
processes. A general method is based on a well-known L\'{e}vy-Khintchine
formula (\ref{lk}) which gives a representation of the characteristic
exponent $\mathbf{\psi }$ of any L\'{e}vy process $\mathbf{X}_{t}$ on $%
\mathbb{R}^{n}$. The characteristic function $\Phi \left( \mathbf{x}%
,t\right) $ of any L\'{e}vy process on $\mathbb{R}^{n}$ can be formally
defined as
\[
\Phi \left( \mathbf{\cdot },t\right) =\mathbb{E}\left[ \exp \left(
i\left\langle \mathbf{\cdot },\mathbf{X}_{t}\right\rangle \right) \right]
=\exp \left( -t\mathbf{\psi }\left( \mathbf{\cdot }\right) \right) ,
\]%
where $\mathbf{\psi }$ is the characteristic exponent of $\mathbf{X}_{t}$
which is uniquely determined. Then the density function $p_{t}$ can be
expressed as%
\[
p_{t}\left( \cdot \right) =\left( 2\pi \right) ^{-n}\int_{\mathbb{R}%
^{n}}\exp \left( -i\left\langle \cdot ,\mathbf{x}\right\rangle -t\mathbf{%
\psi }\left( \mathbf{x}\right) \right) d\mathbf{x.}
\]%
According to the L\'{e}vy-Khintchine formula, for any L\'{e}vy process $%
\mathbf{X}_{t}$ the characteristic exponent $\mathbf{\psi }$ admits the
representation%
\[
\mathbf{\psi }\left( \mathbf{\cdot }\right) =\left\langle \mathbf{A}\cdot ,%
\mathbf{\cdot }\right\rangle -i\left\langle \mathbf{h},\cdot \right\rangle
\]%
\begin{equation}
-\int_{\mathbb{R}^{n}}\left( 1-\exp \left( i\left\langle \cdot ,\mathbf{x}%
\right\rangle \right) -i\left\langle \cdot ,\mathbf{x}\right\rangle \chi
_{D}\left( \mathbf{x}\right) \right) \Pi \left( d\mathbf{x}\right) ,
\label{lk111}
\end{equation}%
where $\chi _{D}$ is the characteristic function of the unit ball in $%
\mathbb{R}^{n}$, $\mathbf{h\in }\mathbb{R}^{n}$, $\mathbf{A}$ is a symmetric
nonnegative-definite matrix and $\Pi \left( d\mathbf{x}\right) $ is a
measure such that%
\[
\int_{\mathbb{R}^{n}}\min \left\{ 1,\left\langle \mathbf{x,x}\right\rangle
\right\} \Pi \left( d\mathbf{x}\right) <\infty \text{, \ }\Pi \left( \left\{
\mathbf{0}\right\} \right) =0.
\]%
The triplet $\left( \mathbf{A,}\Pi ,\mathbf{h}\right) $ in (\ref{lk111}) is
called the \textit{generating triplet} (or the \textit{L\'{e}vy triplet}).
Selecting different L\'{e}vy densities $\Pi $ in the representation (\ref%
{lk111}) we get the set of characteristic exponents of L\'{e}vy processes
(see e.g. \cite{bl1}, p. 200). However, this approach is connected with
numerical computation of integrals over manifolds. For instance, a known
class of high-dimensional L\'{e}vy models is based on so-called KoBoL family
which is defined by
\[
\Pi \left( d\mathbf{x}\right) =\rho ^{-\nu -1}\exp \left( -\lambda \left(
\mathbf{\phi }\right) \rho \right) d\rho d\mathbf{\phi ,}
\]%
where $d\mathbf{\phi }$ is a normalised rotation invariant measure on $%
\mathbb{S}^{n-1}\subset \mathbb{R}^{n}$ and $\lambda $ is a continuous
positive function on $\mathbb{S}^{n-1}$. It is possible to show that the
associated characteristic exponent $\mathbf{\psi }$ has the form%
\[
\mathbf{\psi }\left( \cdot \right) =-i\left\langle \mathbf{\mu },\cdot
\right\rangle +\Gamma \left( -\nu \right) \int_{\mathbb{S}^{n-1}}\left(
\lambda ^{\nu }\left( \mathbf{\phi }\right) -\left( \lambda \left( \mathbf{%
\phi }\right) -i\left\langle \mathbf{A\xi ,\phi }\right\rangle \right) ^{\nu
}\right) d\mathbf{\phi }
\]%
if $\nu \in \left( 0,1\right) \cup \left( 0,2\right) $, $\mathbf{\mu }\in
\mathbb{R}^{n}$ and $\mathbf{A}$ is a positive-definite matrix \cite{bl1},
p. 200. Note that similar models can be obtained if instead of $\mathbb{S}%
^{n-1}\subset \mathbb{R}^{n}$ we consider a homogeneous infinitely smooth $m$%
-dimensional (in the sense of the Lebesgue-Brower dimension) Riemannian
manifold $\mathbb{M}^{m}\subset \mathbb{R}^{n}$, $m<n$. We shall not discuss
here this line of research. Observe that some very specific approaches of
modeling the dependence structure of multivariate L\'{e}vy processes were
discussed in \cite{deelstra1}.

We will adapt a general and practical approach which still allows to get
explicit approximation formulas for pricing of spread options without
involving of numerical methods. This allows application of analytic methods
in our analysis. To model return processes we introduce a class of
stochastic systems of the form
\begin{equation}
\mathbf{U}_{t}=\mathbf{X}_{t}+\mathbf{BZ}_{t},\,\,\,\mathbf{B=}\left( b_{m,k},1\leq m,k\leq
n\right) \label{system11}
\end{equation}%
where $\mathbf{X}_{t}=\left( X_{1,t},\cdot \cdot \cdot ,X_{n,t}\right) $ and
$\mathbf{Z}_{t}=\left( Z_{1,t},\cdot \cdot \cdot ,Z_{n,t}\right) $ have
independent components defined by their characteristic exponents $\psi
_{s}^{\left( 1\right) },$ $1\leq s\leq n$ and $\psi _{m}^{\left( 2\right) },$
$1\leq m\leq n$ respectively and $\mathbf{U}_{t}=\left( U_{1,t},\cdot \cdot
\cdot ,U_{n,t}\right) $. The matrix $\mathbf{B}$
reflects the dependence between the processes $U_{1,t},\cdot
\cdot \cdot ,U_{n,t}$. As a linear combination of L\'{e}vy processes $%
\mathbf{U}_{t}$ is a L\'{e}vy process (see e.g. \cite{sato}, p. 65) and
return process is%
\begin{equation}
\mathbf{S}_{t}=\left\{ S_{j,t}=S_{j,0}\exp \left( U_{j,t}\right) ,1\leq
j\leq n\right\} .  \label{stokc1}
\end{equation}%
Empirical studies show that the stock prices are highly correlated (which is
modeled by the matrix $\mathbf{B}$) if the market is in crisis (see e.g.

http://www.economicsofcrisis.com/lit.html for more information). We give an
explicit form of the characteristic function $\Phi \left( \mathbf{z}%
,t\right) $ \ of $\mathbf{U}_{t}$,%
\[
\Phi \left( z_{1},\cdot \cdot \cdot ,z_{n},t\right) =\Phi \left( \mathbf{z}%
,t\right) =\exp \left( -t\mathbf{\psi }\left( \mathbf{z}\right) \right) ,
\]%
where%
\begin{equation}
\mathbf{\psi }\left( \mathbf{z}\right) :=\sum_{s=1}^{n}\psi _{s}^{\left(
1\right) }\left( z_{s}\right) +\sum_{k=1}^{n}\psi _{m}^{\left( 2\right)
}\left( \sum_{k=1}^{n}b_{k,m}z_{k}\right) \text{.}  \label{exponent1}
\end{equation}%
We specify sufficient equivalent martingale measure conditions for our model
(\ref{stokc1}). Under the equivalent martingale measure $\mathbb{Q}$ all
assets have the same expected rate of return which is a risk free rate $r$.
This means that under no-arbitrage conditions the risk preferences of
investors acting on the market do not enter into valuation decisions. It is
known \cite{DelSch} that the existence of equivalent martingale measure $%
\mathbb{Q}$ is equivalent to the no-arbitrage condition. Remark that $%
\mathbb{Q}$ is absolutely continuous with respect to $\mathbb{P}$, the
historic measure inferred from the observations of returns (see Appendix I
for details). We show that under equivalent martingale measure condition $%
\mathbf{\psi }$ must satisfy the condition $\mathbf{\psi }^{\mathbb{Q}%
}\left( -i\mathbf{e}_{s}\right) =-r$, $1\leq s\leq n$, where $\left\{
\mathbf{e}_{s},1\leq s\leq n\right\} $ is the standard basis in $\mathbb{R}%
^{n}$. In general, $\mathbb{Q}$ is not unique. Moreover, the class of
equivalent martingale measures is sufficiently large to generate option
prices from some dense set of an interval which depends on model parameters.
One mathematically tractable choice is the so-called Esscher equivalent
measure (see Appendix III for more information). We assume that $\mathbb{Q}$
has been fixed and all expectations have been calculated with respect to
this measure. Also, we shall not be concerned here with the problem of model
calibration (see \cite{carmona} for more information). One-dimensional
characteristic exponents $\psi _{s}^{\left( 1\right) }$ and $\psi
_{m}^{\left( 2\right) }$ in (\ref{exponent1}) are building blocks of our
model. Selecting different $\psi _{s}^{\left( 1\right) }$ and $\psi
_{m}^{\left( 2\right) }$ and $\mathbf{B=}\left( b_{m,k}\right) $ in (\ref%
{exponent1}) we get a wide range of high-dimensional jump-diffusion models.

As a motivating example we consider a popular among practitioners class of
models, so-called KoBoL family. Characteristic exponents $\psi $ of such
models have been considered in \cite{bl1, bl22, s1, s2, s3, my-phd} and can
be obtained directly from the one-dimensional L\'{e}vy-Khintchine formula (%
\ref{lk111}),
\[
\psi \left( \xi \right) =-i\mu \xi +c_{-}\Gamma \left( -\nu \right) \left(
\left( -\lambda _{-}\right) ^{\nu }-\left( -\lambda _{-}-i\xi \right) ^{\nu
}\right)
\]%
\begin{equation}
+c_{+}\Gamma \left( -\nu \right) \left( \lambda _{+}^{\nu }-\left( \lambda
_{+}+i\xi \right) ^{\nu }\right) ,  \label{exp-repr}
\end{equation}%
where $\nu \in \left( 0,1\right) $, $\mu \in \mathbb{R}$, $c_{+},c_{-}>0$, $%
\lambda _{-}<0<\lambda _{+}$ are one-dimensional model parameters. Observe
that the parameters $\left( \nu ,\mu ,c_{+},c_{-},\lambda _{+},\lambda
_{-}\right) $ determine the probability density. Larger values of $\nu $ and
$c_{+}$, $c_{-}$ produce a larger peak of the probability distribution while
the parameters $c_{+}$, $c_{-}$ control asymmetry and $\lambda _{-}$, $%
\lambda _{+}$ determine the rate of exponential decay as $\left\vert \xi
\right\vert \rightarrow \infty $. For our applications is sufficient to
notice that the function $\psi \left( \xi \right) $ defined by (\ref%
{exp-repr}) is analytic in the domain $\mathbb{C\setminus }\left\{ \left(
-i\infty ,i\lambda _{-}\right] \cup \left[ i\lambda _{+},+i\infty \right)
\right\} $ and%
\[
\left\vert \Phi \left( \xi ,t\right) \right\vert =\left\vert \exp \left(
-t\psi \left( \xi \right) \right) \right\vert \asymp \exp \left(
-C\left\vert \xi \right\vert ^{\nu }\right) ,
\]%
as $\left\vert \xi \right\vert \rightarrow \infty $, $\left\vert {\rm Im}%
\xi \right\vert \in \left( \lambda _{-},\lambda _{+}\right) $, $\nu \in
\left( 0,1/2\right) $. Here $C>0$ is an absolute constant since $\Gamma
\left( -\nu \right) <0$ and $\cos \left( \nu \pi /2\right) >0$ if $\nu \in
\left( 0,1\right) $ and $c_{+},c_{-}>0$. Hence applying Cauchy theorem in
the strip $\kappa _{-}\leq {\rm Im}\xi \leq \kappa _{+}$, $\lambda
_{-}<\kappa _{-}<0<\kappa _{+}<\lambda _{+}$ we get%
\begin{equation}
p_{t}^{\mathbb{Q}}\left( x\right) =\mathrm{M}\left( x\right) \mathrm{N}%
\left( x,t\right) ,  \label{pttt}
\end{equation}%
where%
\[
\mathrm{M}\left( x\right) :=\frac{1}{2\pi \left( \exp \left( \kappa
_{-}x\right) +\exp \left( \kappa _{+}x\right) \right) }
\]%
and
\[
\mathrm{N}\left( x,t\right) :=\int_{\mathbb{R}}\exp \left( -ix\xi \right)
\left( \Phi ^{\mathbb{Q}}\left( \xi -i\kappa _{-},t\right) +\Phi ^{\mathbb{Q}%
}\left( \xi -i\kappa _{+},t\right) \right) d\xi
\]%
is a bounded function on $\mathbb{R}$. Observe if $\int_{\mathbb{R}}M\left(
x\right) H\left( x\right) dx<\infty $ then our model process is adapted to
the reward function $H$. In particular, if $H$ is European call reward
function (\ref{payoff}) then $H\left( x\right) \sim \exp \left( x\right) $,
as $x\rightarrow \infty $ and we should assume $\lambda _{+}>1$ to guarantee
convergence of pricing integral (\ref{price11}). In general, if $\Phi ^{%
\mathbb{Q}}\left( \xi ,t\right) $, $\left( \xi ,t\right) \in \mathbb{R\times
R}_{+}$ does not admit analytic extension with respect to $\xi $ then we may
apply stationary phase approximation to establish asymptotic for $p_{t}^{%
\mathbb{Q}}\left( x\right) $ as $x\rightarrow \infty $ to select admissible
reward functions. In the multidimensional settings we have a similar
situation. Assume, for simplicity, that characteristic function $\Phi ^{%
\mathbb{Q}}\left( \mathbf{z},t\right) $, $\left( \mathbf{z},t\right) \in
\mathbb{R}^{n}\mathbb{\times R}_{+}$ admits analytic extension with respect
to each variable $z_{k}$, $1\leq k\leq n$ into the strips
$\left\vert {\rm Im}z_{k}\right\vert \in \left[ -b_{k},b_{k}\right] $, where $%
\lim_{\left\vert z_{k}\right\vert \rightarrow \infty }\left\vert \Phi ^{%
\mathbb{Q}}\left( z_{1},\cdot \cdot \cdot ,z_{n},t\right) \right\vert =0$, $%
1\leq k\leq n$. Then we show that
\[
p_{t}^{\mathbb{Q}}\left( \mathbf{x}\right) =\mathrm{M}\left( \mathbf{x}%
\right) \mathrm{N}\left( \mathbf{x},t\right) ,
\]%
where%
\begin{equation}
\mathrm{M}\left( \mathbf{x}\right) =2^{-2n}\pi ^{-n}\left(
\prod_{k=1}^{n}\cosh \left( b_{k}x_{k}\right) \right) ^{-1}
\label{pois}
\end{equation}%
and $\mathrm{N}\left( \mathbf{x},t\right) $ is a bounded function,
\[
\mathrm{N}\left( \mathbf{x},t\right) =\int_{\mathbb{R}^{n}}\exp \left(
-i\left\langle \mathbf{x,z}\right\rangle \right) \Phi _{n}^{\mathbb{Q}%
}\left( \mathbf{z},t\right) d\mathbf{z,}
\]%
where the function $\Phi _{n}^{\mathbb{Q}}\left( \mathbf{z},t\right) $ is
defined as,
\[
\Phi _{1}^{\mathbb{Q}}\left( \mathbf{z},t\right) :=\Phi ^{\mathbb{Q}}\left(
\mathbf{z+}i\mathbf{e}_{1}b_{1},t\right) +\Phi ^{\mathbb{Q}}\left( \mathbf{z-%
}i\mathbf{e}_{1}b_{1},t\right) ,
\]%
\[
\Phi _{k}^{\mathbb{Q}}\left( \mathbf{z},t\right) :=\Phi _{k-1}^{\mathbb{Q}%
}\left( \mathbf{z+}i\mathbf{e}_{k}b_{k},t\right) +\Phi _{k-1}^{\mathbb{Q}%
}\left( \mathbf{z-}i\mathbf{e}_{k}b_{k},t\right) \text{, }2\leq k\leq n\text{%
.}
\]%
Our method of reconstruction of density functions $p_{t}^{\mathbb{Q}}$ is
based on the Poisson summation formula justified by (\ref{pois}). Let $P$ be
a truncation parameter. Application of the Poisson summation allows us to
construct a periodic extension
\begin{equation}
\widetilde{p}_{t}^{\mathbb{Q}}\left( \mathbf{x}\right) \approx \sum_{\mathbf{%
m}\in \mathbb{Z}^{n}}\Phi ^{\mathbb{Q}}\left( -\frac{2\pi }{P}\mathbf{m}%
,t\right) \exp \left( \frac{2\pi i}{P}\left\langle \mathbf{m,x}\right\rangle
\right)  \label{series111}
\end{equation}%
of $p_{t}^{\mathbb{Q}}$ of the same smoothness as $p_{t}^{\mathbb{Q}}$.
Observe that the characteristic function $\Phi ^{\mathbb{Q}}$ is known
explicitly and $\Phi ^{\mathbb{Q}}\left( \mathbf{x}\right) $ decays
exponentially fast as $\left\vert \mathbf{x}\right\vert \rightarrow \infty $
in many cases of practical interest. Hence the series (\ref{series111})
converges absolutely and represents an infinitely differentiable function on
$2^{-1}PQ_{n}$, where $Q_{n}$ is the unit cube in $\mathbb{R}^{n}$. For
example, if the characteristic exponent $\mathbf{\psi }$ is defined by (\ref%
{exponent1}) then $\left\vert p_{t}^{\mathbb{Q}}\left( \mathbf{x}\right) -%
\widetilde{p}_{t}^{\mathbb{Q}}\left( \mathbf{x}\right) \right\vert \ll \exp
\left( -2^{-1}Pb\right) $, $P\rightarrow \infty $ for any $\mathbf{x}\in
2^{-1}PQ_{n}$, where $b$ is a model parameter. Next, we approximate $%
\widetilde{p}_{t}^{\mathbb{Q}}$ by the Fourier projection with the spectrum
in the domain $\Omega _{1/R}^{^{\prime }}\subset \mathbb{R}^{n}$ defined in
the Theorem 25. Observe that $\Omega _{1/R}^{^{\prime }}$ has the
shape of an exponential hyperbolic cross whose shape depends on model
parameters. We show that $\Omega _{1/R}^{^{\prime }}$ contains $m\asymp
P^{n}\left( \ln R\right) ^{\nu }$, $R\rightarrow \infty $ points with
integer components, where $\nu $ is a model parameter and
\[
\left\vert \widetilde{p}_{t}^{\mathbb{Q}}\left( \mathbf{x}\right) -\sum_{%
\mathbf{m}\in \mathbb{Z}^{n}\cap \Omega _{1/R}^{^{\prime }}}\Phi ^{\mathbb{Q}%
}\left( -\frac{2\pi }{P}\mathbf{m},t\right) \exp \left( \frac{2\pi i}{P}%
\left\langle \mathbf{m,x}\right\rangle \right) \right\vert
\]%
\[
\ll \left( mP^{-n}\right) ^{1-\nu ^{-1}}\exp \left( -\left( mP^{-n}\right)
^{\nu ^{-1}}\right) ,\text{ }m,P\rightarrow \infty
\]%
for any $\mathbf{x}\in 2^{-1}PQ_{n}$.

We give a detailed treatment of the problem of comparison of numerical
methods. To show that the method of approximation given by (\ref{series111})
is an optimal in the exponential scale we use $m$-widths instead of a
commonly used tabulation approach. This allows us to compare a wide range of
methods of approximation and reconstruction (including nonlinear). All
technical details are presented in Section 5.4 and Appendix IV.

Applying this approach for any concrete model process we can construct
almost optimal method of recovery of $\widetilde{p}_{t}^{\mathbb{Q}}\left(
\mathbf{x}\right) $ (which reflects the course of dimensionality).

Final Chapter 6 deals with the problem of option pricing. We give a detailed
proof of Hurd-Zhou theorem which is important in our applications. On the
basis of this theorem and results developed in Chapter 5 we construct
explicit approximation formulas for the price $V$ of a spread option given
by (\ref{price11}) in the case when all one-dimensional L\'{e}vy processes $%
\psi _{s}^{\left( 1\right) }$ and $\psi _{m}^{\left( 2\right) }$, which
define the characteristic exponent $\mathbf{\psi }\left( \mathbf{v}\right) $
in (\ref{exponent1}), are KoBoL processes. Theorem 36
indicates
an exponential rate of convergence of approximation formulas presented. A
similar analysis is applicable for general jump-diffusion models.

In the Appendices I-IV we collected all necessary results which we use in
the text. In Appendix I we introduce $L_{p}$ spaces, present Fubini and
Radon-Nikodym theorems which are important in our applications. In Appendix
II we collect fundamental facts from Harmonic Analysis which are useful in
Pricing Theory, such as Plancherel, Riesz and Riesz-Thorin theorems.
Appendix III introduces martingales and presents two results on martingale
conversion, the Doob-Meyer decomposition and Girsanov's theorem which are
important ingredients of the Theory of Pricing. Also, we present basic
properties of the set of equivalent martingale measures. Appendix IV
contains results on optimal approximation which are important in comparison
of numerical algorithms.

The results obtained have been presented and discussed on Applied and
Financial Mathematics seminars of the Department of Mathematics, University
of Leicester 2010-2014, International Workshop-Radial Basis Functions, 2014,
Birkbeck College, University of London, seminar of the Department of
Economics, Mathematics and Statistics, Birkbeck College, University of
London, 2014, Actuarial Teachers and Researchers Conference 2012, European
Numerical Mathematics and Applications-2011, Leicester, Festivals of PhD
students, University of Leicester, 2011 and 2012, British Mathematical
Colloquium-2011, Leicester, and many other National and International
meetings. I\ thank all the participants of these meetings for providing me
with opportunities to talk on my research and to learn from their talks.

This book may be considered as a research report mostly based on results of
the author and his colleagues. We review the basic material which is needed
and give proofs of new results and of assertions not available in relevant
books. In this sense we have tried to present a self-contained treatment,
accessible to non-specialists. We hope that the book may now be reasonably
free from error, in spite of the mass detail which it contains.

\bigskip

\bigskip

\chapter{\protect\bigskip Remarks on notation}


\bigskip

$\mathbb{N}$, $\mathbb{Z}$, $\mathbb{R}$ and $\mathbb{C}$ are, respectively,
the sets of all positive integers, all integers, all real numbers, and all
complex numbers. $\mathbb{Z}_{+}$ and $\mathbb{R}_{+}$ are the collections
of nonnegative elements of $\mathbb{Z}$ and $\mathbb{R},$ respectively. $%
\mathbb{R}^{n}$ is the $n$-dimensional Euclidean space with the canonical
basis $\mathbf{e}_{1},\cdot \cdot \cdot ,\mathbf{e}_{n}$. Its elements $%
\mathbf{x}=\left( x_{1},\cdot \cdot \cdot ,x_{n}\right) $ and $\mathbf{y}%
=\left( y_{1},\cdot \cdot \cdot ,y_{n}\right) $ are vectors with $n$ real
components. The inner product in $\mathbb{R}^{n}$ is $\left\langle \mathbf{%
x,y}\right\rangle =\sum_{j=1}^{n}x_{j}y_{j}$; the norm is $\left\vert
\mathbf{x}\right\vert =\left( \sum_{j=1}^{n}x_{j}^{2}\right) ^{1/2}$. \

$\mathbb{C}^{n}$ is the $n$-dimensional complex space. Its elements $\mathbf{%
z}=\left( z_{1},\cdot \cdot \cdot ,z_{n}\right) $ are vectors with $n$
complex components. Similarly we define $\mathbb{N}^{n}$ and $\mathbb{Z}^{n}$%
.

For a matrix $\mathbf{A}=(a_{j,k})$, $\mathbf{A}^{\mathrm{T}}=(a_{k,j})$
means its transpose.

Let $X$ be a vector space over reals. Let $x_{1},\cdot \cdot \cdot \cdot
,x_{m}\in X$. By $\mathrm{lin}\left\{ x_{1},\cdot \cdot \cdot \cdot
,x_{m}\right\} $ and \textrm{aff}$\left\{ x_{1},\cdot \cdot \cdot \cdot
,x_{m}\right\} $ we denote the linear span and affine combination of $%
x_{1},\cdot \cdot \cdot \cdot ,x_{m}$ respectively. $\mathrm{lin}\left\{
A,B\right\} $ means the linear span of $A,B\subset X$.

For $A,B\subset X$, $\mathbf{z}\in X$, and $c\in \mathbb{R}$, $A+\mathbf{z}%
=\left\{ \mathbf{x+z}\left\vert \mathbf{x}\in A\right. \right\} $, $A-%
\mathbf{z}=\left\{ \mathbf{x-z}\left\vert \mathbf{x}\in A\right. \right\} $,
$cA=\left\{ c\mathbf{x}\left\vert \mathbf{x}\in A\right. \right\} $, $%
-A=\left\{ -\mathbf{x}\left\vert \mathbf{x}\in A\right. \right\} $, $%
A\setminus B=\left\{ \mathbf{x}\left\vert \mathbf{x}\in A\&\mathbf{x}\notin
B\right. \right\} $. Minkovski's sum and difference of $A\subset X$ and $%
B\subset X$ are defined as%
\begin{equation}
A+B=\left\{ \mathbf{x+y}\left\vert \mathbf{x}\in A,\mathbf{y}\in B\right.
\right\}  \label{minkowski}
\end{equation}%
and $A-B=\left\{ \mathbf{x-y}\left\vert \mathbf{x}\in A,\mathbf{y}\in
B\right. \right\} $ respectively. For sets $A$ and $B$, $A\times B$ denotes
the Cartesian product.

Let $\left( X,\vartheta \right) $ be a metric space. The open ball $B\left(
x,r\right) $ of radius $r>0$ abot $x\in X$ is the set $B\left( x,r\right)
=\left\{ y\in X\left\vert \vartheta \left( x,y\right) <r\right. \right\} $.
A subset $U\subset X$ is called open if for every $x\in U$ there exists an $%
r>0$ such that $B\left( x,r\right) \subset U$. The complement $X\diagdown U$
of an open set $U$ is called closed. The interior, the closure and the
boundary of a set $U\subset X$ are denoted by $\mathrm{int}U$, $\overline{U}$
and $\partial U$ respectively.

Let $(\Omega ,\mathcal{F},\upsilon )$ be a measure space. If $\mathcal{F}$
is the Lebesgue $\sigma $-algebra $\mathcal{L}$ then we write $(\Omega ,%
\mathcal{L},\upsilon )$. $\mathrm{Vol}_{n}\left( B\right) $ is the Lebesgue
measure of a set $B\subset \mathbb{R}^{n}$. $\chi _{B}$ is the indicator
function of a set $B$, that is, $\chi _{B}\left( \mathbf{x}\right) =1$ for $%
\mathbf{x}\in B$ and $0$ for $\mathbf{x}\notin B$.

The abbreviation a.s. denotes almost surely, that is, with \ probability $1$%
. The abbreviation a.e. denotes almost everywhere, or almost surely, with
respect to the Lebesgue measure. Similarly, $\upsilon $-a.e. denotes almost
everywhere, or almost every, with respect to a measure $\upsilon $.

The symbol $\delta _{\mathbf{a}}$ represents the probability measure
concentrated at $\mathbf{a\in }\mathbb{R}^{n}$. If $\mathbf{a}=\mathbf{0}$
then we shall write $\delta _{\mathbf{a}}=\delta $. The expression $\upsilon
_{1}\ast \upsilon _{2}$ represents the convolution of finite measures $%
\upsilon _{1}$ and $\upsilon _{2}$; $\upsilon ^{\left( m\right) }$ is the $m$%
-fold convolution of $\upsilon $. When $m=0$, $\upsilon ^{\left( m\right) }$
is understood to be $\delta _{\mathbf{0}}$. $C(\mathbb{R}^{n})$ be the space
of continuous functions on $\mathbb{R}^{n}$ and $L_{p}(\mathbb{R}^{n})$ be
the usual space of $p$-integrable functions $f:\mathbb{R}^{n}\mapsto \mathbb{%
R}$ (or $f:\mathbb{R}^{n}\mapsto \mathbb{C}$) equipped with the norm
\[
\Vert f\Vert _{p}=\Vert f\Vert _{L_{p}(\mathbb{R}^{n})}:=\left\{
\begin{array}{cc}
\left( \int_{\mathbb{R}^{n}}\left\vert f(\mathbf{x})\right\vert ^{p}d\mathbf{%
x}\right) ^{1/p}, & 1\leq p<\infty , \\
\mathrm{ess}\,\,\sup_{\mathbf{x}\in \mathbb{R}^{n}}|f(\mathbf{x})|, &
p=\infty .%
\end{array}%
\right.
\]%
Let $f:\mathbb{R}^{n}\rightarrow \mathbb{R}$ be an integrable function, $%
f\in L_{1}\left( \mathbb{R}^{n}\right) $. Define the Fourier transform
\[
\mathbf{F}\left( f\right) \left( \mathbf{y}\right) =\int_{\mathbb{R}%
^{n}}\exp \left( -i\left\langle \mathbf{x,y}\right\rangle \right) f\left(
\mathbf{x}\right) d\mathbf{x}
\]%
and its formal inverse as
\[
\mathbf{F}^{-1}\left( f\right) \left( \mathbf{y}\right) =\left( 2\pi \right)
^{-n}\int_{\mathbb{R}^{n}}\exp \left( i\left\langle \mathbf{x,y}%
\right\rangle \right) f\left( \mathbf{x}\right) d\mathbf{x.}
\]

$\mathbb{P}\left( A\right) $ is the probability of an event $A$. $\mathbb{E}%
\left[ X\right] $ is the expectation of a random variable $X$.

$\mathbf{I}$ is the identity matrix. $\mathbf{A}^{\mathrm{T}}$ and $\mathbf{A%
}^{\ast }$ are, respectively, the transpose and the conjugate of a matrix $%
\mathbf{A}$.

Let $X$ be a Banach space and $f$ be a function, $f\in X$. The notation $%
\left\Vert f\left( \cdot ,\mathbf{\alpha }\right) \right\Vert _{X}$ means
that we are taking the norm of $f\left( \cdot ,\mathbf{\alpha }\right) $
with respect to the argument denoted by $\left( \cdot \right) $. Let $X$ and
$Y$ be Banach spaces. The norm $\left\Vert \boldsymbol{A}\right\Vert :=\sup
\left\{ \left\Vert \boldsymbol{A}x\right\Vert _{Y}\left\vert \left\Vert
x\right\Vert _{X}\leq 1\right. \right\} $ of a linear operator $\boldsymbol{A%
}:X\rightarrow Y$ is denoted by $\left\Vert \boldsymbol{A}\left\vert
X\rightarrow Y\right. \right\Vert $ and the space of bounded linear
operators $\boldsymbol{A}$ is denoted by $\mathrm{L}\left( X,Y\right) $. Let
$X$, $Y$ and $Z$ be Banach spaces $\boldsymbol{A}\in \mathrm{L}\left(
X,Y\right) $ and $\boldsymbol{B}\in \mathrm{L}\left( Y,Z\right) $ then the
composition of $\boldsymbol{A}$ and $\boldsymbol{B}$ is denoted by $%
\boldsymbol{B}\circ \boldsymbol{A}:X\rightarrow Z$.

The expression $f\left( x\right) \sim g\left( x\right) $ means that $%
\lim_{x\rightarrow \infty }f\left( x\right) /g\left( x\right) =1$. We shall
write $f\left( x\right) \lesssim g\left( x\right) $ if $\lim_{x\rightarrow
\infty }f\left( x\right) /g\left( x\right) \leq 1$ and $A\approx B_{n}$, $%
n\in \mathbb{N}$ if $B_{n\text{ }}$ is a sequence of formal approximants to $%
A$ without regard to any type of convergence.

Different positive universal constants are mostly denoted by the letter $C$.
We did not carefully distinguish between the different constants, neither \
did we try to get good estimates for them. The same letter will be used to
denote different universal constants. For the easy of notation we put $%
a_{m}\gg b_{m}$ for two sequences, if $a_{m}\geq Cb_{m}$ and $a_{m}\asymp
b_{m}$ if $C_{1}b_{m}\leq a_{m}\leq C_{2}b_{m}$ for all $m\in \mathbb{N}$
and some constants $C$, $C_{1}$, and $C_{2}$. Through the text $\left[ a%
\right] $ means integer part of $a\in \mathbb{R}$.

\bigskip


\bigskip

\chapter{\protect\bigskip General definitions}

\label{general definitions} 

\section{Market and derivative instruments}

A \textit{market }is a system of institutions, procedures, social relations
and infrastructures where parties engage in exchange. \textit{Market
participants} consist of all the \textit{buyers} and \textit{sellers} of a
\textit{good} who influence its \textit{price}. A market allows any tradable
item to be evaluated and priced. In general, the structure of a
well-functioning market can be approximated as following:

\begin{enumerate}
\item Many small buyers and sellers.

\item Buyers and sellers have equal access to information.

\item Products are comparable.
\end{enumerate}

An \textit{investor} is someone who puts money into something with the
expectation of a financial return. \textit{Assets} are economic resources,
i.e. value of ownership which has a positive economic value and that can be
converted into cash. \textit{Finance} is the study of how investors allocate
their assets over time under conditions of certainty and uncertainty. A
\textit{derivative instrument} is a contract between two parties that
specifies conditions under which payments are to be made between the
parties. We say that a financial contract is a \textit{derivative security}
(or a \textit{contingent claim}) if its value at expiration date $T$ is
determined exactly by the market price of the underlying cash instrument at
time $0$. An \textit{option} (in finance) is a derivative instrument that
specifies a contract between two parties for a future transaction on an
\textit{asset} (commonly a \textit{stock}, a \textit{bond}, a currency or a
futures contract) at a reference price (the \textit{strike}). A \textit{stock%
} represents the original capital invested in the business by its founders.
A \textit{bond} is a negotiable certificate that acknowledges the
indebtedness of the bond issuer to the holder. A \textit{forward contract}
is an obligation to buy (or sell) an underlying asset at a fixed price (%
\textit{forward price}) on a known date $T$. A \textit{European call option}
on a security $S_{t}$ is the right to buy the security at a fixed strike
price $K$ at the expiration date $T$. The call option can be purchased for a
price $C_{t}$ (called the premium) at time $t<T$. A \textit{European put
option} gives the owner the right to sell an asset at a specified price at
expiration $T$. Instead, \textit{American options} can be exercised at any
time \ $t$, $0<t\leq T.$ Before the option is first written at time $t$, its
value $C_{t}$ is unknown. That is why it is important to get some estimates
of what this price will be if the option is written. Hence, the problem is
to get a good approximation for $C_{t}$ as a function of the underlying
assets price and the relevant market parameters. The \textit{bid-ask spread}
is the difference between the bid and ask price.

To simplify our model we assume that our market is such that:

\begin{enumerate}
\item There are no commissions and fees (the price of an asset in trade is
much bigger then commissions and fees).

\item The bid-ask spreads on $S_{t}$ and $C_{t}$ are zero (the market is in
equilibrium).
\end{enumerate}

With these assumptions we have the following two possibilities. If $%
S_{T}\leq K$ (the option is out-of-money) then the option will have no
value. Hence, $C_{T}=0$. Otherwise, if $S_{T}>K$ (the option is in-the
money) then (since by our assumption, there are no commissions and bid-ask
spreads) the net profit will be $C_{T}=S_{T}-K>0$. Joining these
possibilities we get $C_{T}=\max \left\{ S_{T}-K,0\right\} =\left(
S_{T}-K\right) _{+},$ where
\[
\left( a\right) _{+}:=\left\{
\begin{array}{cc}
a, & a>0, \\
0, & a\leq 0.%
\end{array}%
\right.
\]

A state of nature is said to be \textit{insurable state} when there exists a
portfolio which has a non-zero return in that state. For a market where
every state is insurable, a price vector can be uniquely determined. Hence,
a \textit{complete market} can be defined as a market in which all the
contingent claims are attainable.

A complete market can be defined with respect to the concept of a viable
financial market. If any strategy which is implemented at the initial time
with a zero cost has a zero terminal payoff then we have the absence of
\textit{riskless arbitrage opportunities}. A \textit{viable financial market}
is defined as a market where there is no profitable riskless arbitrage
opportunities. Note that there is an important relationship between
arbitrage and the martingale property of securities prices. It means that
the best estimation of the future price is derived from the latest
information, i.e only the most recent information matters. A financial
market is \textit{viable} iff there is a probability $\mathbb{Q}$ which is
equivalent to a historical probability $\mathbb{P}$, under which the
discounted asset prices have the martingale property. We say that a viable
market is \textit{complete} iff there is such a probability $\mathbb{Q}$.

\section{The time value of money}

The \textit{time value of money} is one of the central concepts in finance
theory which states that a unit of currency today is worth more than the
same unit of currency tomorrow due to its potential earning capacity. In
other words, $\pounds 1$ paid now is worth more than $\pounds 1$ paid in a
year because by depositing $\pounds 1$ in the bank today, one gets more then
a pound in a year. \textit{Present value} (or present discounted value) is a
future value of an asset that has been discounted to reflect its value
today. Similarly, \textit{future value} is the value of an asset in the
future which is equivalent to a specified sum at present. For a fixed time
period $[T_{1},T_{2}]$, \textit{interest} is the additional gain between the
beginning $T_{1}$ and the end $T_{2}$ of the time period. Present value $P$
of a future sum $F$ can be obtained using \textit{continuous compound
interest rate} $r$ as $P=\exp \left( -r\left( T_{2}-T_{1}\right) \right) F$.
In general, if $r=r\left( t\right) $ is a function of $t$, then
\[
P=F\exp \left( -\int_{T_{1}}^{T_{2}}r(t)dt\right) .
\]


\section{Arbitrage theorem}

All known methods of pricing derivatives employ the notion of \textit{%
arbitrage}. An \textit{arbitrage} can be defined as a way to make guaranteed
profit from nothing by selling an asset at time $T_{1}$ and then settling
accounts at $T_{2}$. An existence of arbitrage provides an investment
opportunity with infinite rate of return. Hence, investors would try to use
arbitrage to make money without putting up anything at time $T_{1}$.
Consequently, to eliminate this possibility we need to introduce so-called
\textit{Efficient Market Hypothesis} which are essentially are:

\begin{enumerate}
\item All known information is reflected on prices of all securities.

\item The current prices are the best estimates of the values of securities.

\item The prices will instantaneously adjust according to any new
information.

\item An investor cannot outperform the market price using all known
information.
\end{enumerate}

To give an analytic definition of an arbitrage consider a simple model with
two time points $T_{1}$ and $T_{2}$, $T_{1}<T_{2}$ and zero interest. Let $a$
be the value of $S(T)$ at $T_{2}$ with probability $p$ and $b$ be the value
of $S(T)$ at $T_{2}$ with probability $1-p$, $a<b$. By this way we specify $%
\mathbb{P}$ on $(\Omega ,\mathcal{F})$, where $\Omega =\left\{ a,b\right\} $
and $\mathcal{F}=\left\{ \varnothing \text{, }\left\{ a\right\} \text{, }%
\left\{ b\right\} \text{, }\left\{ a,b\right\} \right\} $. Hence, we get a
probability space $(\Omega ,\mathcal{F},\mathbb{P})$. Consider a portfolio $%
(N,MS(T))$ consisting of $N$ units of money and $M$ units of stocks. The
value $V\left( T\right) $ of this portfolio at $T_{1}$ is $V\left(
T_{1}\right) =N+MS(T_{1})$ and at $T_{2}$ is $V\left( T_{2}\right)
=N+MS(T_{2})$. We say that there exists an \textit{arbitrage opportunity} if
there exists a portfolio $(N,MS(T))$ such that $V\left( T_{1}\right) =0$, $%
V\left( T_{2}\right) \geq 0$ and $\mathbb{P}\left( V\left( T_{2}\right)
>0\right) >0$. It is possible to show that there exist no arbitrage
opportunities iff $a<S(T_{1})<b$ \cite{Alison}.

{\bf Theorem 1}
\begin{em}
\label{theorem of asset pricing} (Fundamental theorem of asset pricing)
There exist no arbitrage opportunities iff there exist a probability measure
$\mathbb{Q}$ equivalent to the original probability measure $\mathbb{P}$
such that the stock price process $(S(T_{1}),S(T_{2}))$ satisfies $\mathbb{E}%
^{\mathbb{Q}}\left[ S(T_{2})\left\vert S(T_{1})\right. \right] =S(T_{1}).$
\end{em}

A probability measure $\mathbb{Q}$ is called an \textit{equivalent
martingale measure}. Observe that Theorem 1
explicitly relates a fundamental notion of arbitrage to a far advanced
theory of martingales. In the case of multi-period model we have a similar
result \cite{Harry van Zanten}.

{\bf Theorem 2}
\begin{em}
There are no arbitrage opportunities in the multi-period model iff for every
$t$, the one-period model $(S_{t}, S_{t+1})$, with respect to the filtration
$(\mathcal{F}_{t},\mathcal{F}_{t+1})$, admits no arbitrage opportunities.
\end{em}

See Appendix III for more information. \ Consider the case of
continuous-time settings (see, e.g. \cite{Harry van Zanten} ).

{\bf Definition 3}
\begin{em}
A probability measure $\mathbb{Q}$ on a measure space $(\Omega ,\mathcal{F})$
is called \textit{equivalent martingale measure} if it is equivalent to $%
\mathbb{P}$ and $S_{t}$ is a martingale with respect to $\mathbb{Q}$. The
collection of all equivalent martingale measures on the measure space $%
(\Omega ,\mathcal{F})$ is denoted by $\mathcal{M}_{S}(\Omega ,\mathcal{F})$.
\end{em}

The change of measure spaces $(\Omega ,\mathcal{F},\mathbb{P})\rightarrow
(\Omega ,\mathcal{F},\mathbb{Q})$ is based on the Radon-Nikodim theorem (see
Appendix I, Theorem 41

{\bf Theorem 4}
\begin{em}
There are no arbitrage opportunities iff there exists an equivalent
martingale measure.
\end{em}

The proof of this statement is based on the Hahn-Banach theorem for locally
convex topological vector spaces and Banach-Alaoglu theorem which we shall
not discuss here.

If a probability measure $\mathbb{P}$ is estimated using historical return
data for the underlying stock, the measure is referred to as the market
measure (or the physical measure, or historical measure). Asset prices are
modeled by stochastic processes $(S_{t})_{t>0}$ whose evolutions are
determined by a fixed probability measure. In the theory of arbitrage
pricing there exists a risk neutral probability measure under which asset
prices are arbitrage free. The absence of arbitrage is equivalent to the
existence of a risk neutral equivalent martingale measure $\mathbb{Q}$ for $%
(S_{t})_{t>0}$ making the underlying process become a martingale. Under the
equivalent martingale measure all assets have the same expected rate of
return which is the risk free rate. It means that under no-arbitrage
conditions the risk preferences of investors acting on the market do not
enter into valuation decisions \cite{Delbaen-Schachermayer}. For a general
overview on financial derivatives from a mathematical and an economic point
of view we refer to \cite{Hull, Briys, Alison, Elliot1}.

\bigskip

\chapter{L\'{e}vy processes and characteristic exponents}

\label{Levy}


\section{Introduction}

\label{Introduction-Levy processes and characteristic exponents}

\bigskip

In this section we present important for our applications properties of L%
\'{e}vy processes on $\mathbb{R}^{n}$. We introduce a class of stochastic
systems to model return processes which will be studied in the later
chapters and develop sufficient equivalent martingale measure conditions for
such kind of models. The building blocks of our model are one-dimensional
processes. To make our results more specific, we assume that all
one-dimensional components are KoBoL processes with different parameters.
For such kind of processes we give a complete proof for the representation
of characteristic exponent for a particular choice of parameters. We
introduce the notion of $\left( \lambda _{-},\lambda _{+}\right) $%
-analyticity which is a useful tool in study density functions. It is shown
that any KoBoL process of order $\nu \in \left( 0,1/2\right) $ is $\left(
0,\lambda _{+}\right) $-analytic. It allows us to consider a general class
of contour deformations in representations of density functions. Observe
that a very specific class of contour deformations has been considered in
\cite{bl22}.

\bigskip

\bigskip

\bigskip


\section{\protect\bigskip Basic results}

We start with basic definitions and results. A \textit{probability space} $%
\left( \Omega ,\mathcal{F},\mathbb{P}\right) $ is \ a triplet of a set $%
\Omega $, an admissible family $\mathcal{F}$ of subsets, $\mathcal{F}\subset
\left\{ \emptyset ,2^{\Omega }\right\} $ and a mapping $\mathbb{P}:\mathcal{F%
}\longrightarrow \left[ 0,1\right] $ such that

\begin{enumerate}
\item $\Omega \in \mathcal{F}$ and $\emptyset \in \mathcal{F}.$

\item If $A_{n}\in \mathcal{F}$ for any $n\in \mathbb{N},$ then
\[
\bigcup_{n=1}^{\infty }A_{n}\in \mathcal{F}\text{, }%
\bigcap_{n=1}^{\infty }A_{n}\in \mathcal{F}.
\]

\item If $A\in \mathcal{F}$, then $A^{c}\in \mathcal{F}.$

\item $0\leq \mathbb{P}\left( A\right) \leq 1,$ $\mathbb{P}\left( \Omega
\right) =1,$ and $\mathbb{P}\left( \emptyset \right) =0.$

\item If $A_{n}\in \mathcal{F}$ for any $n\in \mathbb{N}$ and $A_{n}\cap
A_{m}=\emptyset $, $\forall n,m\in \mathbb{N}$, $n\neq m,$ then
\[
\mathbb{P}\left( \bigcup_{n=1}^{\infty }A_{n}\right)
=\sum_{n=1}^{\infty }\mathbb{P}\left( A_{n}\right) .
\]
\end{enumerate}

A family $\mathcal{F}\subset \left\{ \emptyset ,2^{\Omega }\right\} $
satisfying 1,2 and 3 is called a $\sigma $\textit{-algebra} and a mapping $%
\mathbb{P}$ with the properties 4 and 5 is called a \textit{probability
measure. }Let $\left( \Omega ,\mathcal{F},\mathbb{P}\right) $ be a
probability space. Let $\mathcal{B}\left( \mathbb{R}^{n}\right) $ be the
collection of all \textit{Borel sets} on $\mathbb{R}^{n}$ which is the $%
\sigma -$algebra generated by all open sets in $\mathbb{R}^{n},$ i.e. the
smallest $\sigma -$algebra that contains all open sets in $\mathbb{R}^{n}.$
A real valued function is called \textit{measurable} (Borel measurable) if
it is $\mathcal{B}\left( \mathbb{R}^{n}\right) $ measurable. A mapping $%
\mathbf{X:}$ $\Omega $ $\longrightarrow \mathbb{R}^{n}$ is an $\mathbb{R}%
^{n}-$valued \textit{random variable} if it is $\mathcal{F}$-measurable,
i.e. for any $B\in \mathcal{B}\left( \mathbb{R}^{n}\right) $ we have $%
\left\{ \mathbf{\omega }|\mathbf{X}(\mathbf{\omega })\in B\right\} \in
\mathcal{F}.$ A \textit{stochastic process} $\mathbf{X}=\{\mathbf{X}%
_{t}\}_{t\in \mathbb{R}_{+}}$ is a one-parametric family of random variables
on a common probability space $\left( \Omega ,\mathcal{F},\mathbb{P}\right) $%
. The\textit{\ trajectory} of the process $\mathbf{X}$ is a map
\[
\begin{array}{ccc}
\mathbb{R}_{+} & \longrightarrow & \mathbb{R}^{n} \\
t & \longmapsto & \mathbf{X}_{t}\left( \mathbf{\omega }\right) ,%
\end{array}%
\]%
where $\mathbf{\omega }\in \Omega $ and $\mathbf{X}_{t}=\left( X_{1,t},\cdot
\cdot \cdot ,X_{n,t}\right) $. For a fixed $0\leq t_{0}<t_{1}<\cdot \cdot
\cdot <t_{m},$ $m\in \mathbb{N}$ and Borel measurable sets $B_{k}\subset
\mathbb{R}^{n},$ $0\leq k\leq m$ consider the map
\[
\begin{array}{ccc}
\mathcal{B}\left( \mathbb{R}^{mn}\right) & \longrightarrow & \mathbb{R}_{+}
\\
\prod_{1\leq k\leq m}B_{k} & \longmapsto & \mathbb{P}\left[ \mathbf{X}%
_{t_{1}}\in B_{1},\cdot \cdot \cdot ,\mathbf{X}_{t_{m}}\in B_{m}\right] ,%
\end{array}%
\]%
which defines a probability measure on $\mathcal{B}\left( \mathbb{R}%
^{mn}\right) .$ The \textit{system of finite-dimensional distributions} of $%
\mathbf{X}$ is the family of all such measures over all choices $0\leq
t_{0}<t_{1}<\cdot \cdot \cdot <t_{m},$ $m\in \mathbb{N}$. Two stochastic
processes $\mathbf{X}$ and $\mathbf{Y}$ are \textit{identical in law},
written as $\mathbf{X}\overset{d}{=}\mathbf{Y}$ (or $\mathbf{X=Y}\mathrm{mod}%
\left( law\right) $) if the systems of their finite-dimensional
distributions are identical.

Consider the $\sigma $-algebra $\mathcal{F}$ generated by the cylinder sets,
known as Kolmogorov's $\sigma $-algebra.

{\bf Theorem 5}
\begin{em}
(Kolmogorov's extension theorem) Suppose that for any $0\leq t_{1}\leq \cdot
\cdot \cdot <t_{m}$ and $m\in \mathbb{N}$ a distribution $\upsilon
_{t_{1},\cdot \cdot \cdot ,t_{m}}$ is given. If for any $B_{1},\cdots
,B_{m}\in \mathcal{B}(\mathbb{R}^{n})$ we have
\[
\upsilon _{t_{1},\cdot \cdot \cdot ,t_{m}}\left(
\prod_{s=1}^{m}B_{s}\right) =\upsilon _{t_{1},\cdot \cdot \cdot
,t_{k-1},t_{k+1},\cdot \cdot \cdot ,t_{m}}\left( \prod_{1\leq s\leq
m,s\neq k}B_{s}\right) ,\text{ \ }B_{k}=\mathbb{R}^{n}
\]%
then there exists a unique probability measure $\mathbb{P}$ on $\mathcal{F}$
that has $\left\{ \upsilon _{t_{1},\cdot \cdot \cdot ,t_{m}}\right\} $ as
its system of finite-dimensional distributions.
\end{em}

Different proofs of this statement can be found in \cite{kolmogorov} and
\cite{breiman}.

$\mathbf{X}=\{\mathbf{X}_{t}\}_{t\in \mathbb{R}_{+}}$ is called a \textit{L%
\'{e}vy process} (process with stationary independent increments) if

\begin{enumerate}
\item The random variables $\mathbf{X}_{t_{0}},\mathbf{X}_{t_{1}}-\mathbf{X}%
_{t_{0}},\cdots ,\mathbf{X}_{t_{m}}-\mathbf{X}_{t_{m-1}}$, for any $0\leq
t_{0}<t_{1}<\cdots <t_{m}$ and $m\in \mathbb{N}$ are independent (\textit{%
independent increment property}).

\item $\mathbf{X}_{0}=\mathbf{0}$ a.s.

\item The distribution of $\mathbf{X}_{t+\tau }-\mathbf{X}_{t}$ is
independent of $\tau $ (\textit{temporal homogeneity or stationary
increments property}).

\item It is \textit{stochastically continuous}, i.e.
\[
\lim_{\tau \rightarrow t}\mathbb{P}\left[ |\mathbf{X}_{\tau }-\mathbf{X}%
_{t}|>\epsilon \right] =0
\]%
for any $\epsilon >0$ and $t\geq 0$.

\item There is $\Omega _{0}\in \mathcal{F}$ with $\mathbb{P}\left( \Omega
_{0}\right) =1$ such that, for any $\mathbf{\omega }\in \Omega _{0},$ $%
\mathbf{X}_{t}\left( \mathbf{\omega }\right) $ is right-continuous on $%
\mathbb{[}0,\infty )$ and has left limits on $\mathbb{(}0,\infty ).$
\end{enumerate}

A process satisfying 1-4 is called a \textit{L\'{e}vy process in law. An
additive process }is a stochastic process which satisfies 1,2,4,5 and an%
\textit{\ additive process in law} satisfies 1,2,4.

Let $\mathbf{x},\mathbf{y}\in \mathbb{R}^{n}$, $\mathbf{x=}\left(
x_{1},...,x_{n}\right) $, $\mathbf{y=}\left( y_{1},...,y_{n}\right) $, $%
\left\langle \mathbf{x,y}\right\rangle $ be the usual scalar product in $%
\mathbb{R}^{n}$, i.e.
\[
\left\langle \mathbf{x,y}\right\rangle =\sum_{k=1}^{n}x_{k}y_{k}\in \mathbb{R%
}
\]%
and $\left\vert \mathbf{x}\right\vert :=\left\langle \mathbf{x,y}%
\right\rangle ^{1/2}.$ Let $C(\mathbb{R}^{n})$ be the space of continuous
functions on $\mathbb{R}^{n}$ and $L_{p}(\mathbb{R}^{n})$ be the usual space
of $p$-integrable functions $f:\mathbb{R}^{n}\mapsto \mathbb{R}$ (or $f:%
\mathbb{R}^{n}\mapsto \mathbb{C}$) equipped with the norm
\[
\Vert f\Vert _{p}=\Vert f\Vert _{L_{p}(\mathbb{R}^{n})}:=\left\{
\begin{array}{cc}
\left( \int_{\mathbb{R}^{n}}\left\vert f(\mathbf{x})\right\vert ^{p}d\mathbf{%
x}\right) ^{1/p}, & 1\leq p<\infty , \\
\mathrm{ess}\,\,\sup_{\mathbf{x}\in \mathbb{R}^{n}}|f(\mathbf{x})|, &
p=\infty .%
\end{array}%
\right.
\]

For a finite measure $\upsilon $ on $\mathbb{R}^{n}$ (i.e. if $%
\upsilon \left( \mathbb{R}^{n}\right) <\infty $) we define its Fourier
transform as
\[
\mathbf{F}\left( \upsilon \right) \left( \mathbf{y}\right) =\int_{\mathbb{R}%
^{n}}\exp \left( -i\left\langle \mathbf{x,y}\right\rangle \right) \upsilon
\left( d\mathbf{x}\right)
\]%
and its formal inverse%
\[
\upsilon \left( d\mathbf{x}\right) =\mathbf{F}^{-1}\circ \mathbf{F}\left(
\upsilon \right) \left( d\mathbf{x}\right) =\frac{1}{\left( 2\pi \right) ^{n}%
}\int_{\mathbb{R}^{n}}\exp \left( i\left\langle \mathbf{x,y}\right\rangle
\right) \mathbf{F}\left( \upsilon \right) \left( \mathbf{y}\right) d\mathbf{y%
}.
\]%
The \textit{convolution} $\upsilon =\upsilon _{1}\ast \upsilon _{2}$ of two
measures $\upsilon _{1}$ and $\upsilon _{2}$ on $\mathbb{R}^{n}$ is defined
as
\[
\upsilon \left( B\right) =\int_{\mathbb{R}^{n}\mathbb{\times R}^{n}}\chi
_{B}(\mathbf{x}+\mathbf{y})\upsilon _{1}\left( d\mathbf{x}\right) \upsilon
_{2}\left( d\mathbf{y}\right) <\infty ,
\]%
where
\[
\chi _{B}(\mathbf{x}):=\left\{
\begin{array}{cc}
1, & \mathbf{x\in }B\mathbf{,} \\
0, & \mathbf{x\notin }B%
\end{array}%
\right.
\]%
is the \textit{characteristic function} of a Borel (Lebesgue) measurable set
$B\subset \mathbb{R}^{n}$. A probability measure $\upsilon $ is called
\textit{infinitely divisible} if for any $m\in \mathbb{N}$ there is a
probability measure $\upsilon _{\left( m\right) }$ such that
\[
\upsilon =\underbrace{\upsilon _{\left( m\right) }\ast \cdot \cdot \cdot
\ast \upsilon _{\left( m\right) }}_{m}.
\]

It is known that if $\upsilon $ is infinitely divisible then there exists a
unique continuous function $\phi :\mathbb{R}^{n}\mathbb{\rightarrow C}$ such
that $\phi \left( \mathbf{0}\right) =0$ and $\exp \left( \phi \left( \mathbf{%
y}\right) \right) =\mathbf{F}\left( \upsilon \right) \left( \mathbf{y}%
\right) $ (see, e.g. \cite{sato}, p. 37).

The \textit{characteristic function }$\Phi \left( \mathbf{x},t\right) $ of
the distribution of $\mathbf{X}_{t}$ of any L\'{e}vy process can be formally
defined as
\[
\Phi \left( \mathbf{x},t\right) :=\mathbb{E}\left[ \exp \left( i\left\langle
\mathbf{x,X}_{t}\right\rangle \right) \right] =\exp \left( -t\mathbf{\psi }%
\left( \mathbf{x}\right) \right)
\]%
\[
=\left( 2\pi \right) ^{n}\mathbf{F}^{-1}\left( p_{t}\right) \left( \mathbf{x}%
\right) ,
\]%
where $p_{t}\left( \mathbf{x}\right) $ is the density function of $\mathbf{X}%
_{t}$, $\mathbf{x}\in \mathbb{R}^{n}$, $t\in \mathbb{R}_{+}$ and the
function $\mathbf{\psi }\left( \mathbf{x}\right) $ \ is uniquely determined.
This function is called the \textit{characteristic exponent}. Vice versa, a L%
\'{e}vy process $\mathbf{X}=\{\mathbf{X}_{t}\}_{t\in \mathbb{R}_{+}}$ is
determined uniquely by its characteristic exponent $\mathbf{\psi }\left(
\mathbf{x}\right) $. In particular, $p_{t}$ can be expressed as
\[
p_{t}\left( \cdot \right) =\left( 2\pi \right) ^{-n}\int_{\mathbb{R}%
^{n}}\exp \left( -i\left\langle \cdot \mathbf{,x}\right\rangle -t\mathbf{%
\psi }\left( \mathbf{x}\right) \right) d\mathbf{x}
\]%
\[
=\left( 2\pi \right) ^{-n}\mathbf{F}\left( \exp \left( -t\mathbf{\psi }%
\left( \mathbf{x}\right) \right) \right) \left( \cdot \right) .
\]%
We say that a matrix $\mathbf{A}$ is \textit{nonnegative-definite} (or
positive-semidefinite) if $\mathbf{x}^{\ast }\mathbf{Ax}\geq 0$ for all $%
\mathbf{x}\in \mathbb{C}^{n}$ (or for all $\mathbf{x}\in \mathbb{R}^{n}$ for
the real matrix), where $\mathbf{x}^{\ast }$ is the conjugate transpose. A
matrix $\mathbf{A}$ is nonnegative-definite iff it arises as the \textit{Gram%
} matrix of some set of vectors $v_{1},\cdots ,v_{n}$, i.e. $\mathbf{A}%
=(a_{i,j})=\left\langle v_{j},v_{i}\right\rangle $.

The following classical result, plays a key role in our analysis.

{\bf Theorem 6}
\begin{em}
(L\'{e}vy-Khintchine formula) Let $\mathbf{X}=\{\mathbf{X}_{t}\}_{t\in
\mathbb{R}_{+}}$ be a L\'{e}vy process on $\mathbb{R}^{n}$. Then its
characteristic exponent admits the representation
\[
\mathbf{\psi }(\mathbf{y})=-\frac{1}{2}\langle \mathbf{Ay},\mathbf{y}\rangle
-i\langle \mathbf{h},\mathbf{y}\rangle
\]%
\begin{equation}
-\int_{\mathbb{R}^{n}}\left( 1-\exp \left( i\langle \mathbf{y},\mathbf{x}%
\rangle \right) -i\langle \mathbf{y},\mathbf{x}\rangle \chi _{D}(\mathbf{x}%
)\right) \Pi (d\mathbf{x}),  \label{lk}
\end{equation}%
where $\chi _{D}(\mathbf{x})$ is the characteristic function of $D:=\{%
\mathbf{x}\in \mathbb{R}^{n},\,\,|\mathbf{x}|\leq 1\}$, $\mathbf{A}$ is a
symmetric nonnegative-definite $n\times n$ matrix, $\mathbf{h}\in \mathbb{R}%
^{n}$ and $\Pi (d\mathbf{x})$ is a measure on $\mathbb{R}^{n}$ such that
\begin{equation}
\int_{\mathbb{R}}\min \{1,\left\vert \mathbf{x}\right\vert ^{2}\}\Pi (d%
\mathbf{x})<\infty ,\,\,\Pi (\{\mathbf{0}\})=0.  \label{ppp}
\end{equation}
\end{em}

The density of $\Pi $ is known as the \textit{L\'{e}vy density} and $\mathbf{%
A}$ is the \textit{covariance matrix}. In particular, if $\mathbf{A}=0$ (or $%
\mathbf{A}=(a_{j,k})_{1\leq j,k\leq n}$, $a_{j,k}=0$) then the L\'{e}vy
process is a \textit{pure non-Gaussian process} and if $\Pi =0$ the process
is \textit{Gaussian}.

{\bf Definition 7}
\begin{em}
We say that the L\'{e}vy process has\textit{\ bounded variation} if its
sample paths have bounded variation on every compact time interval.
\end{em}

A L\'{e}vy process has bounded variation iff $\mathbf{A}=0$ and
\[
\int_{\mathbb{R}^{n}}\min \left\{ \left\vert \mathbf{x}\right\vert
,1\right\} \Pi \left( d\mathbf{x}\right) <\infty ,\,\,\Pi \left( \left\{
\mathbf{0}\right\} \right) =0
\]%
(see e.g. \cite{bertoin}, p. 15).

The systematic exposition of the theory of L\'{e}vy processes can be found
in \cite{gs1, gs2, gs3, sato, Applebaum, McKean}.

\section{\protect\bigskip A class of stochastic systems}

In this section we introduce a class of stochastic systems to model
multidimensional return processes. Let $X_{1,t},\cdot \cdot \cdot ,X_{n,t}$
and $Z_{1,t},\cdot \cdot \cdot ,Z_{n,t}$ be independent random variables,
with the density functions $p_{1,t}^{\left( 1\right) }\left( x_{1}\right)
,\cdot \cdot \cdot ,p_{n,t}^{\left( 1\right) }\left( x_{n}\right) $ and $%
p_{1,t}^{\left( 2\right) }\left( x_{1}\right) ,\cdot \cdot \cdot
,p_{n,t}^{\left( 2\right) }\left( x_{n}\right) $ and characteristic
exponents $\psi _{s}^{\left( 1\right) }$ and $\psi _{m}^{\left( 2\right)
},1\leq s,m\leq n$ respectively. Let $\mathbf{X}_{t}=\left( X_{1,t},\cdot
\cdot \cdot ,X_{n,t}\right) ^{\mathrm{T}}$, $\mathbf{Z}_{t}=\left(
Z_{1,t},\cdot \cdot \cdot ,Z_{n,t}\right) ^{\mathrm{T}}$ and $\mathbf{B}%
=\left( b_{j,k}\right) $ be a real matrix of size $n\times n$. Consider
random vector $\mathbf{U}_{t}=\left( U_{1,t},\cdot \cdot \cdot
,U_{n,t}\right) ^{\mathrm{T}},$
\begin{equation}
\mathbf{U}_{t}=\mathbf{X}_{t}+\mathbf{BZ}_{t}.  \label{system}
\end{equation}%
A matrix $\mathbf{B}$ reflects dependence between the processes $%
U_{1,t},\cdot \cdot \cdot ,U_{n,t}$ in our model. Assume for simplicity that
$\mathbb{E}\left[ X_{s,t}\right] =0$ and $\mathbb{E}\left[ Z_{s,t}\right]
=0, $ $1\leq s\leq n$, $\mathrm{var}\left( X_{s,t}\right) =\mathrm{var}%
\left( Z_{s,t}\right) =v_{t}$ and $b_{s,k}=1,$ $1\leq s,k\leq n$. It is easy
to check that for any $s$ and $l$, $1\leq s\neq l\leq n$ the correlation
coefficient
\[
\rho \left( U_{s,t},U_{l,t}\right) :=\frac{\mathbb{E}\left[ U_{s,t}U_{l,t}%
\right] }{\left( \mathbb{E}\left[ U_{s,t}^{2}\right] \mathbb{E}\left[
U_{l,t}^{2}\right] \right) ^{1/2}}
\]%
between $U_{s,t}$ and $U_{l,t}$, where
\[
U_{s,t}=X_{s,t}+\sum_{k=1}^{n}b_{s,k}Z_{k,t},U_{l,t}=X_{l,t}+%
\sum_{k=1}^{n}b_{l,k}Z_{k,t}
\]%
is $\rho \left( U_{s,t},U_{l,t}\right) =n(n+1)^{-1}$. This reflects our
empirical experience: if the market is in crisis then the prices of stocks
are highly correlated \newline
(see http://www.economicsofcrisis.com/lit.html for more information).

The next statement gives us an explicit form of the characteristic function
of the return process $\mathbf{U}_{t}.$

{\bf Theorem 8}
\begin{em}
\label{tce} Let $\mathbf{U}_{t}=\mathbf{X}_{t}+\mathbf{BZ}_{t},$ $\mathbf{B}%
=\left( b_{m,k}\right) .$ Then in our notation the characteristic function $%
\Phi \left( \mathbf{v,}t\right) $ of $\ \mathbf{U}_{t}$ has the form%
\[
\Phi \left( \mathbf{v,}t\right) =\left( 2\pi \right) ^{n}\left(
\prod_{s=1}^{n}\mathbf{F}^{-1}\left( p_{s,t}^{\left( 1\right)
}\right) \right) \left( \mathbf{v}\right) \mathbf{F}^{-1}\left(
\prod_{m=1}^{n}p_{m,t}^{\left( 2\right) }\right) \left( \mathbf{B}^{%
\mathrm{T}}\mathbf{v}\right) ,
\]%
\[
\text{ \ \ \ \ \ \ \ \ \ }=\exp \left( -t\mathbf{\psi }\left( \mathbf{v}%
\right) \right) ,
\]%
where
\[
\mathbf{\psi }\left( \mathbf{v}\right) =\sum_{s=1}^{n}\psi _{s}^{\left(
1\right) }\left( v_{s}\right) +\sum_{m=1}^{n}\psi _{m}^{\left( 2\right)
}\left( \sum_{k=1}^{n}b_{k,m}v_{k}\right) .
\]
\end{em}

{\bf Proof.}
 Consider the transformation $\mathbb{R}^{2n}\rightarrow \mathbb{R}%
^{2n}$ defined as
\[
\begin{array}{c}
\mathbf{U}_{t}=\mathbf{X}_{t}+\mathbf{BZ}_{t}, \\
\mathbf{Z}_{t}=\mathbf{Z}_{t}.%
\end{array}%
\]%
The inverse is given by
\[
\begin{array}{c}
\mathbf{X}_{t}=\mathbf{U}_{t}-\mathbf{BZ}_{t}, \\
\mathbf{Z}_{t}=\mathbf{Z}_{t},%
\end{array}%
\]%
or
\[
\left(
\begin{array}{c}
\mathbf{X}_{t} \\
\mathbf{Z}_{t}%
\end{array}%
\right) =\left(
\begin{array}{cc}
\mathbf{I} & -\mathbf{B} \\
\mathbf{0} & \mathbf{I}%
\end{array}%
\right) \left(
\begin{array}{c}
\mathbf{U}_{t} \\
\mathbf{Z}_{t}%
\end{array}%
\right)
\]%
and the Jacobian $J$ of this transformation is
\[
J=\det \left(
\begin{array}{cc}
\mathbf{I} & -\mathbf{B} \\
\mathbf{0} & \mathbf{I}%
\end{array}%
\right) =1,
\]%
where $\mathbf{I}=\mathbf{I}_{n\times n}$ is an identity. The density
function%
\[
\widetilde{p}_{t}\left( \mathbf{u},\mathbf{z}\right) =\widetilde{p}%
_{t}\left( u_{1},\cdot \cdot \cdot ,u_{n},z_{1},\cdot \cdot \cdot
z_{n}\right)
\]%
is given by
\[
\widetilde{p}_{t}\left( \mathbf{u},\mathbf{z}\right)
=\prod_{s=1}^{n}p_{s,t}^{\left( 1\right) }\left(
u_{s}-\sum_{m=1}^{n}b_{s,m}z_{m}\right)
\prod_{l=1}^{n}p_{l,t}^{\left( 2\right) }\left( z_{l}\right) .
\]%
This means that the density function $p_{t}\left( \mathbf{u}\right) $ of $%
\mathbf{U}_{t}$ is
\[
p_{t}\left( \mathbf{u}\right) =\int_{\mathbb{R}^{n}}\widetilde{p}_{t}\left(
\mathbf{u},\mathbf{z}\right) d\mathbf{z}
\]%
and the characteristic function has the form%
\[
\Phi \left( \mathbf{v,}t\right) :=\mathbb{E}\left[ \exp \left( i\left\langle
\mathbf{U}_{t},\mathbf{v}\right\rangle \right) \right] :=\exp \left( -t%
\mathbf{\psi }\left( \mathbf{v}\right) \right)
\]%
\[
\text{ \ \ \ \ \ \ \ \ \ \ }=\int_{\mathbb{R}^{n}}\exp \left( i\left\langle
\mathbf{u},\mathbf{v}\right\rangle \right) p_{t}\left( \mathbf{u}\right) d%
\mathbf{u}
\]%
\[
\text{ \ \ \ \ \ \ \ \ \ \ }=\int_{\mathbb{R}^{n}}\exp \left( i\left\langle
\mathbf{u},\mathbf{v}\right\rangle \right) \left( \int_{\mathbb{R}^{n}}%
\widetilde{p}_{t}\left( \mathbf{u},\mathbf{z}\right) d\mathbf{z}\right) d%
\mathbf{u}
\]%
\[
=\int_{\mathbb{R}^{n}}\exp \left( i\left\langle \mathbf{u},\mathbf{v}%
\right\rangle \right) \left( \int_{\mathbb{R}^{n}}\prod_{s=1}^{n}p_{s,t}^{\left( 1\right) }\left(
u_{s}-\sum_{m=1}^{n}b_{s,m}z_{m}\right)
\prod_{m=1}^{n}p_{m,t}^{\left( 2\right) }\left( z_{m}\right) d\mathbf{%
z}\right) d\mathbf{u}
\]%
\begin{equation}
=\int_{\mathbb{R}^{n}}\left( \prod_{s=1}^{n}\int_{\mathbb{R}%
}p_{s,t}^{\left( 1\right) }\left( u_{s}-\sum_{m=1}^{n}b_{s,m}z_{m}\right)
\exp \left( iu_{s}v_{s}\right) du_{s}\right)
\prod_{m=1}^{n}p_{m,t}^{\left( 2\right) }\left( z_{m}\right) d\mathbf{%
z.}  \label{2013}
\end{equation}%
In the last line we applied Fubini theorem (see Appendix I, Theorem 39
{Fubini's theorem}). Let $\xi _{s}=u_{s}-\sum_{m=1}^{n}b_{s,m}z_{m},1\leq
s\leq n.$ Then
\[
\int_{\mathbb{R}}p_{s,t}^{\left( 1\right) }\left(
u_{s}-\sum_{m=1}^{n}b_{s,m}z_{m}\right) \exp \left( iu_{s}v_{s}\right) du_{s}
\]%
\[
=\int_{\mathbb{R}}p_{s,t}^{\left( 1\right) }\left( \xi _{s}\right) \exp
\left( i\left( \xi _{s}+\sum_{m=1}^{n}b_{s,m}z_{m}\right) v_{s}\right) d\xi
_{s}
\]%
\[
=\exp \left( iv_{s}\sum_{m=1}^{n}b_{s,m}z_{m}\right) \int_{\mathbb{R}%
}p_{s,t}^{\left( 1\right) }\left( \xi _{s}\right) \exp \left( i\xi
_{s}v_{s}\right) d\xi _{s}
\]%
\begin{equation}
=\exp \left( iv_{s}\sum_{m=1}^{n}b_{s,m}z_{m}\right) 2\pi \mathbf{F}%
^{-1}\left( p_{s,t}^{\left( 1\right) }\right) \left( v_{s}\right) .
\label{2014}
\end{equation}%
Comparing (\ref{2013}) and (\ref{2014}) we get
\[
\Phi \left( \mathbf{v,}t\right) =\int_{\mathbb{R}^{n}}\left(
\prod_{s=1}^{n}\exp \left( iv_{s}\sum_{m=1}^{n}b_{s,m}z_{m}\right)
2\pi \mathbf{F}^{-1}\left( p_{s,t}^{\left( 1\right) }\right) \left(
v_{s}\right) \right) \prod_{m=1}^{n}z_{m,t}\left( z_{m}\right) d%
\mathbf{z}
\]%
\[
=\prod_{s=1}^{n}2\pi \mathbf{F}^{-1}\left( p_{s,t}^{\left( 1\right)
}\right) \left( v_{s}\right) \int_{\mathbb{R}^{n}}\left(
\prod_{s=1}^{n}\exp \left( iv_{s}\sum_{m=1}^{n}b_{s,m}z_{m}\right)
\right) \prod_{m=1}^{n}p_{m,t}^{\left( 2\right) }\left( z_{m}\right) d%
\mathbf{z}
\]%
\[
=\prod_{s=1}^{n}2\pi \mathbf{F}^{-1}\left( p_{s,t}^{\left( 1\right)
}\right) \left( v_{s}\right) \int_{\mathbb{R}^{n}}\exp \left(
i\sum_{s=1}^{n}\left( v_{s}\sum_{m=1}^{n}b_{s,m}z_{m}\right) \right)
\prod_{m=1}^{n}p_{m,t}^{\left( 2\right) }\left( z_{m}\right) d\mathbf{%
z}
\]%
\[
=\prod_{s=1}^{n}2\pi \mathbf{F}^{-1}\left( p_{s,t}^{\left( 1\right)
}\right) \left( v_{s}\right) \int_{\mathbb{R}^{n}}\exp \left( i\left\langle
\mathbf{v},\mathbf{Bz}\right\rangle \right) \left(
\prod_{m=1}^{n}p_{m,t}^{\left( 2\right) }\left( z_{m}\right) \right) d%
\mathbf{z}
\]%
\[
=\prod_{s=1}^{n}2\pi \mathbf{F}^{-1}\left( p_{s,t}^{\left( 1\right)
}\right) \left( v_{s}\right) \int_{\mathbb{R}^{n}}\exp \left( i\left\langle
\mathbf{B}^{\mathrm{T}}\mathbf{v},\mathbf{z}\right\rangle \right) \left(
\prod_{m=1}^{n}p_{m,t}^{\left( 2\right) }\left( z_{m}\right) \right) d%
\mathbf{z}
\]%
\[
=\prod_{s=1}^{n}2\pi \mathbf{F}^{-1}\left( p_{s,t}^{\left( 1\right)
}\right) \left( v_{s}\right) \mathbf{F}^{-1}\left(
\prod_{m=1}^{n}2\pi p_{m,t}^{\left( 2\right) }\right) \left( \mathbf{B%
}^{\mathrm{T}}\mathbf{v}\right) ,
\]%
where $\mathbf{A}^{\mathrm{T}}=\left( a_{k,j}\right) $ is the transpose of $%
\mathbf{A}$. Hence
\[
\Phi \left( \mathbf{v,}t\right) =\prod_{s=1}^{n}\exp \left( -t\psi
_{s}^{\left( 1\right) }\left( v_{s}\right) \right)
\prod_{m=1}^{n}\exp \left( -t\psi _{m}^{\left( 2\right) }\left(
\sum_{k=1}^{n}b_{k,m}v_{k}\right) \right)
\]%
\[
\text{ }=\exp \left( -t\left( \sum_{s=1}^{n}\psi _{s}^{\left( 1\right)
}\left( v_{s}\right) +\sum_{m=1}^{n}\psi _{m}^{\left( 2\right) }\left(
\sum_{k=1}^{n}b_{k,m}v_{k}\right) \right) \right) .
\]
$\square$

\section{Sufficient equivalent martingale measure conditions for basket
options}

In this section we specify the equivalent martingale measure condition for
our model. Under the equivalent martingale measure all assets have the same
expected rate of return which is a risk free rate. This means that under
no-arbitrage conditions the risk preferences of investors acting on the
market do not enter into valuation decisions. Recall that in general, $%
\mathbb{Q}$ is not unique. It was shown in \cite{44-2015} that the class of
equivalent martingale measures is so rich that every price in some interval $%
\left[ a,b\right] $ can be obtained by a particular martingale measure. We
assume that $\mathbb{Q}$ has been fixed and all expectations will be
computed with respect to this measure (see Appendix III for more
information).

Consider a frictionless market consisting of a riskless bond $B$ and stock $%
S $. In this market $S$ is modeled by an exponential L\'{e}vy process $%
S=S_{t}=S_{0}\exp \left( X_{t}\right) $ under a chosen equivalent martingale
measure $\mathbb{Q}$. Assume that the riskless rate $r$ is constant.

{\bf Theorem 9}
\begin{em}
\label{1demm} Let $D$ be the domain of $\psi ^{\mathbb{Q}}(\xi )$ and $%
\mathbb{R}\cup \{-i\}\subset D$, then in our notations $\psi ^{\mathbb{Q}%
}(-i)=-r$.
\end{em}

{\bf Proof.}
The discounted price process which is given by
\[
\tilde{S_{t}}=\exp (-rt)S_{t}=\exp (-rt)S_{0}\exp (X_{t})
\]%
must be a martingale under a chosen equivalent martingale measure $\mathbb{Q}
$, i.e. for any $0\leq l<t\leq T$ the martingale condition must hold,
\[
\tilde{S_{l}}=\mathbb{E}^{\mathbb{Q}}\left[ \tilde{S_{t}}|\mathcal{F}_{l}%
\right]
\]%
(see Appendix III for more information). Without loss of generality we may
assume $l=0$. Then for any $t\in (0,T]$ we have
\[
\tilde{S_{0}}=S_{0}\exp (-r0)=S_{0}=\mathbb{E}^{\mathbb{Q}}\left[ S_{0}\exp
(-rt)\exp (X_{t})|\mathcal{F}_{0}\right]
\]%
\[
=\mathbb{E}^{\mathbb{Q}}\left[ S_{0}\exp (-rt)\cdot \exp (X_{t})\right]
\]%
\[
=S_{0}\mathbb{E}^{\mathbb{Q}}\left[ \exp (-rt)\cdot \exp (X_{t})\right] .
\]%
Since $S_{0}>0$ then
\[
\mathbb{E}^{\mathbb{Q}}\left[ \exp (-rt)\cdot \exp (X_{t})\right] =1,
\]%
or
\begin{equation}
\exp (rt)=\mathbb{E}^{\mathbb{Q}}\left[ \exp (X_{t})\right] .  \label{zzz}
\end{equation}%
Since $\psi (-i)\subset D$ then by the definition of the characteristic
exponent
\[
\exp (-t\psi (-i))=\mathbb{E}^{\mathbb{Q}}\left[ \exp (i(-i)X_{t})\right] =%
\mathbb{E}^{\mathbb{Q}}\left[ \exp (X_{t})\right] .
\]%
Hence, since $t>0$, then from (\ref{zzz}) it follows that $r=-\psi (-i)$.
$\square$

A commonly used condition on $\psi ^{\mathbb{Q}}(\xi )$ is that it admits
the analytic continuation into the strip $\{z|-1\leq \Im z\leq 0\}$ (see,
e.g. \cite{levendorskii 2011} p. 83).

We specify now the equivalent martingale measure condition for the system (%
\ref{system}).

{\bf Theorem 10}
\begin{em}
Let the stock prices be modeled by%
\[
S_{s,t}=S_{s,0}\exp \left( U_{s,t}\right) ,\text{ \ \ }1\leq s\leq n,
\]%
and the domain $D\subset \mathbb{R}^{n}\mathbb{+}i\mathbb{R}^{n}$ of the
characteristic exponent $\mathbf{\psi }^{\mathbb{Q}}$ contains $\mathbb{R}%
^{n}\cup \left( \cup _{k=1}^{n}\left\{ -i\mathbf{e}_{k}\right\} \right) $\
where\ $\left\{ \mathbf{e}_{k},1\leq k\leq n\right\} $\ is the standard
basis in \ $\mathbb{R}^{n}.$ Then \
\begin{equation}
\mathbf{\psi }^{\mathbb{Q}}\left( -i\mathbf{e}_{s}\right) =-r,\text{ \ \ }%
1\leq s\leq n.  \label{EMM condition}
\end{equation}
\end{em}

{\bf Proof.}
Observe that for any $1\leq s\leq n$ the discount price process $S_{s,t}$
must be a martingale under a chosen equivalent martingale measure $\mathbb{Q}%
.$ Let $\psi _{s}^{\mathbb{Q}}\left( x_{s}\right) $ be the characteristic
exponent of $U_{s,t}$. Then
\[
\exp \left( -t\psi _{s}^{\mathbb{Q}}\left( x_{s}\right) \right) =\mathbb{E}^{%
\mathbb{Q}}\left[ \exp \left( ix_{s}U_{s,t}\right) \right]
\]%
\[
=\mathbb{E}^{\mathbb{Q}}\left[ \exp \left( \left\langle i\mathbf{x,}U_{s,t}%
\mathbf{e}_{s}\right\rangle \right) \right] =\exp \left( -t\mathbf{\psi }^{%
\mathbb{Q}}\left( x_{s}\mathbf{e}_{s}\right) \right) .
\]%
Thus by Theorem 9
we get $r=-\psi _{s}^{\mathbb{Q}}\left(
-i\right) $, which gives a system of $n$ equations
\[
\mathbf{\psi }^{\mathbb{Q}}\left( -i\mathbf{e}_{s}\right) =-r,\text{ \ }%
1\leq s\leq n.
\]
$\square$

Observe that in general riskless interest rate may depend on $s$. In this
case we get the system $\mathbf{\psi }^{\mathbb{Q}}\left( -i\mathbf{e}%
_{s}\right) =-r_{s},$ $1\leq s\leq n.$

\section{KoBoL family}

\label{KoBoL}

In this section we study characteristic exponents of so-called KoBoL family.
The idea is based on a simple observation. From the L\'{e}vy-Khintchine
formula (\ref{lk}) it follows that it is possible to find $\psi (\xi )$
explicitly if we can compute explicitly the inverse Fourier transform of $%
\Pi (dx).$ Therefore, it was suggested by the authors of \cite{bl1} to
consider the following form of $\Pi (dx),$%
\[
\Pi (dx)=\left\vert x\right\vert ^{\alpha }\exp \left( -\beta \left\vert
x\right\vert \right) dx,
\]%
where $\alpha $ and $\beta $ are fixed parameters. Let $\lambda
_{-}<0<\lambda _{+}$,
\[
\Pi ^{+}(\nu ,\lambda _{+},dx)=\left( \max \left\{ x,0\right\} \right)
^{-\nu -1}\exp \left( -\lambda _{+}x\right) dx
\]%
and%
\[
\Pi ^{-}(\nu ,\lambda _{-},dx)=\left( \max \left\{ -x,0\right\} \right)
^{-\nu -1}\exp \left( -\lambda _{-}x\right) dx,
\]%
where $\nu <2.$

{\bf Definition 11}
\begin{em}
A L\'{e}vy process is called a \textit{KoBoL process} of order $\nu <2$ if
it is purely non-Gaussian with the L\'{e}vy measure of the form
\[
\Pi (dx)=c_{+}\Pi ^{+}(\nu ,\lambda _{+},dx)+c_{-}\Pi ^{-}(\nu ,\lambda
_{-},dx),
\]%
where $c_{+}>0,$ $c_{-}>0,$ $\lambda _{-}<0<\lambda _{+}.$
\end{em}

We call $\nu $ the \textit{order of the process}, $\lambda _{+}$ and $%
\lambda _{-}$ \ the \textit{steepness parameters} and $c_{+}$ and $c_{-}$
the \textit{intensity parameters} of the process. The parameter $\lambda
_{-} $ ($\lambda _{+}$ respectively) determines the rate of the exponential
decay of the right (left respectively) tail of the density function. It is
easy to see that the condition (\ref{ppp}) is satisfied, i.e.
\[
\int_{\mathbb{R}}\min \left\{ 1,x^{2}\right\} \left( c_{+}\Pi ^{+}(\nu
,\lambda _{+},dx)+c_{-}\Pi ^{-}(\nu ,\lambda _{-},dx)\right) <\infty .
\]%
Moreover, if $\nu <1$ then
\[
\int_{\mathbb{R}}\min \left\{ 1,\left\vert x\right\vert \right\} \left(
c_{+}\Pi ^{+}(\nu ,\lambda _{+},dx)+c_{-}\Pi ^{-}(\nu ,\lambda
_{-},dx)\right) <\infty ,
\]%
i.e. a KoBoL process is of finite variation iff $\nu <1.$

{\bf Lemma 12}
\begin{em}
\label{lemma KoBoL} (\cite{bl1}, p. 70) If $\nu \in \left( 0,1\right) \cup
\left( 1,2\right) $ then
\[
\psi \left( \xi \right) =-i\mu \xi +c_{-}\Gamma \left( -\nu \right) \left(
\left( -\lambda _{-}\right) ^{\nu }-\left( -\lambda _{-}-i\xi \right) ^{\nu
}\right)
\]%
\begin{equation}
+c_{+}\Gamma \left( -\nu \right) \left( \lambda _{+}^{\nu }-\left( \lambda
_{+}+i\xi \right) ^{\nu }\right) .  \label{representation-k}
\end{equation}%
If $\nu =0,$ then
\[
\psi \left( \xi \right) =-i\mu \xi +c_{-}\left[ \ln \left( -\lambda
_{-}-i\xi \right) -\ln \left( -\lambda _{-}\right) \right]
\]%
\[
+c_{+}\left[ \ln \left( \lambda _{+}+i\xi \right) -\ln \lambda _{+}\right] .
\]%
If $\nu =1,$ then
\[
\psi \left( \xi \right) =-i\mu \xi +c_{-}\left[ \left( -\lambda _{-}\right)
\ln \left( -\lambda _{-}\right) -\left( -\lambda _{-}-i\xi \right) \ln
\left( -\lambda _{-}-i\xi \right) \right]
\]%
\[
+c_{+}\left[ \lambda _{+}\ln \lambda _{+}-\left( \lambda _{+}+i\xi \right)
\ln \left( \lambda _{+}+i\xi \right) \right] ,
\]%
where $\mu \in \mathbb{R},c_{\pm }>0,$ and $\lambda _{-}<0<\lambda _{+}$.
\end{em}

The proof of Lemma 12
 presented in \cite{bl1} is incomplete.
The next statement gives a complete proof of the representation (\ref%
{representation-k}) which is important in our applications.

{\bf Theorem 13}
\begin{em}
\label{theorem KoBoL} Let $\nu \in \left( 0,1\right) $ then in our notation
\[
\psi \left( \xi \right) =-i\mu \xi +c_{-}\Gamma \left( -\nu \right) \left(
\left( -\lambda _{-}\right) ^{\nu }-\left( -\lambda _{-}-i\xi \right) ^{\nu
}\right)
\]%
\begin{equation}  \label{exp11}
+c_{+}\Gamma \left( -\nu \right) \left( \lambda _{+}^{\nu }-\left( \lambda
_{+}+i\xi \right) ^{\nu }\right) ,
\end{equation}%
where $\mu $ is a real parameter.
\end{em}

{\bf Proof.}
It is sufficient to prove the statement just for the $\Pi ^{+}\left( \nu
,\lambda ,dx\right) $, i.e. to find
\[
-\psi ^{+}\left( \xi \right) :=\int_{\mathbb{R}}\left( \exp \left( ix\xi
\right) -1-ix\xi \chi _{\left[ -1,1\right] }\left( x\right) \right) \Pi
^{+}\left( dx\right)
\]%
\[
=\int_{\mathbb{R}}\left( \exp \left( ix\xi \right) -1-ix\xi \chi _{\left[
-1,1\right] }\left( x\right) \right) \max \left\{ x,0\right\} ^{-\nu -1}\exp
\left( -\lambda x\right) dx
\]%
\[
=\int_{0}^{\infty }\left( \exp \left( ix\xi \right) -1-ix\xi \chi _{\left[
-1,1\right] }\left( x\right) \right) x^{-\nu -1}\exp \left( -\lambda
x\right) dx
\]%
\[
=\int_{0}^{\infty }\left( \exp \left( ix\xi \right) -1\right) x^{-\nu
-1}\exp \left( -\lambda x\right) dx
\]%
\[
-i\xi \int_{0}^{1}x^{-\nu }\exp \left( -\lambda x\right) dx
\]%
\[
:=I_{1}\left( \xi ,\nu ,\lambda \right) -i\xi \mathrm{D}\left( \nu ,\lambda
\right) ,
\]%
where $\mathrm{D}\left( \nu ,\lambda \right) :=\int_{0}^{1}x^{-\nu }\exp
\left( -\lambda x\right) dx$ and
\[
I_{1}\left( \xi ,\nu ,\lambda \right) =-\frac{1}{\nu }\int_{0}^{\infty
}\left( \exp \left( -\left( \lambda -i\xi \right) x\right) -\exp \left(
-\lambda x\right) \right) dx^{-\nu }
\]%
\[
=-\frac{1}{\nu }\left( \left( \exp \left( -\left( \lambda -i\xi \right)
x\right) -\exp \left( -\lambda x\right) \right) x^{-\nu }\right) \left\vert
_{0}^{\infty }\right.
\]%
\[
-\left( -\frac{1}{\nu }\right) \int_{0}^{\infty }\left( -\left( \lambda
-i\xi \right) \exp \left( -\left( \lambda -i\xi \right) x\right) +\lambda
\exp \left( -\lambda x\right) \right) x^{-\nu }dx
\]%
\[
=-\frac{\lambda -i\xi }{\nu }\int_{0}^{\infty }\exp \left( -\left( \lambda
-i\xi \right) x\right) x^{-\nu }dx-\lambda ^{\nu }\Gamma \left( -\nu \right)
:=I_{2}-\lambda ^{\nu }\Gamma \left( -\nu \right) .
\]%
Making change of variable $z=\left( \lambda -i\xi \right) x$ in $I_{2}$ we
get%
\[
I_{2}=-\frac{\left( \lambda -i\xi \right) ^{\nu }}{\nu }\int_{\gamma }\exp
\left( -z\right) z^{-\nu }dz,
\]%
where $\gamma $ is the ray $\left\{ z\left\vert z=\left( \lambda -i\xi
\right) x,\lambda >0,\xi \in \mathbb{R}\right. \right\} $, $\lambda $ and $%
\xi $ are fixed parameters and $x\geq 0$. Assume that $\xi \geq 0$. The case
$\xi \leq 0$ can be treated similarly. Consider the contour $\eta :=\gamma
_{1}\cup \gamma _{2}\cup \gamma _{3}\cup \gamma _{4}$, where
\[
\gamma _{1}:=\left\{ z=\rho \exp \left( i\theta \right) \left\vert 0\leq
\theta \leq \arg \left( \lambda -i\xi \right) ,\lambda >0,\xi \in \mathbb{R}%
\right. \right\} ,
\]%
\[
\gamma _{2}:=\left\{ z\left\vert \rho \leq z\leq R,z\in \mathbb{R}\right.
\right\} ,
\]%
\[
\gamma _{3}:=\left\{ z=R\exp \left( i\theta \right) \left\vert 0\leq \theta
\leq \arg \left( \lambda -i\xi \right) ,\text{ }\lambda >0,\text{ }\xi \in
\mathbb{R}\right. \right\} ,
\]%
\[
\gamma _{4}:=\left\{ z\left\vert z=\left( \lambda -i\xi \right) x,\rho \leq
\left\vert z\right\vert \leq R\right. \right\} .
\]%
The function $\exp \left( -z\right) z^{-\nu }$ is analytic in the domain
bounded by $\eta $, hence from the Cauchy theorem it follows that
\[
\oint_{\eta }\exp \left( -z\right) z^{-\nu }dz=0
\]%
and since $\xi \geq 0$ then for some $\delta >0$ we get $-\pi /2+\delta \leq
\arg \left( \lambda -i\xi \right) \leq 0.$ Hence
\[
\lim_{R\rightarrow \infty }\left\vert \int_{\gamma _{3}}\exp \left(
-z\right) z^{-\nu }dz\right\vert
\]%
\[
=\lim_{R\rightarrow \infty }\left\vert \int_{0}^{\arg \left( \lambda -i\xi
\right) }\exp \left( -R\text{ }\exp \left( i\theta \right) \right) R^{-\nu
}\exp \left( -i\nu \theta \right) Ri\exp \left( i\theta \right) d\theta
\right\vert
\]%
\[
\leq \frac{\pi }{2}\lim_{R\rightarrow \infty }\exp \left( -R\cos \delta
\right) R^{1-\nu }=0.
\]%
Observe that%
\[
\lim_{\rho \rightarrow 0}\left\vert \int_{\gamma _{1}}\exp \left( -z\right)
z^{-\nu }dz\right\vert
\]%
\[
\leq \lim_{\rho \rightarrow 0}\left\vert \int_{0}^{2\pi }\exp \left( -\rho
\text{ }\mathrm{exp}\left( i\theta \right) \right) \rho ^{-\nu }\exp \left(
-i\nu \theta \right) \rho i\exp \left( i\theta \right) d\theta \right\vert
\]%
\[
\leq 2\pi \lim_{\rho \rightarrow 0}\rho ^{-\nu +1}=0.
\]%
Hence%
\[
\int_{\gamma }\exp \left( -z\right) z^{-\nu }dz=\int_{\mathbb{R}_{+}}\exp
\left( -z\right) z^{-\nu }dz=\Gamma \left( -\nu +1\right) =-\nu \Gamma
\left( -\nu \right) .
\]%
Consequently,
\[
I_{2}=-\frac{\left( \lambda -i\xi \right) ^{\nu }}{\nu }\int_{\gamma }\exp
\left( -y\right) y^{-\nu }dy=\Gamma \left( -\nu \right) \left( \lambda -i\xi
\right) ^{\nu }
\]%
and%
\[
\psi ^{+}\left( \xi \right) =\Gamma \left( -\nu \right) \left( \lambda ^{\nu
}-\left( \left( \lambda -i\xi \right) ^{\nu }\right) \right) +i\xi \mathrm{D}%
\left( \nu ,\lambda \right) .
\]%
Finally, the term $i\xi \mathrm{D}\left( \nu ,\lambda \right) $ can be
considered as a part of $i\mu \xi $, where $\mu \in \mathbb{R}$ is a free
parameter.
$\square$

Observe that the parameters $\left( \nu ,c_{+},c_{-},\lambda _{+},\lambda
_{-}\right) $ determine the probability density. For larger $\nu $ and $%
c_{\pm }$ we get a larger peak of the probability distribution. The
parameters $c_{+}$ and $c_{-}$ control asymmetry of the probability
distribution while $\lambda _{-}$ and $\lambda _{+}$ determine the rate of
exponential decay as $\xi \rightarrow \pm \infty .$

Consider the asymptotic behavior of KoBoL exponent $\psi \left( \xi \right) $
in the strip $\Im \xi \in \left( \lambda _{-},\lambda _{+}\right) $ as $%
\left\vert \xi \right\vert \rightarrow \infty $. In what follows we shall
adapt the standard notations, $z^{\nu }=\exp \left( \nu \ln z\right) $,
where $\nu ,z\in \mathbb{C}$ such that $z\not\in (-\infty ,0]$ and $\ln z$
denotes the branch of $\ln z$ defined on $\mathbb{C}\setminus (-\infty ,0]$
and such that that $\ln (1)=0$.

{\bf Lemma 14}
\begin{em}
\label{asymp-exponent} Let $c_{+}=c_{-}=c>0$, $\xi =\rho \exp \left( i\phi
\right) $ and ${\rm Im}\xi \in \left( \lambda _{-},\lambda _{+}\right) $.
Then%
\[
{\rm Re}\psi \left( \xi \right) \sim -\rho ^{\nu }2c\Gamma \left( -\nu
\right) \cos \left( \frac{\pi \nu }{2}\right)
\]%
if ${\rm Re}\xi \rightarrow \infty $ and
\[
{\rm Re}\psi \left( \xi \right) \sim -\rho ^{\nu }2c\Gamma \left( -\nu
\right) \cos \left( \frac{\pi \nu }{2}\right) \cos \left( \nu \pi \right)
\]%
if ${\rm Re}\xi \rightarrow -\infty $.
\end{em}

{\bf Proof.}
Clearly
\[
\left( -\lambda _{-}\right) ^{\nu }-\left( -\lambda _{-}-i\xi \right) ^{\nu
}\sim -\rho ^{\nu }\exp \left( i\left( -\frac{\pi }{2}+\phi \right) \nu
\right)
\]%
and%
\[
\lambda _{+}^{\nu }-\left( \lambda _{+}+i\xi \right) ^{\nu }\sim -\rho ^{\nu
}\exp \left( i\left( \frac{\pi }{2}+\phi \right) \nu \right) \text{ \ }
\]%
as $\rho \rightarrow \infty $. Since $c_{-}=c_{+}=c$ then
\[
\psi \left( \xi \right) =-i\mu \xi +c_{-}\Gamma \left( -\nu \right) \left(
\left( -\lambda _{-}\right) ^{\nu }-\left( \lambda _{-}-i\xi \right) ^{\nu
}\right)
\]%
\[
+c_{+}\Gamma \left( -\nu \right) \left( \lambda _{+}^{\nu }-\left( \lambda
_{+}+i\xi \right) ^{\nu }\right)
\]%
\[
\sim -i\mu \rho \exp \left( i\phi \right) -2c\Gamma \left( -\nu \right) \exp
\left( i\nu \phi \right) \cos \left( \frac{\pi \nu }{2}\right) \rho ^{\nu }.
\]%
Hence
\begin{equation}
{\rm Re}\psi \left( \xi \right) \sim \rho \mu \sin \phi +\rho ^{\nu
}2c\left( -\Gamma \left( -\nu \right) \cos \left( \frac{\pi \nu }{2}\right)
\right) \cos \left( \nu \phi \right) .  \label{asymp-kobol}
\end{equation}%
To complete the proof we remark that $\phi \rightarrow 0$ if ${\rm Re}\xi
\rightarrow \infty $ and $\phi \rightarrow \pi $ if ${\rm Re}\xi
\rightarrow -\infty $ in the strip ${\rm Im}\xi \in \left( \lambda
_{-},\lambda _{+}\right) $.
$\square$

{\bf Corollary 15}
\begin{em}
\label{cor1} Let $\nu \in \left( 0,1/2\right) $, $\xi =\rho \exp \left(
i\phi \right) $ and ${\rm Im}\xi \in \left( \lambda _{-},\lambda
_{+}\right) $. Then the respective characteristic function $\Phi \left( \xi
,t\right) =\exp \left( -t\psi \left( \xi \right) \right) $ can be estimated
as
\[
\left\vert \Phi \left( \xi ,t\right) \right\vert =\left\vert \exp \left(
-t\psi \left( \xi \right) \right) \right\vert
\]%
\[
\leq \exp \left( -t{\rm Re}\psi \left( \xi \right) \right) \lesssim \exp
\left( 2t\rho ^{\nu }c\Gamma \left( -\nu \right) \cos \left( \frac{\pi \nu }{%
2}\right) \right)
\]%
if ${\rm Re}\xi \rightarrow \infty $ and
\[
\left\vert \Phi \left( \xi ,t\right) \right\vert \lesssim \exp \left( 2t\rho
^{\nu }c\Gamma \left( -\nu \right) \cos \left( \frac{\pi \nu }{2}\right)
\cos \left( \nu \pi \right) \right)
\]%
if ${\rm Re}\xi \rightarrow -\infty $. In particular, if $\nu \in \left(
0,1/2\right) $ then%
\[
tc\Gamma \left( -\nu \right) \cos \left( \frac{\pi \nu }{2}\right) \cos
\left( \nu \pi \right) <0
\]
and
\[
\left\vert \Phi \left( \xi ,t\right) \right\vert \ll \exp \left(
-Ct\left\vert \xi \right\vert ^{\nu }\right) \text{, \ }\left\vert \xi
\right\vert \rightarrow \infty \text{, }{\rm Im}\xi \in \left( \lambda
_{-},\lambda _{+}\right) .
\]
\end{em}

{\bf Example 16}
\begin{em}
At this point we present two more important examples of characteristic
exponents $\psi (\xi )$ which are of practical interest in empirical studies
of financial markets. Remark that Madan and collaborators \cite{m1}, \cite%
{m3} were first who applied Variance Gamma processes in studies of financial
markets. The respective characteristic exponent has the form
\[
\psi (\xi )=-i\mu \xi +c_{+}[\ln (-\lambda _{-}-i\xi )-\ln (-\lambda
_{-})]+c_{-}[\ln (\lambda _{+}+i\xi )-\ln (\lambda _{+})],
\]%
where $\lambda _{-}<0<\lambda _{+}$, $c>0$ and $\mu \in \mathbb{R}$. A
Variance Gamma process with these parameters is also a L\'{e}vy process of
exponential type $(\lambda _{-},\lambda _{+})$.

So-called Normal Inverse Gaussian processes were introduced and studied by
Barndorff-Nielsen \cite{bn1}-\cite{bn6}. The respective characteristic
exponent is
\[
\psi (\xi )=-i\mu \xi +\delta \left[ \left( \alpha ^{2}-(\beta +i\xi
)^{2}\right) ^{\nu /2}-(\alpha ^{2}-\beta ^{2})^{\nu /2}\right] .
\]
\end{em}

\section{Representations of density functions of KoBoL processes}

{\bf Definition 17}
\begin{em}
\label{lambda-def} \bigskip
For a fixed $R>0$ consider two piecewise smooth
curves%
\[
\lambda _{+}\left( x\right) :=x+i\left( \alpha _{+}+a_{+}\left( x\right)
\right) :\left[ -R,R\right] \rightarrow \left\{ z\left\vert {\rm Im}%
z>0\right. \right\} \text{,}
\]%
and%
\[
\lambda _{-}\left( x\right) :=x+i\left( \alpha _{-}+a_{-}\left( x\right)
\right) :\left[ -R,R\right] \rightarrow \left\{ z\left\vert {\rm Im}%
z<0\right. \right\} \text{,}
\]%
where $\alpha _{+}>0$, $a_{+}\left( x\right) \geq 0$ is an even function
increasing on $\left[ 0,R\right] $ and decreasing on $\left[ -R,0\right] $.
Similarly, $\alpha _{-}<0$, $a_{-}\left( x\right) \leq 0$ is an even
function increasing on $\left[ -R,0\right] $ and decreasing on $\left[ 0,R%
\right] $. Consider six contours,%
\[
\gamma _{1}\left( R\right) :=\left\{ z\left\vert z=\left\vert \lambda
_{+}\left( R\right) \right\vert \mathrm{\exp }\left( i\phi \right) \text{, \
}\phi \in \left[ \arg \left( R+i\left( \alpha _{+}+a_{+}\left( R\right)
\right) \right) ,0\right] \right. \right\} ,
\]%
\[
\gamma _{2}\left( R\right) :=\left[ \left\vert \lambda _{+}\left( R\right)
\right\vert ,\left\vert \lambda _{-}\left( R\right) \right\vert \right] ,
\]%
\[
\gamma _{3}\left( R\right) :=\left\{ z\left\vert z=\left\vert \lambda
_{-}\left( R\right) \right\vert \mathrm{\exp }\left( i\phi \right) \text{, \
}\phi \in \left[ 0,\arg \left( R+i\left( \alpha _{-}+a_{-}\left( R\right)
\right) \right) \right] \right. \right\} ,
\]%
\[
\gamma _{4}\left( R\right) :=\left\{ z\left\vert z=\left\vert \lambda
_{-}\left( -R\right) \right\vert \mathrm{\exp }\left( i\phi \right) \text{,
\ }\phi \in \left[ \arg \left( -R+i\left( \alpha _{-}+a_{-}\left( -R\right)
\right) \right) ,-\pi \right] \right. \right\} ,
\]%
\[
\gamma _{5}\left( R\right) :=\left[ \left\vert \lambda _{-}\left( -R\right)
\right\vert ,\left\vert \lambda _{+}\left( -R\right) \right\vert \right] ,
\]%
\[
\gamma _{6}\left( R\right) :=\left\{ z\left\vert z=\left\vert \lambda
_{+}\left( -R\right) \right\vert \mathrm{\exp }\left( i\phi \right) \text{,
\ }\phi \in \left[ \arg \left( -\pi ,-R+i\left( \alpha _{+}+a_{+}\left(
-R\right) \right) \right) \right] \right. \right\} .
\]

We say that a L\'{e}vy process $X=\left\{ X_{t}\text{, }t>0\right\} $ is $%
\left( \lambda _{-},\lambda _{+}\right) $-analytic if for any $R>0$ its
characteristic exponent $\psi \left( z\right) $ admits analytic extension
into the domain $\Omega _{R}$ bonded by
\[
\lambda _{-}\left( \cdot \right) \cup \lambda _{+}\left( \cdot \right) \cup
\bigcup_{k=1}^{6}\gamma _{k}\left( R\right)
\]%
and%
\[
\lim_{R\rightarrow \infty }\int_{\gamma _{k}\left( R\right) }\exp \left(
iyz-t\psi \left( z\right) \right) dz=0\text{, \ }1\leq k\leq 6\text{, \ }%
t,y>0
\]%
for any $t>0$, $y>0$.
\end{em}

Recall that in the case of European call option the reward function has the
form $H\left( y\right) =\max \left\{ S_{0}\exp \left( y\right) -K,0\right\} $%
. Hence we need just to consider the case $0<\ln \left( K/S_{0}\right) <y$.
Observe that usually $K>S_{0}$ because of potential earning capasity of the
stock $S_{t}$ during the time interval $\left( 0,T\right] $.

A useful representation of density functions is given by the following
statement.

{Theorem 18}
\begin{em}
Let $X=\left\{ X_{t}\text{, }t>0\right\} $ be a $\left( \lambda _{-},\lambda
_{+}\right) $-analytic process. Then%
\[
p_{t}\left( y\right) =\frac{1}{2\pi \left( \exp \left( \alpha _{+}y\right)
+\exp \left( \alpha _{-}y\right) \right) }
\]%
\[
\times \int_{\mathbb{R}}\exp \left( iy\left( x+ia_{+}\left( x\right) \right)
\right) N\left( x\right) +\exp \left( iy\left( x+ia_{-}\left( x\right)
\right) \right) M\left( x\right) dx,
\]%
where%
\[
N\left( x\right) :=\exp \left( -t\psi \left( x+i\left( \alpha
_{+}+a_{+}\left( x\right) \right) \right) \right) \left( 1+i\dot{a}%
_{+}\left( x\right) \right)
\]%
and%
\[
M\left( x\right) :=\exp \left( -t\psi \left( x+i\left( \alpha
_{-}+a_{-}\left( x\right) \right) \right) \right) \left( 1+i\dot{a}%
_{-}\left( x\right) \right) .
\]
\end{em}

{\bf Proof.}
Let $\gamma _{7}\left( R\right) :=\left\{ z\left\vert z=x+i\lambda
_{+}\left( x\right) \text{, \ }x\in \left[ -R,R\right] \right. \right\} $.
Since the process $X=\left\{ X_{t}\text{, }t\geq 0\right\} $ is $\left(
\lambda _{-},\lambda _{+}\right) $-analytic then using Cauchy theorem we get
\[
\int_{\gamma _{1}\cup \left[ \left\vert \lambda _{+}\left( R\right)
\right\vert ,-\left\vert \lambda _{+}\left( -R\right) \right\vert \right]
\cup \gamma _{6}\left( R\right) \cup \gamma _{7}\left( R\right) }\exp \left(
iyz-t\psi \left( z\right) \right) dz=0.
\]%
Applying $\left( \lambda _{-},\lambda _{+}\right) $-analyticity and letting $%
R\rightarrow \infty $ we get%
\[
p_{t}\left( y\right) =\frac{1}{2\pi }\int_{\mathbb{R}}\exp \left( iy\xi
-t\psi \left( \xi \right) \right) d\xi
\]%
\[
=\frac{1}{2\pi }\lim_{R\rightarrow \infty }\int_{\left[ -R,R\right] }\exp
\left( iy\xi -t\psi \left( \xi \right) \right) d\xi
\]%
\[
=\frac{\exp \left( -\alpha _{+}y\right) }{2\pi }
\]%
\[
\times \int_{\mathbb{R}}\exp \left( iyx-ya_{+}\left( x\right) -t\psi \left(
x+i\alpha _{+}+ia_{+}\left( x\right) \right) \right) \left( 1+i\dot{a}%
_{+}\left( x\right) \right) dx.
\]%
Similarly,
\[
p_{t}\left( y\right) =\frac{\exp \left( -\alpha _{-}y\right) }{2\pi }
\]%
\[
\times \int_{\mathbb{R}}\exp \left( iyx-ya_{-}\left( x\right) -t\psi \left(
x+i\alpha _{-}+ia_{-}\left( x\right) \right) \right) \left( 1+i\dot{a}%
_{-}\left( x\right) \right) dx.
\]%
The proof follows from the last two representations of $p_{t}$.
$\square$

In particular, if $X=\left\{ X_{t}\text{, }t>0\right\} $ is a $\left(
0,\lambda _{+}\right) $-analytic process then the respective density
function $p_{t}\left( y\right) $ can be represented as%
\[
p_{t}\left( y\right) =\frac{\exp \left( -\alpha _{+}y\right) }{2\pi }
\]%
\[
\times \int_{\mathbb{R}}\exp \left( iyx-ya_{+}\left( x\right) -t\psi \left(
x+i\alpha _{+}+ia_{+}\left( x\right) \right) \right) \left( 1+i\dot{a}%
_{+}\left( x\right) \right) dx.
\]

The following statement gives a wide range of examples of $\left( 0,\lambda
_{+}\right) $-analytic processes.

{Theorem 19}
\begin{em}
Any KoBoL process with parameters $\mu \geq 0$, $c_{+}=c_{-}=c>0$ and $\nu
\in \left( 0,1/2\right) $ is $\left( 0,\lambda _{+}\right) -$analytic L\'{e}%
vy process.
\end{em}

{\bf Proof.}
Clearly, characteristic exponent $\psi \left( \xi \right) $ given by (\ref%
{exp11}) is analytic in the domain
\[
\mathbb{C\setminus }\left\{ \left[ i\lambda _{+},i\infty \right) \cup \left[
i\lambda _{-},-i\infty \right) \right\} .
\]%
Hence it is sufficient to show that
\[
\lim_{R\rightarrow \infty }I_{+}\left( R,y,t\right) =\lim_{R\rightarrow
\infty }I_{-}\left( R,y,t\right) =0,
\]%
where%
\begin{equation}
I_{+}\left( R,y,t\right) :=\left\vert \int_{\xi \in \Gamma _{R}^{+}}\exp
\left( iy\xi -t\psi \left( \xi \right) \right) d\xi \right\vert ,  \label{l+}
\end{equation}%
\begin{equation}
I_{-}\left( R,y,t\right) :=\left\vert \int_{\xi \in \Gamma _{R}^{-}}\exp
\left( iy\xi -t\psi \left( \xi \right) \right) d\xi \right\vert ,  \label{1-}
\end{equation}%
\[
\Gamma _{R}^{+}=\left\{ \xi \left\vert \xi =\left\vert \lambda _{+}\left(
R\right) \right\vert \exp \left( i\phi \right) \text{, \ }\phi \in \left[
0,\arg \left( R+ia_{+}\left( R\right) \right) \right] \right. \right\}
\]%
and%
\[
\Gamma _{R}^{-}=\left\{ \xi \left\vert \xi =\left\vert \lambda _{+}\left(
-R\right) \right\vert \exp \left( i\phi \right) \text{, \ }\phi \in \left[
\pi ,\arg \left( -R+ia_{+}\left( -R\right) \right) \right] \right. \right\} .
\]%
Let us present estimates for the integral (\ref{l+}) first. Clearly,%
\[
I_{+}\left( R,y,t\right)
\]%
\[
=\left\vert \int_{0}^{\arg \left( R+i\alpha _{+}+ia_{+}\left( R\right)
\right) }\exp \left( iyR\exp \left( i\phi \right) -t\psi \left( R\exp \left(
i\phi \right) \right) \right) Ri\exp \left( i\phi \right) d\phi \right\vert
\]%
\[
\leq R\left\vert \int_{0}^{\arg \left( R+i\alpha _{+}+ia_{+}\left( R\right)
\right) }\exp \left( iyR\exp \left( i\phi \right) -t\psi \left( R\exp \left(
i\phi \right) \right) \right) d\phi \right\vert
\]%
\begin{equation}
\leq R\int_{0}^{\arg \left( R+i\alpha _{+}+ia_{+}\left( R\right) \right)
}\chi \left( y,R,\nu ,\phi \right) d\phi ,  \label{re-exp2}
\end{equation}%
where%
\[
\chi \left( y,R,\nu ,\phi ,t\right) :=\exp \left( {\rm Re}\left( iyR\exp
\left( i\phi \right) -t\psi \left( R\exp \left( i\phi \right) \right)
\right) \right) .
\]%
Since $y\geq 0$, $\mu \geq 0$, $c>0$, $t>0$ and $-\Gamma \left( -\nu \right)
\cos \left( \pi \nu /2\right) >0$ if $\nu \in \left( 0,1/2\right) $ then
applying (\ref{asymp-kobol}) we get%
\[
\chi \left( y,R,\nu ,\phi ,t\right)
\]%
\[
\leq C\exp \left( -yR\sin \phi -tR\mu \sin \phi -2ctR^{\nu }\left( -\Gamma
\left( -\nu \right) \cos \left( \frac{\pi \nu }{2}\right) \right) \cos
\left( \nu \phi \right) \right)
\]%
\begin{equation}
\leq C\exp \left( -2ctR^{\nu }\left( -\Gamma \left( -\nu \right) \cos \left(
\frac{\pi \nu }{2}\right) \right) \cos \left( \nu \phi \right) \right)
\label{re-exp3}
\end{equation}%
Comparing and ({\ref{re-exp2}) and (\ref{re-exp3}) we get
\[
I_{+}\left( R,y,t\right) \leq CR\int_{0}^{\pi /2}\exp \left( -2ctR^{\nu
}\left( -\Gamma \left( -\nu \right) \cos \left( \frac{\pi \nu }{2}\right)
\right) \left( 1-\frac{2\nu }{\pi }\phi \right) \right) d\phi ,
\]%
where we used the fact that }$\cos \left( \nu \phi \right) \geq 1-2\phi \nu
/\pi $ if $\phi \in \left[ 0,\pi /2\right] $. This means that
\[
I_{+}\left( R,y,t\right) \leq CR\exp \left( -2ctR^{\nu }\left( -\Gamma
\left( -\nu \right) \cos \left( \frac{\pi \nu }{2}\right) \right) \right)
\]%
\[
\times \int_{0}^{\pi /2}\exp \left( 2ctR^{\nu }\left( -\Gamma \left( -\nu
\right) \cos \left( \frac{\pi \nu }{2}\right) \right) \frac{2\nu }{\pi }\phi
\right) d\phi
\]%
\[
=\frac{\pi CR^{1-\nu }}{4ct\nu \left( -\Gamma \left( -\nu \right) \cos
\left( \frac{\pi \nu }{2}\right) \right) }\exp \left( -2ctR^{\nu }\left(
-\Gamma \left( -\nu \right) \cos \left( \frac{\pi \nu }{2}\right) \right)
\right)
\]%
\[
\times \left( \exp \left( 2ctR^{\nu }\left( -\Gamma \left( -\nu \right) \cos
\left( \frac{\pi \nu }{2}\right) \right) \nu \right) -1\right)
\]%
\[
\leq \frac{\pi CR^{1-\nu }}{4ct\nu \left( -\Gamma \left( -\nu \right) \cos
\left( \frac{\pi \nu }{2}\right) \right) }\exp \left( -2ctR^{\nu }\left(
-\Gamma \left( -\nu \right) \cos \left( \frac{\pi \nu }{2}\right) \right)
\left( 1-\nu \right) \right) .
\]%
From the last line we see that
\[
I_{+}\left( R,y,t\right) \ll R^{1-\nu }\exp \left( -CtR^{\nu }\right) ,\text{
\ }R\rightarrow \infty
\]%
for any $y>0$. Now we get estimates of the integral (\ref{1-}). In this case
$\phi \in \left[ \arg \left( -R+ia_{+}\left( -R\right) \right) ,\pi \right]
\subset \left[ \pi /2,\pi \right] $. Hence $\left( -\Gamma \left( -\nu
\right) \cos \left( \pi \nu /2\right) \right) \cos \left( \nu \phi \right)
>0 $ since $\nu \in \left( 0,1/2\right) $. Recall that $y>0$, $\mu >0$ and $%
t>0$. Applying the same line of arguments as above we get
\[
I_{-}\left( R,y,t\right) \leq CR\exp \left( -2ctR^{\nu }\left( -\Gamma
\left( -\nu \right) \cos \left( \frac{\pi \nu }{2}\right) \right) \right)
\]%
\[
\times \int_{\pi /2}^{\pi }\exp \left( 2ctR^{\nu }\left( -\Gamma \left( -\nu
\right) \cos \left( \frac{\pi \nu }{2}\right) \right) \frac{2\nu }{\pi }\phi
\right) d\phi
\]%
\[
=\frac{\pi CR^{1-\nu }}{4ct\nu \left( -\Gamma \left( -\nu \right) \cos
\left( \frac{\pi \nu }{2}\right) \right) }\exp \left( 2ctR^{\nu }\Gamma
\left( -\nu \right) \cos \left( \frac{\pi \nu }{2}\right) \right)
\]%
\[
\times \left( \exp \left( -4\nu ctR^{\nu }\Gamma \left( -\nu \right) \cos
\left( \frac{\pi \nu }{2}\right) \right) -\exp \left( -2\nu ctR^{\nu }\Gamma
\left( -\nu \right) \cos \left( \frac{\pi \nu }{2}\right) \right) \right)
\]%
\[
\ll R^{1-\nu }\exp \left( -Ct\left( 1-2\nu \right) R^{\nu }\right) ,\text{ \
}R\rightarrow \infty .
\]
$\square$

\bigskip

\bigskip

\chapter{Recovery of density functions in jump-diffusion models}

\label{High-dimensional density}

\bigskip

\section{\protect\bigskip Introduction}

Recall that the pricing formula has the form $V=\exp \left( -rT\right)
\mathbb{E}^{\mathbb{Q}}\left[ H\right] $, where $\mathbb{Q}$ is a fixed
equivalent martingale measure. Since the reward function $H$ has usually a
simple structure the main problem is to approximate the respective
risk-neutral density function $p_{T}^{\mathbb{Q}}$, where $T>0$ is a
maturity time. Hence it is important to construct a simple, saturation free
and adapted to the course of dimensionality method of approximation of
density functions which are important in the theory of spread options. Our
method is based on the Poisson summation formula and approximation of
density functions by harmonics in the respective exponential hyperbolic
cross. The advantage of this approach is that the application of the Poisson
summation formula gives a periodic extension of the density function of the
same smoothness as the original function \cite{55, 555} instead of known
approaches discussed in \cite{fang, fang1}. Also, this approach allows us to
get approximation formulas without application of numerical methods.

Approximation of smooth functions by subspaces of entire functions of
exponential type and $sk$-splines was considered in \cite{3, 5,
kl-april-2014, kushpel8, izv, kll, kushpel-grand, kushpel-b1, kushpel-b2,
kushpel9}. These methods are saturation free on a wide range of sets of
smooth functions (including analytic and entire functions) and give almost
optimal rate of convergence in the sense of respective $m$-widths. However,
application of these methods requires use of numerical methods. We shall not
discuss this line of research here.

\bigskip

\section{\protect\bigskip Representations of density functions}

Assume for simplicity that all characteristic exponents $\psi _{s}^{\left(
1\right) ,\mathbb{Q}},$ $1\leq s\leq n$ and $\psi _{m}^{\left( 2\right) ,%
\mathbb{Q}},$ $1\leq m\leq n$ in the Theorem 8
correspond to a KoBoL
process and hence are analytically extendable into the strips ${\rm Im}%
z_{s}\in \left[ \kappa _{s,-},\kappa _{s,+}\right] $ and ${\bf Im}z_{m}\in %
\left[ \kappa _{m,-},\kappa _{m,+}\right] $ respectively, where $\lambda
_{s,-}<\kappa _{s,-}<0<\kappa _{s,+}<\lambda _{s,+}$ and $\lambda
_{m,-}^{^{\prime }}<\kappa _{m,-}^{^{\prime }}<0<\kappa _{m,+}^{^{\prime
}}<\lambda _{m,+}^{^{\prime }}$, $1\leq s,m\leq n$. Let $b_{k,m}\geq 0$, $%
1\leq k,m\leq n$. It is easy to check that the function $\mathbf{z=}\left(
z_{1},\cdot \cdot \cdot ,z_{n}\right) \rightarrow \mathbf{\psi }\left(
\mathbf{z}\right) ,$
\[
\mathbf{\psi }\left( \mathbf{z}\right) =\sum_{s=1}^{n}\psi _{s}^{\left(
1\right) ,\mathbb{Q}}\left( z_{s}\right) +\sum_{m=1}^{n}\psi _{m}^{\left(
2\right) ,\mathbb{Q}}\left( \sum_{k=1}^{n}b_{k,m}z_{k}\right)
\]%
defined in Theorem 8
is analytically extendable into the domain%
\[
T_{n}:=\left( \bigcup_{s=1}^{n}\left\{ {\rm Im}z_{s}\in \left[
\kappa _{s,-},\text{ }\kappa _{s,+}\right] \right\} \right)
\]%
\begin{equation}
\text{ \ \ \ \ }\cap \left( \bigcup_{s=1}^{n}\left\{ {\rm Im}%
z_{s}\in \left[ \kappa _{s,-}^{^{\prime }}\left(
\sum_{k=1}^{n}b_{k,m}\right) ^{-1},\text{ }\kappa _{s,+}^{^{\prime }}\left(
\sum_{k=1}^{n}b_{k,m}\right) ^{-1}\right] \right\} \right) ,  \label{dom2015}
\end{equation}%
or $b_{-,s}\leq {\rm Im}z_{s}\leq b_{+,s},$ where%
\[
b_{-,s}:=\max \left\{ \kappa _{s,-},\kappa _{s,-}^{^{\prime }}\left(
\sum_{k=1}^{n}b_{k,m}\right) ^{-1}\right\} ,\text{ \ \ }
\]%
\[
b_{+,s}:=\min \left\{ \kappa _{s,+},\kappa _{s,+}^{^{\prime }}\left(
\sum_{k=1}^{n}b_{k,m}\right) ^{-1}\right\} ,
\]%
$1\leq s\leq n$ and $b_{k,m}\geq 0$, $1\leq k,m\leq n$. In this case $\Phi ^{%
\mathbb{Q}}\left( \mathbf{z,}t\right) =\Phi ^{\mathbb{Q}}\left( z_{1},\cdot
\cdot \cdot ,z_{n}\mathbf{,}t\right) $ admits an analytic extension into the
same domain $T_{n}\subset \mathbb{C}^{n}.$ Let $\mathbf{b}_{+}:=\left(
b_{+,1},\cdot \cdot \cdot ,b_{+,n}\right) $ and $\mathbf{b}_{-}:=\left(
b_{-,1},\cdot \cdot \cdot ,b_{-,n}\right) $.

{Theorem 20}
\begin{em}
\label{representation2014} Let $\psi _{s}^{\left( 1\right) ,\mathbb{Q}},$ $%
1\leq s\leq n$ and $\psi _{m}^{\left( 2\right) ,\mathbb{Q}},$ $1\leq m\leq n$
be defined by (\ref{representation-k}), i.e.%
\[
\psi _{s}^{\left( 1\right) ,\mathbb{Q}}\left( \xi _{s}\right) =-i\mu _{s}\xi
_{s}+c_{s}\Gamma \left( -\nu _{s}\right) \left( \left( -\lambda
_{-,s}\right) ^{\nu _{s}}-\left( -\lambda _{-,s}-i\xi _{s}\right) ^{\nu
_{s}}\right)
\]%
\[
\text{ \ \ \ \ \ \ \ \ \ \ }+c_{s}\Gamma \left( -\nu _{s}\right) \left(
\lambda _{+,s}^{\nu _{s}}-\left( \lambda _{+,s}+i\xi _{s}\right) ^{\nu
_{s}}\right) \text{, }\nu _{s}\in \left( 0,1/2\right)
\]%
and%
\[
\psi _{m}^{\left( 2\right) ,\mathbb{Q}}\left( \xi _{m}\right) =-i\mu _{m}\xi
_{m}+c_{m}\Gamma \left( -\nu _{m}\right) \left( \left( -\lambda
_{-,m}\right) ^{\nu _{m}}-\left( -\lambda _{-,m}-i\xi _{m}\right) ^{\nu
_{m}}\right)
\]%
\[
\text{ \ \ \ \ \ \ \ \ \ \ \ }+c_{m}\Gamma \left( -\nu _{m}\right) \left(
\lambda _{+,m}^{\nu _{m}}-\left( \lambda _{+,m}+i\xi _{m}\right) ^{\nu
_{m}}\right) \text{, \ }\nu _{m}\in \left( 0,1/2\right) ,
\]%
where $c_{s},c_{m}>0$, $\nu _{s},\nu _{m}\in \left( 0,1/2\right) $. Then the
respective density function $p_{T}^{\mathbb{Q}}\left( \cdot \right) $ can be
represented as
\[
p_{T}^{\mathbb{Q}}\left( \cdot \right) =\frac{1}{\left( 2\pi \right)
^{n}\left( \exp \left( \left\langle \cdot ,\mathbf{b}_{+}\right\rangle
\right) +\exp \left( \left\langle \cdot ,\mathbf{b}_{-}\right\rangle \right)
\right) }
\]%
\[
\text{ \ \ \ \ \ \ \ }\times \int_{\mathbb{R}^{n}}\exp \left( -i\left\langle
\cdot \mathbf{,z}\right\rangle \right) \left( \Phi ^{\mathbb{Q}}\left(
\mathbf{z-}i\mathbf{b}_{+}\mathbf{,}T\right) +\Phi ^{\mathbb{Q}}\left(
\mathbf{z-}i\mathbf{b}_{-}\mathbf{,}T\right) \right) d\mathbf{z}.
\]%
In particular, if $-\mathbf{b}_{-}=\mathbf{b}_{+}:=\mathbf{b}$ then
\[
p_{T}^{\mathbb{Q}}\left( \cdot \right) =\frac{1}{2\left( 2\pi \right) ^{n}}%
\left( \cosh \left( \left\langle \cdot ,\mathbf{b}\right\rangle \right)
\right) ^{-1}
\]%
\begin{equation}
\text{ \ \ \ \ \ \ \ }\times \int_{\mathbb{R}^{n}}\exp \left( -i\left\langle
\cdot \mathbf{,z}\right\rangle \right) \left( \Phi ^{\mathbb{Q}}\left(
\mathbf{z+}i\mathbf{b,}T\right) +\Phi ^{\mathbb{Q}}\left( \mathbf{z-}i%
\mathbf{b,}T\right) \right) d\mathbf{z}.  \label{pis1}
\end{equation}

Let
\[
\Phi _{1}^{\mathbb{Q}}\left( \mathbf{z,}T\right) :=\Phi ^{\mathbb{Q}}\left(
\mathbf{z+}i\mathbf{e}_{1}b_{1}\mathbf{,}T\right) +\Phi ^{\mathbb{Q}}\left(
\mathbf{z-}i\mathbf{e}_{1}b_{1}\mathbf{,}T\right) ,
\]%
\[
\Phi _{k}^{\mathbb{Q}}\left( \mathbf{z,}T\right) :=\Phi _{k-1}^{\mathbb{Q}%
}\left( \mathbf{z+}i\mathbf{e}_{k}b_{k}\mathbf{,}T\right) +\Phi _{k-1}^{%
\mathbb{Q}}\left( \mathbf{z-}i\mathbf{e}_{k}b_{k}\mathbf{,}T\right) ,\text{ }%
2\leq k\leq n.
\]%
Then
\begin{equation}
p_{T}^{\mathbb{Q}}\left( \mathbf{x}\right) =\frac{1}{2^{2n}\pi ^{n}}\left(
\prod_{s=1}^{n}\cosh \left( b_{s}x_{s}\right) \right) ^{-1}\int_{%
\mathbb{R}^{n}}\exp \left( -i\left\langle \mathbf{x,z}\right\rangle \right)
\Phi _{n}^{\mathbb{Q}}\left( \mathbf{z,}T\right) d\mathbf{z}.  \label{pis2}
\end{equation}
\end{em}

{\bf Proof.}
We shall prove just (\ref{pis1}) since (\ref{pis2}) follows in a similar
manner. In our notation density function can be represented as
\[
p_{T}^{\mathbb{Q}}\left( \cdot \right) =\left( 2\pi \right) ^{-n}\int_{%
\mathbb{R}^{n}}\exp \left( -i\left\langle \cdot \mathbf{,z}\right\rangle
\right) \Phi ^{\mathbb{Q}}\left( \mathbf{z,}T\right) d\mathbf{z}=\left( 2\pi
\right) ^{-n}\mathbf{F}\left( \Phi ^{\mathbb{Q}}\left( \mathbf{z,}T\right)
\right) \left( \cdot \right) .
\]%
Recall that $\psi _{s}^{\left( 1\right) ,\mathbb{Q}}\left( \xi _{s}\right) $%
, $1\leq s\leq n$ admits an analytic extension into the strip ${\rm Im}\xi
_{s}\in \left[ \kappa _{s,-},\kappa _{s,+}\right] $, where $\lambda
_{s,-}<\kappa _{s,-}<0<\kappa _{s,+}<\lambda _{s,+}$, $1\leq s\leq n$ and $%
\psi _{m}^{\left( 2\right) ,\mathbb{Q}}\left( \xi _{m}\right) $, $1\leq
m\leq n$ admits an analytic extension into the strip ${\rm Im}\xi _{m}\in %
\left[ \kappa _{m,-}^{^{\prime }},\kappa _{m,+}^{^{\prime }}\right] $, where
$\lambda _{m,-}^{^{\prime }}<\kappa _{m,-}^{^{\prime }}<0<\kappa
_{m,+}^{^{\prime }}<\lambda _{m,+}^{^{\prime }}$, $1\leq m\leq n.$ From
Corollary 15
 it follows that%
\[
\left\vert \Phi \left( \mathbf{z,}T\right) \right\vert =\left\vert \exp
\left( -T\mathbf{\psi }\left( \mathbf{z}\right) \right) \right\vert
\]%
\[
=\left\vert \exp \left( -T\left( \sum_{s=1}^{n}\psi _{s}^{\left( 1\right)
}\left( z_{s}\right) +\sum_{m=1}^{n}\psi _{m}^{\left( 2\right) }\left(
\sum_{k=1}^{n}b_{k,m}z_{k}\right) \right) \right) \right\vert
\]%
\[
\ll \left\vert \exp \left( -T\sum_{s=1}^{n}\psi _{s}^{\left( 1\right)
}\left( z_{s}\right) \right) \right\vert
\]%
\begin{equation}
\ll \exp \left( -CT\sum_{s=1}^{n}\left\vert z_{s}\right\vert ^{\nu
_{s}}\right) ,  \label{asymp2}
\end{equation}%
where $\left\vert z_{k}\right\vert \rightarrow \infty ,$ $z_{k}\in
T_{n},1\leq k\leq n$, where the domain $T_{n}$ is defined by (\ref{dom2015}%
). Hence, applying Cauchy theorem (see, e.g. \cite{bp11}) $n$ times in the
domain $T_{n}$, which is justified by (\ref{asymp2}), we get
\[
p_{T}^{\mathbb{Q}}\left( \cdot \right) =\left( 2\pi \right) ^{-n}\int_{%
\mathbb{R}^{n}}\exp \left( -i\left\langle \cdot \mathbf{,z}\right\rangle
\right) \Phi ^{\mathbb{Q}}\left( \mathbf{z,}T\right) d\mathbf{z}
\]%
\[
\text{ \ \ \ \ \ \ \ }=\left( 2\pi \right) ^{-n}\int_{\mathbb{R}^{n}+i%
\mathbf{b}_{+}}\exp \left( -i\left\langle \cdot \mathbf{,z}\right\rangle
\right) \Phi ^{\mathbb{Q}}\left( \mathbf{z,}T\right) d\mathbf{z}
\]%
\[
\text{ \ \ \ \ \ \ \ }=\left( 2\pi \right) ^{-n}\int_{\mathbb{R}^{n}}\exp
\left( -i\left\langle \cdot \mathbf{,z+}i\mathbf{b}_{+}\right\rangle \right)
\Phi ^{\mathbb{Q}}\left( \mathbf{z+}i\mathbf{b}_{+}\mathbf{,}T\right) d%
\mathbf{z}
\]%
\[
\text{ \ \ \ \ \ \ \ }=\exp \left( \left\langle \cdot ,\mathbf{b}%
_{+}\right\rangle \right) \left( 2\pi \right) ^{-n}\int_{\mathbb{R}^{n}}\exp
\left( -i\left\langle \cdot \mathbf{,z}\right\rangle \right) \Phi ^{\mathbb{Q%
}}\left( \mathbf{z+}i\mathbf{b}_{+}\mathbf{,}T\right) d\mathbf{z,}
\]%
or%
\begin{equation}
p_{T}^{\mathbb{Q}}\left( \cdot \right) \exp \left( -\left\langle \cdot ,%
\mathbf{b}_{+}\right\rangle \right) =\left( 2\pi \right) ^{-n}\int_{\mathbb{R%
}^{n}}\exp \left( -i\left\langle \cdot \mathbf{,z}\right\rangle \right) \Phi
^{\mathbb{Q}}\left( \mathbf{z+}i\mathbf{b}_{+}\mathbf{,}T\right) d\mathbf{z.}
\label{ptq1}
\end{equation}%
Similarly,
\begin{equation}
p_{T}^{\mathbb{Q}}\left( \cdot \right) \exp \left( -\left\langle \cdot ,%
\mathbf{b}_{-}\right\rangle \right) =\left( 2\pi \right) ^{-n}\int_{\mathbb{R%
}^{n}}\exp \left( -i\left\langle \cdot \mathbf{,z}\right\rangle \right) \Phi
^{\mathbb{Q}}\left( \mathbf{z+}i\mathbf{b}_{-}\mathbf{,}T\right) d\mathbf{z.}
\label{ptq2}
\end{equation}%
Comparing (\ref{ptq1}) and (\ref{ptq2}) we get the proof.
$\square$

\section{Approximation of density functions by Poisson summation}

We will need the following result which is known as the Poisson summation
formula.

{\bf Theorem 21}
\begin{em}
\label{Poisson summation} (\cite{stain} p. 252) Suppose that for some $A>0$
and $\delta >0$ we have
\[
\max \left\{ f\left( \mathbf{x}\right) ,\mathbf{F}f\left( \mathbf{x}\right)
\right\} \leq A\left( 1+\left\vert \mathbf{x}\right\vert \right) ^{-n-\delta
}.
\]%
Then
\[
\sum_{\mathbf{m}\in \mathbb{Z}^{n}}f\left( \mathbf{x}+P\mathbf{m}\right) =%
\frac{1}{P^{n}}\sum_{\mathbf{m}\in \mathbb{Z}^{n}}\mathbf{F}\left( f\right)
\left( \frac{2\pi }{P}\mathbf{m}\right) \exp \left( \frac{2\pi i}{P}%
\left\langle \mathbf{m},\mathbf{x}\right\rangle \right)
\]%
for any $P>0$. The series converges absolutely.
\end{em}

Assume that $\nu _{s},\nu _{m}\in \left( 0,1/2\right) $, $1\leq s,m\leq n$
as before. Put
\[
\text{ \ \ }\widetilde{M}:=\frac{1}{2^{2n}\pi ^{n}}\left\Vert \int_{\mathbb{R%
}^{n}}\exp \left( -i\left\langle \cdot \mathbf{,v}\right\rangle \right) \Phi
_{n}^{\mathbb{Q}}\left( \mathbf{v,}T\right) d\mathbf{v}\right\Vert
_{L_{\infty }\left( \mathbb{R}^{n}\right) },
\]%
Observe that $\widetilde{M}<\infty $ because of the estimate (\ref{asymp2}).
Fix $T>0$, $\epsilon >0$ and select such $P\in \mathbb{N}$ that
\begin{equation}
\widetilde{M}\prod_{s=1}^{n}\sum_{m_{k}\in \mathbb{Z},\mathbf{m}\neq
\mathbf{0}}\left( \cosh \left( b_{s}\frac{2m_{k}-1}{2}P\right) \right)
^{-1}\leq \epsilon ,  \label{epsilon111}
\end{equation}%
where $\mathbf{m}=\left( m_{1},\cdot \cdot \cdot ,m_{n}\right) $ \ and $%
\mathbf{m}\neq \mathbf{0}$ means $\mathbf{m}\neq \left( 0,\cdot \cdot \cdot
,0\right) $. Clearly
\begin{equation}
\epsilon \asymp \exp \left( -\frac{P}{2}\min \left\{ b_{s}\left\vert 1\leq
s\leq n\right. \right\} \right) ,\text{ \ }P\rightarrow \infty .
\label{eps1}
\end{equation}

{\bf Theorem 22}
\begin{em}
\label{desity 111} Let $Q_{n}:=\left\{ \mathbf{x}\left\vert \mathbf{x}%
=\left( x_{1},\cdot \cdot \cdot ,x_{n}\right) \in \mathbb{R}^{n},\text{ }%
\left\vert x_{k}\right\vert \leq 1,\text{ }1\leq k\leq n\right. \right\} $
be the unit cube in $\mathbb{R}^{n}$ and $-\mathbf{b}_{-}=\mathbf{b}_{+}:=%
\mathbf{b}$. Then in our notation%
\[
E_{1}\left( P\right) :=\left\Vert p_{T}^{\mathbb{Q}}\left( \mathbf{x}\right)
-\frac{1}{P^{n}}\sum_{\mathbf{m}\in \mathbb{Z}^{n}}\Phi ^{\mathbb{Q}}\left( -%
\frac{2\pi }{P}\mathbf{m},T\right) \exp \left( \frac{2\pi i}{P}\left\langle
\mathbf{m},\mathbf{x}\right\rangle \right) \right\Vert _{L_{q}\left( \frac{P%
}{2}Q_{n}\right) }
\]%
\[
\ll \exp \left( -\frac{P}{2}\min \left\{ b_{s}\left\vert 1\leq s\leq
n\right. \right\} \right) P^{n/q},\text{ }P\rightarrow \infty ,
\]%
where $1\leq p\leq \infty .$
\end{em}

{Proof.}
Using Theorem 20
we get
\begin{equation}
p_{T}^{\mathbb{Q}}\left( \mathbf{x}\right) =p_{T}^{\mathbb{Q}}\left(
x_{1},\cdot \cdot \cdot ,x_{n}\right) \leq \left(
\prod_{s=1}^{n}\cosh \left( b_{s}x_{s}\right) \right) ^{-1}%
\widetilde{M}.  \label{majorant111}
\end{equation}%
Applying (\ref{majorant111}) we can check that the conditions of Theorem 21
are satisfied. Hence using condition (\ref{epsilon111})
we get%
\[
\left\Vert p_{T}^{\mathbb{Q}}\left( \mathbf{x}\right) -\sum_{\mathbf{m}\in
\mathbb{Z}^{n}}p_{T}^{\mathbb{Q}}\left( \mathbf{x}+P\mathbf{m}\right)
\right\Vert _{L_{\infty }\left( \frac{P}{2}Q_{n}\right) }
\]%
\[
=\left\Vert \sum_{\mathbf{m}\in \mathbb{Z}^{n}\diagdown \left\{ \mathbf{0}%
\right\} }p_{T}^{\mathbb{Q}}\left( \mathbf{x}+P\mathbf{m}\right) \right\Vert
_{L_{\infty }\left( \frac{P}{2}Q_{n}\right) }
\]%
\[
\leq \widetilde{M}\prod_{s=1}^{n}\sum_{m_{k}\in \mathbb{Z},\mathbf{m}%
\neq \mathbf{0}}\left( \cosh \left( b_{s}\frac{2m_{k}-1}{2}P\right) \right)
^{-1}\leq \epsilon .
\]%
Observe that
\[
\Phi ^{\mathbb{Q}}\left( -\mathbf{x,}T\right) =\left( 2\pi \right) ^{n}%
\mathbf{F}^{-1}\left( p_{T}^{\mathbb{Q}}\right) \left( -\mathbf{x}\right)
\]%
\[
\text{ \ \ \ \ \ \ \ \ \ \ \ \ \ \ \ }=\left( 2\pi \right) ^{n}\left( \frac{1%
}{\left( 2\pi \right) ^{n}}\int_{\mathbb{R}^{n}}\exp \left( i\left\langle
\mathbf{x,y}\right\rangle \right) p_{T}^{\mathbb{Q}}\left( -\mathbf{y}%
\right) d\mathbf{y}\right)
\]%
\[
\text{ \ \ \ \ \ \ \ \ \ \ \ \ \ \ \ }=\int_{\mathbb{R}^{n}}\exp \left(
i\left\langle -\mathbf{x,y}\right\rangle \right) p_{T}^{\mathbb{Q}}\left(
\mathbf{y}\right) d\mathbf{y}
\]%
\[
\text{ \ \ \ \ \ \ \ \ \ \ \ \ \ \ \ }=\mathbf{F}\left( p_{T}^{\mathbb{Q}%
}\right) \left( \mathbf{x}\right) .
\]%
Consequently,%
\[
\left\Vert p_{T}^{\mathbb{Q}}\left( \mathbf{x}\right) -\sum_{\mathbf{m}\in
\mathbb{Z}^{n}}p_{T}^{\mathbb{Q}}\left( \mathbf{x}+P\mathbf{m}\right)
\right\Vert _{L_{\infty }\left( \frac{P}{2}Q_{n}\right) }
\]%
\[
=\left\Vert p_{T}^{\mathbb{Q}}\left( \mathbf{x}\right) -\frac{1}{P^{n}}\sum_{%
\mathbf{m}\in \mathbb{Z}^{n}}\mathbf{F}\left( p_{T}^{\mathbb{Q}}\right)
\left( \frac{2\pi }{P}\mathbf{m}\right) \exp \left( \frac{2\pi i}{P}%
\left\langle \mathbf{m},\mathbf{x}\right\rangle \right) \right\Vert
_{L_{\infty }\left( \frac{P}{2}Q_{n}\right) }
\]%
\[
=\left\Vert p_{T}^{\mathbb{Q}}\left( \mathbf{x}\right) -\frac{1}{P^{n}}\sum_{%
\mathbf{m}\in \mathbb{Z}^{n}}\Phi ^{\mathbb{Q}}\left( -\frac{2\pi }{P}%
\mathbf{m},T\right) \exp \left( \frac{2\pi i}{P}\left\langle \mathbf{m},%
\mathbf{x}\right\rangle \right) \right\Vert _{L_{\infty }\left( \frac{P}{2}%
Q_{n}\right) }\leq \epsilon
\]%
and%
\[
\left\Vert p_{T}^{\mathbb{Q}}\left( \mathbf{x}\right) -\frac{1}{P^{n}}\sum_{%
\mathbf{m}\in \mathbb{Z}^{n}}\Phi ^{\mathbb{Q}}\left( -\frac{2\pi }{P}%
\mathbf{m},T\right) \exp \left( \frac{2\pi i}{P}\left\langle \mathbf{m},%
\mathbf{x}\right\rangle \right) \right\Vert _{L_{1}\left( \frac{P}{2}%
Q_{n}\right) }\leq \epsilon P^{n}.
\]%
Finally, applying Riesz-Thorin interpolation theorem (see Appendix II,
Theorem 43
and (\ref{eps1}) we obtain
\[
\left\Vert p_{T}^{\mathbb{Q}}\left( \mathbf{x}\right) -\frac{1}{P^{n}}\sum_{%
\mathbf{m}\in \mathbb{Z}^{n}}\Phi ^{\mathbb{Q}}\left( -\frac{2\pi }{P}%
\mathbf{m},T\right) \exp \left( \frac{2\pi i}{P}\left\langle \mathbf{m},%
\mathbf{x}\right\rangle \right) \right\Vert _{L_{q}\left( \frac{P}{2}%
Q_{n}\right) }
\]%
\[
\leq \epsilon P^{n/q}\asymp \exp \left( -\frac{P}{2}\min \left\{
b_{s}\left\vert 1\leq s\leq n\right. \right\} \right) P^{n/q},\text{ \ }%
P\rightarrow \infty .
\]
$\square$

Observe that according to (\ref{asymp2}) the function $\left\vert \Phi ^{%
\mathbb{Q}}\left( -\frac{2\pi }{P}\mathbf{m},T\right) \right\vert $
exponentially decays as $\left\vert \mathbf{m}\right\vert \rightarrow \infty
$. Hence the series%
\[
\widetilde{p}_{T}^{\mathbb{Q}}\left( \mathbf{x}\right) :=\frac{1}{P^{n}}%
\sum_{\mathbf{m}\in \mathbb{Z}^{n}}\Phi ^{\mathbb{Q}}\left( -\frac{2\pi }{P}%
\mathbf{m},T\right) \exp \left( \frac{2\pi i}{P}\left\langle \mathbf{m},%
\mathbf{x}\right\rangle \right)
\]%
converges absolutely and represents an infinitely differentiable and $\frac{P%
}{2}Q_{n}$-periodic function which will be denoted again by $\widetilde{p}%
_{T}^{\mathbb{Q}}\left( \mathbf{x}\right) $.

{\bf Example 23}
\begin{em}
Let
\[
p\left( x,y\right) =\left( 2\pi \right) ^{-1}\exp \left( -\frac{x^{2}+y^{2}}{%
2}\right) .
\]%
be a Gaussian density then its Fourier transform is $\exp \left(
-2^{-1}\left( x^{2}+y^{2}\right) \right) $. For a fixed $m$ and $P$ consider
the approximant
\[
g\left( p,m,P,x,y\right) :=\frac{1}{P^{2}}\sum_{\left\vert k\right\vert \leq
m}\sum_{\left\vert s\right\vert \leq m}\mathbf{F}\left( p\right) \left( -%
\frac{2\pi k}{P},-\frac{2\pi s}{P}\right) \exp \left( ikx\frac{2\pi }{P}+isy%
\frac{2\pi }{P}\right)
\]%
\[
=\frac{1}{P^{2}}\sum_{\left\vert k\right\vert \leq m}\sum_{\left\vert
s\right\vert \leq m}\exp \left( -\frac{2\pi ^{2}k^{2}}{P^{2}}-\frac{2\pi
^{2}s^{2}}{P^{2}}\right) \exp \left( ikx\frac{2\pi }{P}+isy\frac{2\pi }{P}%
\right) .
\]%
The respective error of approximation is
\[
\varepsilon \left( p,P,m\right) :=\max \left\{ \left\vert p\left( x,y\right)
-g\left( p,m,P,x,y\right) \right\vert ,\text{ }x,y\in \left[ -P/2,P/2\right]
\right\} .
\]%
In particular, let $n=2$, $P=6$ and $m=3$ then $\varepsilon \left(
p,P,m\right) =1.\,\allowbreak 747\,\times 10^{-3}$.
\end{em}

\bigskip

\section{Comparison of methods of approximation}

\bigskip

Theorem 22
shows an exponential rate of convergence of $%
\widetilde{p}_{T}^{\mathbb{Q}}\left( \mathbf{x}\right) $ which is given by
an infinite trigonometric series to $p_{T}^{\mathbb{Q}}\left( \mathbf{x}%
\right) $ if $\mathbf{x\in }\frac{P}{2}Q_{n}$ as $P\rightarrow \infty $. In
this section we discuss the problem of optimal recovery of density functions
in the sense of $m$-widths and $m$-cowidths. This allows us to compare a
wide range of numerical methods and construct a quasi optimal truncation of
the series $\widetilde{p}_{T}^{\mathbb{Q}}\left( \mathbf{x}\right) $.

Let $\left\{ \varphi _{k}\left( \mathbf{x}\right) ,k\in \mathbb{N}\right\} $
be a set of continuous orthonormal and uniformly bounded functions on a
measure space $\left( \Omega ,\mathcal{F},\upsilon \right) $. Let
\[
L:=\sup_{k\in \mathbb{N}}\left\Vert \varphi _{k}\right\Vert _{\infty
}<\infty .
\]%
For any $f\in L_{1}:=L_{1}\left( \Omega ,\mathcal{F},\upsilon \right) $ we
construct a formal Fourier series
\[
\mathtt{s}\left[ f\right] =\sum_{k=1}^{\infty }c_{k}\left( f\right) \varphi
_{k}\mathbf{,}\text{ \ }c_{k}\left( f\right) :=\int_{\Omega }f\overline{%
\varphi }_{k}dx.
\]%
Consider the set of functions
\[
\Lambda :=\left\{ f\left\vert \left\vert c_{k}\left( f\right) \right\vert
\leq \lambda _{k},\text{ }k\in \mathbb{N}\right. \right\} ,
\]%
where $\lambda _{k}>0,$ $k\in \mathbb{N}$. It is easy to check that $\Lambda
$ is a convex and symmetric set. Also, $\Lambda $ is compact in $L_{1}\left(
\Omega ,\mathcal{F},\upsilon \right) $ $\ $(see Appendix I for definitions)
if
\[
\sum_{k=1}^{\infty }\lambda _{k}<\infty .
\]%
Let $\varkappa _{m}$ be one of the widths $d_{m}\left( \Lambda ,L_{q}\left(
\Omega ,\mathcal{F},\upsilon \right) \right) $, $a_{m}\left( \Lambda
,L_{q}\left( \Omega ,\mathcal{F},\upsilon \right) \right) $, $a^{m}\left(
\Lambda ,L_{q}\left( \Omega ,\mathcal{F},\upsilon \right) \right) $, $%
\lambda ^{m}\left( \Lambda ,L_{q}\left( \Omega ,\mathcal{F},\upsilon \right)
\right) $ (see Appendix IV for definitions).

{\bf Theorem 24}
\begin{em}
\label{lower bound} Let $\lambda _{k},$ $k\in \mathbb{N}$ be a nonincreasing
sequence of positive numbers, $\sum_{k=1}^{\infty }\lambda _{k}<\infty $
then in our notation
\[
\varkappa _{m}\geq \eta L^{-1}\left( \int_{\Omega }d\upsilon \right)
^{1/q-1}\lambda _{m+1},\text{ }q\geq 1,
\]%
where $\eta =1$ if $\varkappa _{m}$ is $d_{m}$ or $a_{m}$ and $\eta =2^{-1}$
if $\varkappa _{m}$ is $a^{m}$ or $\lambda ^{m}$.
\end{em}

\bigskip

{\bf Proof.}
Fix
\[
L_{m+1}:=\mathrm{lin}\left\{ \varphi _{k},1\leq k\leq m+1\right\}
\]%
and consider the set
\[
Q_{m+1}:=\left\{ t_{m+1}:=\sum_{k=1}^{m+1}c_{k}\varphi _{k},\text{ }%
\left\vert c_{k}\right\vert \leq 1\right\}
\]%
which is the unit ball in $L_{m+1}$. The respective norm in $L_{m+1}$ is
denoted by $\left\Vert t_{m+1}\right\Vert _{Q_{m+1}}$. Since $\lambda _{k},$
$k\in \mathbb{N}$ is a nonincreasing sequence of positive numbers then
\begin{equation}
\lambda _{m+1}Q_{m+1}\subset \Lambda .  \label{incl1}
\end{equation}%
Applying Riesz's theorem (see Appendix II, Theorem 44
we
get%
\[
\left\Vert t_{m+1}\right\Vert _{L_{1}}\geq L^{-1}\max \left\{ \left\vert
c_{k}\right\vert \mathbf{,}\text{ }1\leq k\leq m+1\right\} .
\]%
This means that%
\[
\left\Vert t_{m+1}\right\Vert _{L_{1}}\geq L^{-1}\left\Vert
t_{m+1}\right\Vert _{Q_{m+1}}=\left\Vert t_{m+1}\right\Vert _{LQ_{m+1}},
\]%
for any $t_{m+1}\subset L_{m+1}$, or
\[
B_{1}\cap L_{m+1}\subset LQ_{m+1},
\]%
where $B_{1}:=\left\{ f\left\vert \left\Vert f\right\Vert _{L_{1}}\leq
1\right. \right\} $. Hence, applying (\ref{incl1}) we get%
\[
L^{-1}\lambda _{m+1}B_{1}\cap L_{m+1}\subset \lambda _{m+1}Q_{m+1}\subset
\Lambda .
\]
From the last line and the definition of Bernstein's $m$-width (see Appendix
IV)
\begin{equation}
b_{m}\left( \Lambda ,L_{1}\left( \Omega ,\mathcal{F},\upsilon \right)
\right) \geq L^{-1}\lambda _{m+1}.  \label{bernst1}
\end{equation}%
From Jensen's inequality (see Appendix I, Theorem 38
it follows that
\[
\left( \int_{\Omega }d\upsilon \right) ^{1-1/q}\left\Vert f\right\Vert
_{L_{1}}\leq \left\Vert f\right\Vert _{L_{q}}
\]%
for any $f\in L_{q}$, $q\geq 1$. Hence by the definition of Bernstein's $n$%
-widths and (\ref{bernst1}) we get%
\begin{equation}
b_{m}\left( \Lambda ,L_{q}\left( \Omega ,\mathcal{F},\upsilon \right)
\right) \geq L^{-1}\left( \int_{\Omega }d\upsilon \right) ^{-1+1/q}\lambda
_{m+1},\text{ }q\geq 1.  \label{bern11}
\end{equation}

Applying Corollary 62
 (see Appendix IV) we get%
\[
d_{m}\left( \Lambda ,L_{q}\left( \Omega ,\mathcal{F},\upsilon \right)
\right) \geq L^{-1}\left( \int_{\Omega }d\upsilon \right) ^{-1+1/q}\lambda
_{m+1},\text{ }q\geq 1.
\]%
The same lower bound for $a_{m}\left( \Lambda ,L_{q}\left( \Omega ,\mathcal{F%
},\upsilon \right) \right) $ follows from Theorem 66
, Appendix IV.

Let us obtain lower bounds for the respective cowidths. From the Theorem \ref%
{ba1} it follows that for any compact and symmetric set $A$ in a Banach
space $X$
\[
b_{m}\left( A,X\right) \leq 2a_{m}\left( A,X\right) ,
\]%
It is known that \cite{tikh2} p. 222,%
\[
a_{m}\left( A,X\right) \leq u_{m}\left( A,X\right)
\]%
and \cite{tikhomirov1} p. 190,%
\[
u_{m}\left( A,X\right) \leq a^{m}\left( A,X\right) .
\]%
Hence
\begin{equation}
b_{m}\left( A,X\right) \leq 2a^{m}\left( A,X\right) .  \label{alexandrov1}
\end{equation}%
Finally, comparing (\ref{bernst1})-(\ref{alexandrov1}) we get%
\[
a^{m}\left( \Lambda ,L_{q}\left( \Omega ,\mathcal{F},\upsilon \right)
\right) \geq 2^{-1}L^{-1}\left( \int_{\Omega }d\upsilon \right)
^{-1+1/q}\lambda _{m+1},\text{ }q\geq 1.
\]

A similar result can be obtained for $\lambda ^{m}\left( \Lambda
,L_{q}\left( \Omega ,\mathcal{F},\upsilon \right) \right) $.
$\square$

Now we apply Theorem 24
to estimate from below the rate of
convergence. First we need the following result.

{\bf Theorem 25}
\begin{em}
\label{m-term1} Let
\[
\Omega \left( \Phi ^{\mathbb{Q}},T,\varrho ,P\right) :=\left\{ \mathbf{z}\in
\mathbb{R}^{n},\left\vert \Phi ^{\mathbb{Q}}\left( \frac{2\pi }{P}\mathbf{z,}%
T\right) \right\vert \geq \varrho \right\} .
\]%
Then
\[
\mathrm{Card}\left( \Omega \left( \Phi ^{\mathbb{Q}},T,\varrho ,P\right)
\cap \mathbb{Z}^{n}\right) \ll P^{n}\left( \ln \varrho ^{-1}\right)
^{\sum_{s=1}^{n}\nu _{s}^{-1}},\text{ }
\]%
for any $\varrho >0$ and fixed $T>0$ and $\nu _{s}\in \left( 0,1/2\right) $,
$1\leq s\leq n$, as $P\rightarrow \infty $. Let%
\[
\Omega _{\varrho }^{^{\prime }}\left( \Phi ^{\mathbb{Q}}\mathbf{,}T,\varrho
,P\right)
\]%
\[
:=\left\{ \mathbf{z=}\left( z_{1},\cdot \cdot \cdot ,z_{n}\right) \in
\mathbb{R}^{n},\text{ }\exp \left( -CT\sum_{s=1}^{n}\left\vert \frac{2\pi }{P%
}z_{s}\right\vert ^{\nu _{s}}\right) \geq \varrho \right\}
\]%
then $\Omega \left( \Phi ^{\mathbb{Q}}\mathbf{,}T,\varrho ,P\right) \subset
\Omega _{\varrho }^{^{\prime }}\left( \Phi ^{\mathbb{Q}}\mathbf{,}T,\varrho
,P\right) $ and
\begin{equation}  \label{ass11}
\mathrm{Card}\left( \Omega ^{^{\prime }}\left( \Phi ^{\mathbb{Q}}\mathbf{,}T,%
\text{ }\varrho ,P\right) \cap \mathbb{Z}^{n}\right) \asymp P^{n}\left( \ln
\varrho ^{-1}\right) ^{\sum_{s=1}^{n}\nu _{s}^{-1}},
\end{equation}%
as $P\rightarrow \infty $.
\end{em}

{Proof.}
From (\ref{asymp2}) it follows that%
\[
\Omega \left( \Phi ^{\mathbb{Q}}\mathbf{,}T,\varrho ,P\right) \subset \Omega
_{\varrho }^{^{\prime }}\left( \Phi ^{\mathbb{Q}}\mathbf{,}T,\varrho
,P\right)
\]%
\[
=\left\{ \mathbf{z}\in \mathbb{R}^{n},\text{ }\sum_{s=1}^{n}\left\vert
\left( \frac{CT}{\ln \varrho ^{-1}}\right) ^{\nu _{s}^{-1}}\frac{2\pi }{P}%
z_{s}\right\vert ^{\nu _{s}}\leq 1\right\} .
\]%
Since the boundary of $\Omega _{\varrho }^{^{\prime }}\left( \Phi ^{\mathbb{Q%
}}\mathbf{,}T,\varrho \right) $ is piecewise smooth then
\[
\mathrm{Card}\left( \Omega _{\varrho }^{^{\prime }}\left( \Phi ^{\mathbb{Q}}%
\mathbf{,}T,\varrho ,P\right) \cap \mathbb{Z}^{n}\right) \sim \mathrm{Vol}%
_{n}\left( \Omega _{\varrho }^{^{\prime }}\left( \Phi ^{\mathbb{Q}}\mathbf{,}%
T,\varrho ,P\right) \right) ,
\]%
as $P\rightarrow \infty $. Hence
\[
\mathrm{Vol}_{n}\left( \Omega _{\varrho }^{^{\prime }}\left( \Phi ^{\mathbb{Q%
}}\mathbf{,}T,\varrho ,P\right) \right) =\int_{\Omega _{\varrho }^{^{\prime
}}\left( \Phi ^{\mathbb{Q}}\mathbf{,}T,\varrho ,P\right) }d\mathbf{z}
\]%
\[
=\prod_{s=1}^{n}\left( \frac{\ln \varrho ^{-1}}{CT}\right) ^{\nu
_{s}^{-1}}\left( \frac{P}{2\pi }\right) ^{n}\mathrm{Vol}_{n}\left( B\left(
\nu _{1},\cdot \cdot \cdot ,\nu _{n}\right) \right) ,
\]%
\[
=\left( CT\right) ^{-\sum_{s=1}^{n}\nu _{s}^{-1}}\left( \frac{P}{2\pi }%
\right) ^{n}\mathrm{Vol}_{n}\left( B\left( \nu _{1},\cdot \cdot \cdot ,\nu
_{n}\right) \right) \left( \ln \varrho ^{-1}\right) ^{\sum_{s=1}^{n}\nu
_{s}^{-1}},
\]%
where%
\[
B\left( \nu _{1},\cdot \cdot \cdot ,\nu _{n}\right) :=\left\{ \mathbf{z}%
=\left( z_{1},\cdot \cdot \cdot ,z_{n}\right) \in \mathbb{R}%
^{n},\sum_{s=1}^{n}\left\vert z_{s}\right\vert ^{\nu _{s}}\leq 1\right\}
\]%
and $\nu _{1}>0,\cdot \cdot \cdot ,\nu _{n}>0.$ It is known \cite{WangX}
that
\[
\mathrm{Vol}_{n}B\left( \nu _{1},\cdot \cdot \cdot ,\nu _{n}\right) =2^{n}%
\frac{\prod_{s=1}^{n}\Gamma \left( 1+\nu _{s}\right) }{\Gamma \left(
1+\sum_{s=1}^{n}\nu _{s}\right) }.
\]%
Hence%
\[
\mathrm{Card}\left( \Omega \left( \Phi ^{\mathbb{Q}}\mathbf{,}T,\text{ }%
\varrho ,P\right) \cap \mathbb{Z}^{n}\right) \ll \mathrm{Card}\left( \Omega
^{^{\prime }}\left( \Phi ^{\mathbb{Q}}\mathbf{,}T,\text{ }\varrho ,P\right)
\cap \mathbb{Z}^{n}\right)
\]%
\begin{equation}
\asymp P^{n}\left( \ln \varrho ^{-1}\right) ^{\sum_{s=1}^{n}\nu _{s}^{-1}},
\label{pppp}
\end{equation}%
as $P\rightarrow \infty $.
$\square$

\bigskip Consider the measure space $\left( P\mathbb{T}^{n},\mathcal{L},d%
\mathbf{x}\right) $, where $P\mathbb{T}^{n}=\mathbb{R}^{n}/P\mathbb{Z}^{n}$
is the $n$-dimensional torus, $\mathcal{L}$ is the Lebesgue $\sigma $%
-algebra and $d\mathbf{x}$ is the Lebesgue measure on $P\mathbb{T}^{n}$.
Define the function class
\[
\Lambda :=\left\{ f\left( \mathbf{x}\right) =\sum_{\mathbf{m}\in \mathbb{Z}%
^{n}}c_{\mathbf{m}}\varphi _{\mathbf{m}}\left( \mathbf{x}\right) \right\} ,%
\text{ }\mathbf{m}=\left( m_{1},\cdot \cdot \cdot ,m_{n}\right) ,
\]%
where $\left\vert c_{\mathbf{m}}\right\vert \leq \lambda _{\mathbf{m}}$ and
\[
\varphi _{\mathbf{m}}\left( \mathbf{x}\right) :=P^{-n/2}\exp \left( \frac{%
2\pi i}{P}\left\langle \mathbf{m},\mathbf{x}\right\rangle \right) ,\text{ }%
\mathbf{m}\in \mathbb{Z}^{n}.
\]%
Observe that the system $\left\{ \varphi _{\mathbf{m}}\left( \mathbf{x}%
\right) ,\text{ }\mathbf{m\in }\mathbb{Z}^{n}\right\} $ is orthonormal and
\[
L=\sup_{\mathbf{m}\in \mathbb{N}}\left\Vert \varphi _{\mathbf{m}}\right\Vert
_{\infty }=P^{-n/2}.
\]%
Let%
\[
\lambda _{\mathbf{m}}=\exp \left( -CT\sum_{s=1}^{n}\left\vert \frac{2\pi }{P}%
m_{s}\right\vert ^{\nu _{s}}\right) .
\]

Recall that
\[
\left\vert \Phi ^{\mathbb{Q}}\left( -\frac{2\pi }{P}\mathbf{m,}T\right)
\right\vert \leq \exp \left( -CT\sum_{s=1}^{n}\left\vert \frac{2\pi }{P}%
m_{s}\right\vert ^{\nu _{s}}\right) .
\]%
Hence
\[
\widetilde{p}_{T}^{\mathbb{Q}}\left( \mathbf{x}\right) =\frac{1}{P^{n}}\sum_{%
\mathbf{m}\in \mathbb{Z}^{n}}\Phi ^{\mathbb{Q}}\left( -\frac{2\pi }{P}%
\mathbf{m},T\right) \exp \left( \frac{2\pi i}{P}\left\langle \mathbf{m},%
\mathbf{x}\right\rangle \right) \in \Lambda .
\]

{\bf Theorem 26}
\begin{em}
\label{widths-lower} In our notation%
\[
\varkappa _{m}\left( \Lambda ,L_{q}\left( P\mathbb{T}^{n},\mathcal{L},d%
\mathbf{x}\right) \right) \gg P^{-1/2+1/q}\exp \left( -\left( P^{-n}m\right)
^{\left( \sum_{s=1}^{n}\nu _{s}^{-1}\right) ^{-1}}\right) ,\text{ }
\]%
as $m\rightarrow \infty $.
\end{em}

{\bf Proof.}
Let $d\mathbf{x}$ be a Lebesgue measure on $\Omega =2^{-1}PQ_{n}$, where%
\[
Q_{n}:=\left\{ \mathbf{x=}\left( x_{1},\cdot \cdot \cdot ,x_{n}\right) \in
\mathbb{R}^{n}\text{, }\max \left\vert x_{k}\right\vert \leq 1\text{, }1\leq
k\leq n\right\}
\]%
is the unit cube in $\mathbb{R}^{n}$. From Theorem 24
we get%
\[
\varkappa _{m}\left( \Lambda ,L_{q}\left( P\mathbb{T}^{n},\mathcal{L},d%
\mathbf{x}\right) \right) \gg L^{-1}\left( \int_{2^{-1}PQ_{n}}d\mathbf{x}%
\right) ^{-1+1/q}\lambda _{m+1}
\]%
\begin{equation}
\gg P^{n/2}P^{n\left( -1+1/q\right) }\lambda _{m+1}=P^{-1/2+1/q}\lambda
_{m+1}.  \label{kka2}
\end{equation}%
Let $m\asymp P^{n}\left( \ln \varrho ^{-1}\right) ^{\sum_{s=1}^{n}\nu
_{s}^{-1}}$ then, by Theorem 25
,
\begin{equation}
\varrho \asymp \exp \left( -\left( P^{-n}m\right) ^{\left( \sum_{s=1}^{n}\nu
_{s}^{-1}\right) ^{-1}}\right)  \label{kka1}
\end{equation}%
and applying (\ref{ass11}) we get $\lambda _{m+1}\asymp \varrho $. Hence,
using (\ref{kka1}) and (\ref{kka2}) we obtain%
\[
\varkappa _{m}\left( \Lambda ,L_{q}\left( P\mathbb{T}^{n},\mathcal{L},d%
\mathbf{x}\right) \right) \gg P^{-1/2+1/q}\exp \left( -\left( P^{-n}m\right)
^{\left( \sum_{s=1}^{n}\nu _{s}^{-1}\right) ^{-1}}\right) ,
\]%
as $m\rightarrow \infty $.
$\square$

\bigskip

\section{Approximation of density functions by $m$-term exponential sums}

The next statements deal with approximation of functions using $m$-term
exponential sums with spectrum in the domain $\Omega _{1/R}^{^{\prime }}$.

{\bf Theorem 27}
\begin{em}
\label{cross11} Let $2\leq q\leq \infty ,$ $1/q+1/q^{^{\prime }}=1$, $\nu
_{s}\in \left( 0,1/2\right) $, $1\leq s\leq n$,%
\[
m:=P^{n}\left( \ln R\right) ^{\sum_{s=1}^{n}\nu _{s}^{-1}}.
\]%
Then in our notation%
\[
E\left( m,P\right) :=\left\Vert \widetilde{p}_{T}^{\mathbb{Q}}\left( \mathbf{%
x}\right) -\frac{1}{P^{n}}\sum_{\mathbf{m}\in \mathbb{Z}^{n}\cap \Omega
_{1/R}^{^{\prime }}}\Phi ^{\mathbb{Q}}\left( -\frac{2\pi }{P}\mathbf{m}%
,T\right) \exp \left( \frac{2\pi i}{P}\left\langle \mathbf{m},\mathbf{x}%
\right\rangle \right) \right\Vert _{L_{q}\left( \frac{P}{2}Q_{n}\right) }
\]%
\[
\ll \left( mP^{-n}\right) ^{\left( 1-\left( \sum_{s=1}^{n}\nu
_{s}^{-1}\right) ^{-1}\right) /q^{^{\prime }}}\exp \left( -\left(
mP^{-n}\right) ^{\left( \sum_{s=1}^{n}\nu _{s}^{-1}\right) ^{-1}}\right) ,
\]%
as $m,$ $P\rightarrow \infty $.
\end{em}

{\bf Proof.}
Recall that the system of functions%
\[
\varphi _{\mathbf{m}}\left( \mathbf{x}\right) :=P^{-n/2}\exp \left( \frac{%
2\pi i}{P}\left\langle \mathbf{m},\mathbf{x}\right\rangle \right) ,\text{ \
\ }\mathbf{m\in }\mathbb{Z}^{n},\text{ \ \ }\mathbf{x\in }\frac{P}{2}Q_{n}
\]%
is uniformly bounded, $\left\vert \varphi _{\mathbf{m}}\left( \mathbf{x}%
\right) \right\vert \leq P^{-n/2},$ $\forall \mathbf{m\in }\mathbb{Z}^{n}$
and orthonormal in $L_{2}\left( \frac{P}{2}Q_{n}\right) $. Let $\rho
\rightarrow \infty $. Then, by (\ref{pppp})
\[
\mathrm{Vol}_{n}\left( \Omega _{1/\rho }^{^{\prime }}\right) \asymp
P^{n}\left( \ln \rho \right) ^{\sum_{s=1}^{n}\nu _{s}^{-1}}:=V\left( \rho
\right) .
\]%
Applying Riesz theorem (see Appendix I, Theorem 44
we get
\[
E\left( m,P\right) =\left\Vert \frac{1}{P^{n}}\sum_{\mathbf{m}\in \left(
\mathbb{R}^{n}\setminus \Omega _{1/R}^{^{\prime }}\right) \cap \mathbb{Z}%
^{n}}\Phi ^{\mathbb{Q}}\left( -\frac{2\pi }{P}\mathbf{m},T\right) \exp
\left( \frac{2\pi i}{P}\left\langle \mathbf{m},\mathbf{x}\right\rangle
\right) \right\Vert _{L_{q}\left( \frac{P}{2}Q_{n}\right) }
\]%
\[
=\left\Vert \frac{1}{P^{n/2}}\sum_{\mathbf{m}\in \left( \mathbb{R}%
^{n}\setminus \Omega _{1/R}^{^{\prime }}\right) \cap \mathbb{Z}^{n}}\Phi ^{%
\mathbb{Q}}\left( -\frac{2\pi }{P}\mathbf{m},T\right) \varphi _{\mathbf{m}%
}\left( \mathbf{x}\right) \right\Vert _{L_{q}\left( \frac{P}{2}Q_{n}\right) }
\]%
\[
\ll P^{-n/2}P^{-\left( n/2\right) \left( 2/q^{^{\prime }}-1\right) }\left(
\int_{R}^{\infty }\rho ^{-q^{^{\prime }}}dV\left( \rho \right) \right)
^{1/q^{^{\prime }}}:=P^{-n/q^{^{\prime }}}\left( I\left( R\right) \right)
^{1/q^{^{\prime }}},
\]%
where%
\[
I\left( R\right) =\int_{R}^{\infty }\rho ^{-q^{^{\prime }}}dV\left( \rho
\right)
\]%
\begin{equation}
=P^{n}\left( \sum_{s=1}^{n}\nu _{s}^{-1}\right) \int_{R}^{\infty }\rho
^{-q^{^{\prime }}-1}\left( \ln \rho \right) ^{\sum_{s=1}^{n}\nu
_{s}^{-1}-1}d\rho .  \label{star1}
\end{equation}%
Observe that $\nu _{s}\in \left( 0,1/2\right) $. Hence $\sum_{s=1}^{n}\nu
_{s}^{-1}-1>0$. Let $\alpha >1$ and $\beta >0$. Then
\[
\int_{R}^{\infty }x^{-\alpha }\left( \ln x\right) ^{\beta }dx
\]%
\[
=\frac{1}{-\alpha +1}x^{-\alpha +1}\left( \ln x\right) ^{\beta }\left\vert
_{R}^{\infty }\right. -\int_{R}^{\infty }\frac{1}{-\alpha +1}x^{-\alpha
+1}\beta \left( \ln x\right) ^{\beta -1}x^{-1}dx
\]%
\[
=\frac{1}{\alpha -1}R^{-\alpha +1}\left( \ln R\right) ^{\beta }+\frac{\beta
}{\alpha -1}\int_{R}^{\infty }x^{-\alpha }\left( \ln x\right) ^{\beta -1}dx
\]%
\begin{equation}
=\frac{1}{\alpha -1}R^{-\alpha +1}\left( \ln R\right) ^{\beta },\text{ \ }%
R\rightarrow \infty ,  \label{star2}
\end{equation}%
since%
\[
\lim_{R\rightarrow \infty }\frac{\int_{R}^{\infty }x^{-\alpha }\left( \ln
x\right) ^{\beta -1}dx}{\int_{R}^{\infty }x^{-\alpha }\left( \ln x\right)
^{\beta }dx}=0.
\]%
Comparing (\ref{star1}) and (\ref{star2}) we get%
\[
I\left( R\right) \ll P^{n}R^{-q^{^{\prime }}}\left( \ln R\right)
^{\sum_{s=1}^{n}\nu _{s}^{-1}-1},\text{ \ }R\rightarrow \infty .
\]%
Hence%
\[
E\left( m,P\right) \ll R^{-1}\left( \ln R\right) ^{\left( \sum_{s=1}^{n}\nu
_{s}^{-1}-1\right) /q^{^{\prime }}},\text{ \ }R\rightarrow \infty .
\]

This means that using%
\[
m=P^{n}\left( \ln R\right) ^{\sum_{s=1}^{n}\nu _{s}^{-1}}
\]%
harmonics from\textrm{\ }$\Omega _{1/R}^{^{\prime }}\cap \mathbb{Z}^{n}$ we
get the error of approximation%
\[
E\left( m,P\right) \ll \left( mP^{-n}\right) ^{\left( 1-\left(
\sum_{s=1}^{n}\nu _{s}^{-1}\right) ^{-1}\right) /q^{^{\prime }}}\exp \left(
-\left( mP^{-n}\right) ^{\left( \sum_{s=1}^{n}\nu _{s}^{-1}\right)
^{-1}}\right) ,
\]

as $m,$ $P\rightarrow \infty $.
$\square$

\bigskip

Comparing Theorem 26
and Theorem 27
we see that
the domain of truncation $\Omega _{1/R}^{^{\prime }}$ is optimal in
exponential scale in the sense of $n$-cowidth.

\bigskip

Applying Theorem 22
and Theorem 27
we get the
following statement.

{\bf Corollary 28}
\begin{em}
\label{collorary-conv}

Let $2\leq q\leq \infty $ and $b:=\min \left\{ b_{s}\left\vert 1\leq s\leq
n\right. \right\} $ Then in our notation
\[
E_{1}\left( P\right) +E\left( m,P\right)
\]%
\[
=\left\Vert p_{T}^{\mathbb{Q}}\left( \mathbf{x}\right) -\frac{1}{P^{n}}\sum_{%
\mathbf{m}\in \Omega _{1/R}^{^{\prime }}\cap \mathbb{Z}^{n}}\Phi ^{\mathbb{Q}%
}\left( -\frac{2\pi }{P}\mathbf{m},T\right) \exp \left( \frac{2\pi i}{P}%
\left\langle \mathbf{m},\mathbf{x}\right\rangle \right) \right\Vert
_{L_{q}\left( \frac{P}{2}Q_{n}\right) }
\]%
\[
\ll \exp \left( -Pb\right) P^{n/q}
\]%
\[
+\left( mP^{-n}\right) ^{\left( 1-\left( \sum_{s=1}^{n}\nu _{s}^{-1}\right)
^{-1}\right) /q^{^{\prime }}}\exp \left( -\left( mP^{-n}\right) ^{\left(
\sum_{s=1}^{n}\nu _{s}^{-1}\right) ^{-1}}\right) ,
\]%
as $m,$ $P\rightarrow \infty $.
\end{em}

\bigskip Let for simplicity $q=\infty $. Let $P$ in Corollary 28
be such that
\[
Pb=\left( mP^{-n}\right) ^{\left( \sum_{s=1}^{n}\nu _{s}^{-1}\right) ^{-1}},
\]%
or%
\begin{equation}
P=\left( b^{-1}m^{\left( \sum_{s=1}^{n}\nu _{s}^{-1}\right) ^{-1}}\right)
^{\left( 1+n\left( \sum_{s=1}^{n}\nu _{s}^{-1}\right) ^{-1}\right) ^{-1}}
\label{pp-pp}
\end{equation}%
then%
\[
E_{1}\left( P\right) +E\left( m,P\right) \ll \exp \left( -am^{k}\right)
m^{h},\text{ }m\rightarrow \infty ,
\]%
where%
\[
a:=b^{1-\left( 1+n\left( \sum_{s=1}^{n}\nu _{s}^{-1}\right) ^{-1}\right)
^{-1}},
\]%
\[
k:=\frac{\left( \sum_{s=1}^{n}\nu _{s}^{-1}\right) ^{-1}}{1+n\left(
\sum_{s=1}^{n}\nu _{s}^{-1}\right) ^{-1}}
\]%
and%
\[
h:=\frac{1-\left( \sum_{s=1}^{n}\nu _{s}^{-1}\right) ^{-1}}{1+n\left(
\sum_{s=1}^{n}\nu _{s}^{-1}\right) ^{-1}}.
\]%
This means that using $m$ harmonics with spectrum in $\Omega
_{1/R}^{^{\prime }}$, where $R=\exp \left( \left( P^{-n}m\right) ^{\left(
\sum_{s=1}^{n}\nu _{s}^{-1}\right) ^{-1}}\right) $ and $P$ is defined by (%
\ref{pp-pp}) we get the error of convergence $\exp \left( -am^{b}\right)
m^{k}$ as $m\rightarrow \infty $.

\bigskip

\bigskip

\chapter{Option pricing}

\label{Pricing of basket options}

\section{Introduction}

\bigskip Pricing of high-dimensional options is a deep problem of Financial
Mathematics. The main aim of this chapter is to develop new simple and
practical methods of pricing of basket options. As a motivating example
consider a frictionless market with no arbitrage opportunities with a
constant riskless interest rate $r>0$. Let $S_{j,t}$, $1\leq j\leq n,t\geq 0$%
, be $n$ asset price processes. The common spread option with maturity $T>0$
and strike $K\geq 0$ is the contract that pays $H=\left(
S_{1,T}-\sum_{j=2}^{n}S_{j,T}-K\right) _{+}$ at time $T$. There is a wide
range of such options traded across different sectors of the financial
markets. Assuming the existence of a risk-neutral equivalent martingale
measure $\mathbb{Q}$ we get the following pricing formula for the value $V$
of the spread option at time $0$,
\[
V=\exp \left( -rT\right) \mathbb{E}^{\mathbb{Q}}\left[ H\right] ,
\]%
where $H$ is a reward function and the expectation is taken with respect to
the equivalent martingale measure.

There is an extensive literature on spread options and their applications.
In particular, if $K=0$ a spread option is the same as an option to exchange
one asset for another. An explicit solution in this case has been obtained
by Margrabe \cite{Margrabe}. Margrabe model assumes that $S_{t,1}$ and $%
S_{t,2}$ follow a geometric Brownian motion whose volatilities $\sigma _{1}$
and $\sigma _{2}$ do not need to be constant, but the volatility $\sigma $
of $S_{t,1}/S_{t,2}$ is a constant, $\sigma =\left( \sigma _{1}^{2}+\sigma
_{2}^{2}-2\sigma _{1}\sigma _{2}\rho \right) ,$ where $\rho $ is the
correlation coefficient of the Brownian motions $S_{1,t}$ and $S_{2,t}$.
Margrabe formula states that
\[
V=\exp \left( -q_{1}T\right) S_{0,1}\mathrm{\Phi }\left( d_{1}\right) -\exp
\left( -q_{2}T\right) S_{0,2}\mathrm{\Phi }\left( d_{2}\right) ,
\]%
where $\mathrm{\Phi }$ denotes the cumulative distribution for a standard
Normal distribution,
\[
d_{1}=\frac{1}{\sigma T^{1/2}}\left( \ln \left( \frac{S_{0,1}}{S_{0,2}}%
\right) +\left( q_{1}-q_{2}+\frac{\sigma }{2}\right) T\right) ,
\]%
$d_{2}=d_{1}-\sigma T^{1/2}$ and $q_{1},q_{2}$ are the constant continuous
dividend yields.

Unfortunately, in the case where $K>0$ and $S_{t,1}$, $S_{t,2}$ are
geometric Brownian motions, no explicit pricing formula is known. In this
case various approximation methods have been developed. There are three main
approaches: Monte Carlo techniques which are most convenient for
high-dimensional situation because the convergence is independent of the
dimension, fast Fourier transform methods studied in \cite{Carr and Madan}
and PDEs. Observe that PDE based methods are suitable if the dimension of
the PDE is low (see, e.g. \cite{Jitse Niesen, Duffy, Tavella, Wilmott} for
more information). The usual PDE's approach is based on numerical
approximation resulting in a large system of ordinary differential equations
which can then be solved numerically.

Approximation formulas usually allow quick calculations. In particular, a
popular among practitioners Kirk formula \cite{Kirk} gives a good
approximation to the spread call (see also Carmona-Durrleman procedure \cite%
{carmona, Li Deng Zhou}). Various applications of the fast Fourier transform
have been considered in \cite{Dempster and Hong, Lord}. Different approaches
of pricing basket options using geometric Brownian motion have been
discussed in \cite{Beisser1999, Levy1992, Ju2002, Mbanefo, Milevsky and
Posner1998}.

It is well-known that the Merton-Black-Scholes theory becomes much more
efficient if additional stochastic factors are introduced. Consequently, it
is important to consider a wider family of L\'{e}vy processes. Stable L\'{e}%
vy processes have been used first in this context by Mandelbrot \cite{man1}
and Fama \cite{f11}. From the 90$^{th}$ L\'{e}vy processes became very
popular (see, e.g. \cite{ms1, ms2, bp1, bl1, Deng} and references therein).
We present here a general pricing formula which is applicable for a wide
range of jump-diffusion models \cite{55, 555}.

\bigskip

\section{\protect\bigskip Hurd-Zhou theorem}

In this section we prove a technical result (see \cite{hurd, hurd1}) which
is important in our applications. Let $\Gamma \left( z\right) $ be the gamma
function,
\[
\Gamma \left( \xi \right) :=\int_{0}^{\infty }x^{\xi -1}\exp \left(
-x\right) dx,\text{ }\xi \in \mathbb{C\setminus }\left\{ -\mathbb{N\cup }%
\left\{ 0\right\} \right\} .
\]

The proof is based on several lemmas.

{\bf Lemma 29}
\begin{em}
\label{l11} Let
\[
H\left( x_{1},x_{2}\right) :=\left( \exp \left( x_{1}\right) -\exp \left(
x_{2}\right) -1\right) _{+}.
\]%
Then for any real numbers $\mathbf{\epsilon }=\left( \epsilon _{1},\epsilon
_{2}\right) $, $\epsilon _{2}>0$, $\epsilon _{1}+\epsilon _{2}<-1$,%
\[
H\left( x_{1},x_{2}\right) =\left( 2\pi \right) ^{-2}\int_{\mathbb{R}^{2}+i%
\mathbf{\epsilon }}\exp \left( i\left\langle \mathbf{u,x}\right\rangle
\right) g\left( \mathbf{u}\right) d\mathbf{u}
\]%
\[
=\left( 2\pi \right) ^{-2}\int_{-\infty +i\epsilon _{1}}^{\infty +i\epsilon
_{1}}\int_{-\infty +i\epsilon _{2}}^{\infty +i\epsilon _{2}}\exp \left(
i\left( x_{1}u_{1}+x_{2}u_{2}\right) \right) g\left( u_{1},u_{2}\right)
du_{1}du_{2},
\]%
where
\[
g\left( u_{1},u_{2}\right) =\frac{\Gamma \left( i\left( u_{1}+u_{2}\right)
-1\right) \Gamma \left( -iu_{2}\right) }{\Gamma \left( iu_{1}+1\right) }.
\]
\end{em}

{\bf Proof.}
Let $\epsilon _{2}>0$ and $\epsilon _{1}+\epsilon _{2}<-1$ then using
definition of $H\left( \mathbf{x}\right) $ it is possible to show that%
\[
\exp \left( \left\langle \mathbf{x,\epsilon }\right\rangle \right) H\left(
\mathbf{x}\right) =\exp \left( x_{1}\epsilon _{1}+x_{2}\epsilon _{2}\right)
H\left( x_{1},x_{2}\right)
\]%
\[
=\exp \left( x_{1}\epsilon _{1}+x_{2}\epsilon _{2}\right) \left( \exp \left(
x_{1}\right) -\exp \left( x_{2}\right) -1\right) _{+}\in L_{2}\left( \mathbb{%
R}^{2}\right) .
\]

Hence, by the Plancherel theorem (see Appendix II, Theorem 42
there is such function $r\left( \mathbf{u}\right) \in L_{2}\left( \mathbb{R}%
^{2}\right) $ that
\[
\exp \left( \left\langle \mathbf{x,\epsilon }\right\rangle \right) H\left(
\mathbf{x}\right) =\left( 2\pi \right) ^{-2}\int_{\mathbb{R}^{2}}\exp \left(
i\left\langle \mathbf{x,u}\right\rangle \right) r\left( \mathbf{u}\right) d%
\mathbf{u.}
\]%
Consequently,%
\[
H\left( \mathbf{x}\right) =\left( 2\pi \right) ^{-2}\int_{\mathbb{R}%
^{2}}\exp \left( i\left\langle \mathbf{x,u}\right\rangle -\left\langle
\mathbf{x,\epsilon }\right\rangle \right) r\left( \mathbf{u}\right) d\mathbf{%
u}
\]%
\[
=\left( 2\pi \right) ^{-2}\int_{\mathbb{R}^{2}}\exp \left( i\left\langle
\mathbf{x,u+}i\mathbf{\epsilon }\right\rangle \right) r\left( \mathbf{u}%
\right) d\mathbf{u}
\]%
\[
=\left( 2\pi \right) ^{-2}\int_{\mathbb{R}^{2}+i\mathbf{\epsilon }}\exp
\left( i\left\langle \mathbf{x,u}\right\rangle \right) r\left( \mathbf{u}-i%
\mathbf{\epsilon }\right) d\mathbf{u.}
\]%
and%
\[
r\left( \mathbf{u}\right) =\int_{\mathbb{R}^{2}}\exp \left( -i\left\langle
\mathbf{x,u}\right\rangle \right) \exp \left( \left\langle \mathbf{%
x,\epsilon }\right\rangle \right) H\left( \mathbf{x}\right) d\mathbf{x}
\]%
\[
=\int_{\mathbb{R}^{2}}\exp \left( -i\left\langle \mathbf{x,u+}i\mathbf{%
\epsilon }\right\rangle \right) H\left( \mathbf{x}\right) d\mathbf{x.}
\]

Let $r\left( \mathbf{u}-i\mathbf{\epsilon }\right) :=g\left( \mathbf{u}%
\right) $ then
\[
g\left( \mathbf{u}\right) =\int_{\mathbb{R}^{2}}\exp \left( -i\left\langle
\mathbf{x,u}\right\rangle \right) H\left( \mathbf{x}\right) d\mathbf{x}
\]%
\[
=\int_{\mathbb{R}^{2}}\exp \left( -i\left( x_{1}u_{1}+x_{2}u_{2}\right)
\right) \left( \exp \left( x_{1}\right) -\exp \left( x_{2}\right) -1\right)
_{+}dx_{1}dx_{2}.
\]%
Clearly, $\left( \exp \left( x_{1}\right) -\exp \left( x_{2}\right)
-1\right) _{+}\geq 0$ if $x_{1}\geq 0$ and $\exp \left( x_{1}\right) -\exp
\left( x_{2}\right) -1\geq 0$. Hence
\[
g\left( u_{1},u_{2}\right)
\]%
\[
=\int_{0}^{\infty }\exp \left( -iu_{1}x_{1}\right)
\]%
\[
\times \left( \int_{-\infty }^{\ln \left( \exp \left( x_{1}\right) -1\right)
}\exp \left( -iu_{2}x_{2}\right) \left( \left( \exp \left( x_{1}\right)
-1\right) -\exp \left( x_{2}\right) \right) dx_{2}\right) dx_{1}
\]%
\[
=\int_{0}^{\infty }\exp \left( -iu_{1}x_{1}\right) \left( \exp \left(
x_{1}\right) -1\right) ^{1-iu_{2}}\left( \left( -iu_{2}\right) ^{-1}-\left(
1-iu_{2}\right) ^{-1}\right) dx_{1}.
\]%
Making change of variable $z=\exp \left( -x_{1}\right) $ we get%
\[
g\left( u_{1},u_{2}\right) =\frac{1}{\left( -iu_{2}\right) \left(
1-iu_{2}\right) }\int_{0}^{1}z^{iu_{1}-1}\left( \frac{1-z}{z}\right)
^{1-iu_{2}}dz
\]%
\[
=\frac{1}{\left( -iu_{2}\right) \left( 1-iu_{2}\right) }\int_{0}^{1}z^{%
\left( i\left( u_{1}+u_{2}\right) -1\right) -1}\left( 1-z\right) ^{\left(
2-iu_{2}\right) -1}dz
\]%
\[
=\frac{1}{\left( -iu_{2}\right) \left( 1-iu_{2}\right) }\mathrm{B}\left(
i\left( u_{1}+u_{2}\right) -1,\left( 2-iu_{2}\right) \right) ,
\]%
where \
\[
\mathrm{B}\left( a,b\right) :=\int_{0}^{1}z^{a-1}\left( 1-z\right) ^{b-1}dz=%
\frac{\Gamma \left( a\right) \Gamma \left( b\right) }{\Gamma \left(
a+b\right) }\text{ }
\]%
is the Beta function which is defined for ${\rm Re}a>0,$ ${\rm Re}b>0$.
Hence,%
\[
g\left( u_{1},u_{2}\right) =\frac{\Gamma \left( i\left( u_{1}+u_{2}\right)
-1\right) \Gamma \left( -iu_{2}+2\right) }{\left( -iu_{2}\right) \left(
1-iu_{2}\right) \Gamma \left( iu_{1}+1\right) }
\]%
\[
=\frac{\Gamma \left( i\left( u_{1}+u_{2}\right) -1\right) \Gamma \left(
-iu_{2}\right) }{\Gamma \left( iu_{1}+1\right) },
\]%
since $\Gamma \left( -iu_{2}+2\right) =\left( 1-iu_{2}\right) \Gamma \left(
-iu_{2}+1\right) =\left( -iu_{2}\right) \left( 1-iu_{2}\right) \Gamma \left(
-iu_{2}+1\right) .$
$\square$

{\bf Lemma 30}
\begin{em}
\label{l12} \bigskip Let $z\in \mathbb{R},$ $\mathbf{x}=\left( x_{1},\cdot
\cdot \cdot ,x_{n}\right) \in \mathbb{R}^{n}$ and $\mathbf{u}=\left(
u_{1},\cdot \cdot \cdot ,u_{n}\right) \in \mathbb{C}^{n}$, ${\rm Im}u_{k}>0$%
, $1\leq k\leq n$. Then%
\[
\int_{\mathbb{R}^{n}}\delta \left( \exp \left( z\right) -\sum_{k=1}^{n}\exp
\left( x_{k}\right) \right) \exp \left( z-i\left\langle \mathbf{u,x}%
\right\rangle \right) d\mathbf{x}
\]%
\[
=\frac{\prod_{k=1}^{n}\Gamma \left( -iu_{k}\right) }{\Gamma \left(
-i\sum_{k=1}^{n}u_{k}\right) }\exp \left( -iz\sum_{k=1}^{n}u_{k}\right) ,
\]%
where $\delta \left( \cdot \right) $ denotes the delta function.
\end{em}

{\bf Proof.}
Making change of variables $\rho =\exp \left( z\right) $ and $\sigma
_{k}=\exp \left( x_{k}\right) $ we get%
\[
I_{n}:=\int_{\mathbb{R}^{n}}\delta \left( \exp \left( z\right)
-\sum_{k=1}^{n}\exp \left( x_{k}\right) \right) \exp \left( z-i\left\langle
\mathbf{u,x}\right\rangle \right) d\mathbf{x}
\]%
\[
=\rho \int_{\rho Q^{n}}\delta \left( \rho -\sum_{k=1}^{n}\sigma _{k}\right)
\prod_{k=1}^{n}\sigma _{k}^{-iu_{k}-1}\prod_{k=1}^{n}d\sigma
_{k},
\]%
where
\[
Q^{n}:=\left\{ \mathbf{x}=\left( x_{1},\cdot \cdot \cdot ,x_{n}\right)
\left\vert 0<x_{k}\leq 1\text{, }1\leq k\leq n\right. \right\}
\]%
since%
\[
\int_{\mathbb{R}^{n}\setminus \rho Q^{n}}\delta \left( \rho
-\sum_{k=1}^{n}\sigma _{k}\right) \prod_{k=1}^{n}\sigma
_{k}^{-iu_{k}-1}\prod_{k=1}^{n}d\sigma _{k}=0.
\]

We proceed by induction. It is easy to check that $I_{1}=\rho ^{-iu_{1}}$,
or Lemma 30
 is true for $n=1$. If Lemma 12
  is true for $m=n$,
then for $m=n+1$ we get%
\[
I_{n+1}=\rho \int_{\rho Q^{n+1}}\delta \left( \left( \rho -\sigma
_{n+1}\right) -\sum_{k=1}^{n}\sigma _{k}\right) \sigma
_{n+1}^{-iu_{n+1}-1}\prod_{k=1}^{n}\sigma _{k}^{-iu_{k}-1}d\sigma
_{n+1}\prod_{k=1}^{n}d\sigma _{k},
\]%
We can rewrite $I_{n+1}$ as
\[
I_{n+1}=\rho \int_{0}^{\rho }\frac{\sigma _{n+1}^{-iu_{n+1}-1}}{\rho -\sigma
_{n+1}}J_{n}\left( \rho ,\sigma _{n+1}\right) d\sigma _{n+1},
\]

where%
\[
J_{n}\left( \rho ,\sigma _{n+1}\right) :=\left( \rho -\sigma _{n+1}\right)
\int_{\rho Q^{n}}\prod_{k=1}^{n}\sigma _{k}^{-iu_{k}-1}\delta \left(
\left( \rho -\sigma _{n+1}\right) -\sum_{k=1}^{n}\sigma _{k}\right)
\prod_{k=1}^{n}d\sigma _{k}.
\]
By the induction hypothesis%
\[
J_{n}\left( \rho ,\sigma _{n+1}\right) =\frac{\prod_{k=1}^{n}\Gamma
\left( -iu_{k}\right) }{\Gamma \left( -i\sum_{k=1}^{n}u_{k}\right) }\exp
\left( -i\sum_{k=1}^{n}u_{k}\ln \left( \rho -\sigma _{n+1}\right) \right)
\]%
\[
=\frac{\prod_{k=1}^{n}\Gamma \left( -iu_{k}\right) }{\Gamma \left(
-i\sum_{k=1}^{n}u_{k}\right) }\left( \rho -\sigma _{n+1}\right)
^{-i\sum_{k=1}^{n}u_{k}}.
\]%
Hence%
\[
I_{n+1}=\frac{\prod_{k=1}^{n}\Gamma \left( -iu_{k}\right) }{\Gamma
\left( -i\sum_{k=1}^{n}u_{k}\right) }\rho \int_{0}^{\rho }\frac{\sigma
_{n+1}^{-iu_{n+1}-1}}{\rho -\sigma _{n+1}}\left( \rho -\sigma _{n+1}\right)
^{-i\sum_{k=1}^{n}u_{k}}d\sigma _{n+1}
\]%
\[
=\frac{\prod_{k=1}^{n}\Gamma \left( -iu_{k}\right) }{\Gamma \left(
-i\sum_{k=1}^{n}u_{k}\right) }\rho ^{-i\sum_{k=1}^{n}u_{k}}\int_{0}^{\rho
}\sigma _{n+1}^{-iu_{n+1}-1}\left( 1-\frac{\sigma _{n+1}}{\rho }\right)
^{-1-i\sum_{k=1}^{n}u_{k}}d\sigma _{n+1}.
\]%
Making change of variables $\xi :=\sigma _{n+1}/\rho ,$ we obtain%
\[
I_{n+1}=\frac{\prod_{k=1}^{n}\Gamma \left( -iu_{k}\right) }{\Gamma
\left( -i\sum_{k=1}^{n}u_{k}\right) }\rho
^{-i\sum_{k=1}^{n}u_{k}}\int_{0}^{1}\left( \rho \xi \right)
^{-iu_{n+1}-1}\left( 1-\xi \right) ^{-1-i\sum_{k=1}^{n}u_{k}}\rho d\xi
\]%
\[
=\frac{\prod_{k=1}^{n}\Gamma \left( -iu_{k}\right) }{\Gamma \left(
-i\sum_{k=1}^{n}u_{k}\right) }\rho ^{-i\sum_{k=1}^{n+1}u_{k}}\int_{0}^{1}\xi
^{-iu_{n+1}-1}\left( 1-\xi \right) ^{-1-i\sum_{k=1}^{n}u_{k}}d\xi
\]%
\[
=\frac{\prod_{k=1}^{n}\Gamma \left( -iu_{k}\right) }{\Gamma \left(
-i\sum_{k=1}^{n}u_{k}\right) }\rho ^{-i\sum_{k=1}^{n+1}u_{k}}\mathrm{B}%
\left( -iu_{n+1},-i\sum_{k=1}^{n}u_{k}\right)
\]%
\[
=\frac{\prod_{k=1}^{n}\Gamma \left( -iu_{k}\right) }{\Gamma \left(
-i\sum_{k=1}^{n}u_{k}\right) }\rho ^{-i\sum_{k=1}^{n+1}u_{k}}\frac{\Gamma
\left( -iu_{n+1}\right) \Gamma \left( -i\sum_{k=1}^{n}u_{k}\right) }{\Gamma
\left( -i\sum_{k=1}^{n}u_{k}-iu_{n+1}\right) }
\]%
\[
=\frac{\prod_{k=1}^{n+1}\Gamma \left( -iu_{k}\right) }{\Gamma \left(
-i\sum_{k=1}^{n+1}u_{k}\right) }\exp \left( -iz\sum_{k=1}^{n+1}u_{k}\right) .
\]
$\square$

{\bf Theorem 31}
\begin{em}
\label{hurd theorem} 
(Hurd-Zhou) Let $n\geq 2$. For any real numbers $\mathbf{\epsilon }=\left(
\epsilon _{1},\cdot \cdot \cdot ,\epsilon _{n}\right) $ with $\epsilon
_{m}>0 $ for $2\leq m\leq n$ and $\epsilon _{1}<-1-\sum_{m=2}^{n}\epsilon
_{m},$%
\[
\left( \exp \left( x_{1}\right) -\sum_{m=2}^{n}\exp \left( x_{m}\right)
-1\right) _{+}=\left( 2\pi \right) ^{-n}\int_{\mathbb{R}^{n}+i\mathbf{%
\epsilon }}\exp \left( i\left\langle \mathbf{u},\mathbf{x}\right\rangle
\right) g\left( \mathbf{u}\right) d\mathbf{u,}
\]%
where $\mathbf{x}=\left( x_{1},\cdot \cdot \cdot ,x_{n}\right) $ and, for $%
\mathbf{u=}\left( u_{1},\cdot \cdot \cdot ,u_{n}\right) \in \mathbb{C}^{n},$%
\begin{equation}
g\left( \mathbf{u}\right) =\frac{\Gamma \left( i\sum_{m=1}^{n}u_{m}-1\right)
\prod_{m=2}^{n}\Gamma \left( -iu_{m}\right) }{\Gamma \left(
iu_{1}+1\right) }.  \label{g}
\end{equation}
\end{em}

{\bf Proof.}
\bigskip We need to show (\ref{g}). Observe that
\[
\int_{\mathbb{R}}\delta \left( \exp \left( z\right) -\sum_{k=2}^{n}\exp
\left( x_{k}\right) \right) \exp \left( z\right) dz=1.
\]%
Hence%
\[
g\left( \mathbf{u}\right)
\]%
\[
=\int_{\mathbb{R}^{n}}\int_{\mathbb{R}}\delta \left( \exp \left( z\right)
-\sum_{k=2}^{n}\exp \left( x_{k}\right) \right) \left( \exp \left(
x_{1}\right) -\sum_{k=2}^{n}\exp \left( x_{k}\right) -1\right) _{+}
\]%
\[
\times \exp \left( z-i\left\langle \mathbf{u,x}\right\rangle \right) dzd%
\mathbf{x}
\]%
\[
=\int_{\mathbb{R}^{2}}\left( \exp \left( x_{1}\right) -\exp \left( z\right)
-1\right) _{+}
\]%
\[
\times \left( \int_{\mathbb{R}^{n-1}}\delta \left( \exp \left( z\right)
-\sum_{k=2}^{n}\exp \left( x_{k}\right) \right) \exp \left( z-i\left\langle
\mathbf{u,x}\right\rangle \right) dx_{2}\cdot \cdot \cdot dx_{n}\right)
dx_{1}dz.
\]%
Applying Lemma 30
 and Lemma 29
  we get%
\[
g\left( \mathbf{u}\right) =\frac{\prod_{k=2}^{n}\Gamma \left(
-iu_{k}\right) }{\Gamma \left( -i\sum_{k=2}^{n}u_{k}\right) }
\]%
\[
\times \int_{\mathbb{R}^{2}}\exp \left( -iu_{1}x_{1}-iz\sum_{k=2}^{n}\exp
\left( u_{k}\right) \right) \left( \exp \left( x_{1}\right) -\exp \left(
z\right) -1\right) _{+}dx_{1}dz
\]%
\[
=\frac{\Gamma \left( i\sum_{k=1}^{n}u_{k}-1\right)
\prod_{k=2}^{n}\Gamma \left( -iu_{k}\right) }{\Gamma \left(
iu_{1}+1\right) }.
\]
$\square$

\bigskip

\bigskip

\section{Approximation formulas}

In applications it is important to construct a pricing theory which includes
a wide range of reward functions $H$. For instance, the reward function for
a spread option which is given by
\[
H=H\left( \mathbf{x}\right) =H\left( x_{1},\cdot \cdot \cdot ,x_{n}\right)
\]%
\[
=\left( S_{0,1}\exp \left( x_{1}\right) -\sum_{j=2}^{n}S_{0,j}\exp \left(
x_{j}\right) -K\right) _{+}
\]%
admits an exponential growth with respect to $x_{1}$ as $x_{1}\rightarrow
\infty $. Hence we need to introduce the following definition.

{\bf Definition 32}
\begin{em}
We say that the model process $\mathbf{S}_{t}=\left\{ S_{j,t},\text{ }1\leq
j\leq n\right\} $ is adapted to the payoff \ $H$ if $\mathbb{E}^{\mathbb{Q}}%
\left[ H\right] <\infty $.
\end{em}

Clearly, if $\mathbb{E}^{\mathbb{Q}}\left[ H\right] =\infty $ then the
option can not be priced. Recall that the operator of expectation is taken
with respect to the density function $p_{t}^{\mathbb{Q}}$\ which satisfies
the equivalent martingale measure condition (\ref{EMM condition}).

The next statement reduces the reward function to a canonical form.

{Lemma 33}
\begin{em}
\label{shift1} In our notation
\[
V=K\exp \left( -rT\right) \int_{\mathbb{R}^{n}}\left( \exp \left(
y_{1}\right) -\sum_{j=2}^{n}\exp \left( y_{j}\right) -1\right) _{+}p_{T}^{%
\mathbb{Q}}\left( \mathbf{y-d}\right) d\mathbf{y,}
\]%
where
\[
\mathbf{d}:=\left( d_{1},\cdot \cdot \cdot ,d_{n}\right) ,\text{ }d_{j}=\ln
\left( \frac{S_{0,j}}{K}\right) ,\text{ \ \ }1\leq j\leq n.
\]
\end{em}

{\bf Proof.}
Recall that $V=\exp \left( -rT\right) \mathbb{E}^{\mathbb{Q}}\left[ H\right]
$. In our case%
\[
H=\left( S_{1,T}-\sum_{j=2}^{n}S_{j,T}-K\right) _{+},
\]%
where
\[
S_{j,T}=S_{j,0}\exp \left( U_{j,T}\right) ,\text{ \ \ }1\leq j\leq n.
\]%
This means that
\[
V=\exp \left( -rT\right) \int_{\mathbb{R}^{n}}\left( S_{0,1}\exp \left(
x_{1}\right) -\sum_{j=2}^{n}S_{0,j}\exp \left( x_{j}\right) -K\right)
_{+}p_{T}^{\mathbb{Q}}\left( \mathbf{x}\right) d\mathbf{x,}
\]%
\[
=K\exp \left( -rT\right)
\]%
\[
\times \int_{\mathbb{R}^{n}}\left( \exp \left( x_{1}+\ln \left( \frac{S_{0,1}%
}{K}\right) \right) -\sum_{j=2}^{n}\exp \left( x_{j}+\ln \left( \frac{S_{0,j}%
}{K}\right) \right) -1\right) _{+}p_{T}^{\mathbb{Q}}\left( \mathbf{x}\right)
d\mathbf{x,}
\]%
where $S_{0,j},$ $\ 1\leq j\leq n$ are the respective spot prices. Making
the change of variables
\[
y_{j}=x_{j}+\ln \left( \frac{S_{0,j}}{K}\right) ,\text{ \ \ }1\leq j\leq n,
\]%
we get%
\[
V=K\exp \left( -rT\right) \int_{\mathbb{R}^{n}}\left( \exp \left(
y_{1}\right) -\sum_{j=2}^{n}\exp \left( y_{j}\right) -1\right) _{+}p_{T}^{%
\mathbb{Q}}\left( \mathbf{y-d}\right) d\mathbf{y,}
\]%
where
\[
\mathbf{d}:=\left( d_{1},\cdot \cdot \cdot ,d_{n}\right) \text{, \ }%
d_{j}=\ln \left( \frac{S_{0,j}}{K}\right) ,\text{ \ \ }1\leq j\leq n.
\]
$\square$

{\bf Theorem 34}
\begin{em}
\label{hurd integral} In our notation, for any $\mathbf{m}=\left(
m_{1},\cdot \cdot \cdot ,m_{n}\right) \mathbf{\in }\mathbb{Z}^{n}$ and $%
\mathbf{\epsilon }=\left( \epsilon _{1},\cdot \cdot \cdot ,\epsilon
_{n}\right) $ with $\epsilon _{m}>0$ for $2\leq m\leq n$ and $\epsilon
_{1}<-1-\sum_{m=2}^{n}\epsilon _{m},$ we have
\[
\int_{\mathbb{R}^{n}}\exp \left( \left\langle \frac{2\pi i}{P}\mathbf{%
m+\epsilon },\mathbf{x}\right\rangle \right) H\left( \mathbf{x}\right) d%
\mathbf{x}
\]%
\[
=\frac{\Gamma \left( -\frac{2\pi i}{P}\sum_{s=1}^{n}m_{s}-\sum_{s=1}^{n}%
\epsilon _{s}-1\right) \prod_{s=2}^{n}\Gamma \left( \frac{2\pi i}{P}%
m_{s}+\epsilon _{s}\right) }{\Gamma \left( -\frac{2\pi i}{P}m_{1}-\epsilon
_{1}+1\right) }.
\]
\end{em}

{\bf Proof.}
Observe that
\[
H\left( \mathbf{x}\right) =\left( 2\pi \right) ^{-n}\int_{\mathbb{R}^{n}+i%
\mathbf{\epsilon }}\exp \left( i\left\langle \mathbf{u},\mathbf{x}%
\right\rangle \right) g\left( \mathbf{u}\right) d\mathbf{u}
\]%
\[
=\left( 2\pi \right) ^{-n}\int_{\mathbb{R}^{n}}\exp \left( i\left\langle
\mathbf{z+}i\mathbf{\epsilon },\mathbf{x}\right\rangle \right) g\left(
\mathbf{z+}i\mathbf{\epsilon }\right) d\mathbf{z}
\]%
\[
=\left( 2\pi \right) ^{-n}\exp \left( -\left\langle \mathbf{\epsilon },%
\mathbf{x}\right\rangle \right) \int_{\mathbb{R}^{n}}\exp \left(
i\left\langle \mathbf{z},\mathbf{x}\right\rangle \right) g\left( \mathbf{z+}i%
\mathbf{\epsilon }\right) d\mathbf{z,}
\]%
where the function $g$ is defined by (\ref{g}). Hence
\[
H\left( \mathbf{x}\right) \exp \left( \left\langle \mathbf{\epsilon },%
\mathbf{x}\right\rangle \right) =\left( 2\pi \right) ^{-n}\int_{\mathbb{R}%
^{n}}\exp \left( i\left\langle \mathbf{z},\mathbf{x}\right\rangle \right)
g\left( \mathbf{z+}i\mathbf{\epsilon }\right) d\mathbf{z.}
\]%
Since $H\left( \mathbf{x}\right) \exp \left( \left\langle \mathbf{\epsilon },%
\mathbf{x}\right\rangle \right) \in L_{2}\left( \mathbb{R}^{n}\right) $
then, applying Plancherel theorem (see Appendix II, Theorem 42
) and Theorem 31
, we get%
\[
\mathbf{F}\left( H\left( \mathbf{x}\right) \exp \left( \left\langle \mathbf{%
\epsilon },\mathbf{x}\right\rangle \right) \right) \left( \mathbf{u}\right)
\]%
\[
=\int_{\mathbb{R}^{n}}\exp \left( -i\left\langle \mathbf{u},\mathbf{x}%
\right\rangle \right) H\left( \mathbf{x}\right) \exp \left( \left\langle
\mathbf{\epsilon },\mathbf{x}\right\rangle \right) d\mathbf{x=}g\left(
\mathbf{u+}i\mathbf{\epsilon }\right)
\]%
\[
=\frac{\Gamma \left( i\left( \left( u_{1}+i\epsilon _{1}\right)
+i\sum_{m=2}^{n}\left( u_{m}+i\epsilon _{m}\right) \right) -1\right)
\prod_{m=2}^{n}\Gamma \left( -i\left( u_{m}+i\epsilon _{m}\right)
\right) }{\Gamma \left( i\left( u_{1}+i\epsilon _{1}\right) +1\right) }
\]%
\[
=\frac{\Gamma \left( i\left( u_{1}+\sum_{m=2}^{n}u_{m}\right)
-\sum_{m=1}^{n}\epsilon _{m}-1\right) \prod_{m=2}^{n}\Gamma \left(
-iu_{m}+\epsilon _{m}\right) }{\Gamma \left( iu_{1}-\epsilon _{1}+1\right) }.
\]%
This means that
\[
\int_{\mathbb{R}^{n}}\exp \left( \frac{2\pi i}{P}\left\langle \mathbf{m},%
\mathbf{x}\right\rangle \right) H\left( \mathbf{x}\right) \exp \left(
\left\langle \mathbf{\epsilon },\mathbf{x}\right\rangle \right) d\mathbf{x=}%
g\left( -\frac{2\pi i}{P}\mathbf{m+}i\mathbf{\epsilon }\right)
\]%
\[
=\frac{\Gamma \left( -\frac{2\pi i}{P}\sum_{s=1}^{n}m_{s}-\sum_{s=1}^{n}%
\epsilon _{s}-1\right) \prod_{s=2}^{n}\Gamma \left( \frac{2\pi i}{P}%
m_{s}+\epsilon _{s}\right) }{\Gamma \left( -\frac{2\pi i}{P}m_{1}-\epsilon
_{1}+1\right) }.
\]
$\square$

The next statement gives a general approximation formula for spread options
which is important in various applications. Observe that it does not show
the rate of convergence. This problem will be discussed later. At this stage
we just explain how to construct the approximation formula.

{Theorem 35}
\begin{em}
\label{approximant 1} Let%
\[
\mathbf{d}:=\left( d_{1},\cdot \cdot \cdot ,d_{n}\right) \text{, }d_{j}=\ln
\left( \frac{S_{0,j}}{K}\right) ,\text{ \ }1\leq j\leq n
\]%
and $\mathbf{\epsilon }=\left( \epsilon _{1},\cdot \cdot \cdot ,\epsilon
_{n}\right) $, $0<\epsilon _{j}\leq b_{+,j}$, $2\leq j\leq n$, $b_{-,1}\leq
\epsilon _{1}<-1-\sum_{m=2}^{n}\epsilon _{m}$. Then the formal approximat $%
\widetilde{V}$ for $V$ can be written as%
\[
\widetilde{V}=\frac{K\exp \left( -rT-\left\langle \mathbf{d,\epsilon }%
\right\rangle \right) }{P^{n}}\sum_{\mathbf{m}\in \Omega _{1/R}^{^{\prime
}}}\Phi ^{\mathbb{Q}}\left( -\frac{2\pi }{P}\mathbf{m}+i\mathbf{\epsilon }%
,T\right) \exp \left( -\frac{2\pi i}{P}\left\langle \mathbf{m},\mathbf{d}%
\right\rangle \right)
\]%
\[
\times \frac{\Gamma \left( -\frac{2\pi i}{P}\sum_{s=1}^{n}m_{s}-%
\sum_{s=1}^{n}\epsilon _{s}-1\right) \prod_{s=2}^{n}\Gamma \left(
\frac{2\pi i}{P}m_{s}+\epsilon _{s}\right) }{\Gamma \left( -\frac{2\pi i}{P}%
m_{1}-\epsilon _{1}+1\right) },\text{ \ }R\rightarrow \infty ,\text{ \ }%
P\rightarrow \infty ,
\]%
where%
\[
\Omega _{1/R}^{^{\prime }}=\left\{ \mathbf{x}\in \mathbb{R}%
^{n},\sum_{s=1}^{n}\left\vert \left( \frac{CT}{\ln R}\right) ^{\nu _{s}^{-1}}%
\frac{2\pi }{P}x_{s}\right\vert ^{\nu _{s}}\leq 1\right\} .
\]
\end{em}

{\bf Proof.}
Applying Lemma 33
we get%
\[
V=\exp \left( -rT\right) \mathbb{E}^{\mathbb{Q}}\left[ H\right]
\]%
\[
=\exp \left( -rT\right) \int_{\mathbb{R}^{n}}\left( S_{0,1}\exp \left(
x_{1}\right) -\sum_{j=2}^{n}S_{0,j}\exp \left( x_{j}\right) -K\right)
_{+}p_{T}^{\mathbb{Q}}\left( \mathbf{x}\right) d\mathbf{x,}
\]%
\[
=K\exp \left( -rT\right) \int_{\mathbb{R}^{n}}\left( \exp \left(
y_{1}\right) -\sum_{j=2}^{n}\exp \left( y_{j}\right) -1\right) _{+}p_{T}^{%
\mathbb{Q}}\left( \mathbf{y-d}\right) d\mathbf{y,}
\]%
where%
\[
\mathbf{d}:=\left( d_{1},\cdot \cdot \cdot ,d_{n}\right) ,\text{ }d_{j}=\ln
\left( \frac{S_{0,j}}{K}\right) ,\text{ \ }1\leq j\leq n.
\]%
For a given $\mathbf{\epsilon }=\left( \epsilon _{1},\cdot \cdot \cdot
,\epsilon _{n}\right) ,$ $b_{-,s}\leq \epsilon _{s}\leq b_{+,s}$, $1\leq
s\leq n$ we can apply Cauchy theorem $n$ times in the domain $T_{n}$ defined
by (\ref{dom2015}), which is justified by (\ref{asymp2}). Hence%
\[
p_{T}^{\mathbb{Q}}\left( \mathbf{y}\right) =\left( 2\pi \right) ^{-n}\int_{%
\mathbb{R}^{n}}\exp \left( -i\left\langle \mathbf{y},\mathbf{x}\right\rangle
\right) \Phi ^{\mathbb{Q}}\left( \mathbf{x},T\right) d\mathbf{x}
\]%
\[
=\left( 2\pi \right) ^{-n}\int_{\mathbb{R}^{n}+i\mathbf{\epsilon }}\exp
\left( -i\left\langle \mathbf{y},\mathbf{x}\right\rangle \right) \Phi ^{%
\mathbb{Q}}\left( \mathbf{x},T\right) d\mathbf{x}
\]%
\[
=\left( 2\pi \right) ^{-n}\int_{\mathbb{R}^{n}}\exp \left( -i\left\langle
\mathbf{y},\mathbf{x}+i\mathbf{\epsilon }\right\rangle \right) \Phi ^{%
\mathbb{Q}}\left( \mathbf{x}+i\mathbf{\epsilon },T\right) d\mathbf{x}
\]%
\[
=\exp \left( \left\langle \mathbf{y},\mathbf{\epsilon }\right\rangle \right)
\left( 2\pi \right) ^{-n}\int_{\mathbb{R}^{n}}\exp \left( -i\left\langle
\mathbf{y},\mathbf{x}\right\rangle \right) \Phi ^{\mathbb{Q}}\left( \mathbf{x%
}+i\mathbf{\epsilon },T\right) d\mathbf{x.}
\]%
Let $\mathbf{y}\in \frac{P}{2}Q_{n}$. Recall that $Q_{n}=\left\{ \mathbf{x}%
\left\vert \mathbf{x=}\left( x_{1},\cdot \cdot \cdot ,x_{n}\right) \mathbf{%
\in }\mathbb{R}^{n}\text{, }\left\vert x_{k}\right\vert \leq 1\right.
\right\} $. Then from Corollary \ref{collorary-conv} we get
\[
p_{T}^{\mathbb{Q}}\left( \mathbf{y}\right) \approx \exp \left( \left\langle
\mathbf{y},\mathbf{\epsilon }\right\rangle \right) \left( \frac{1}{P^{n}}%
\sum_{\mathbf{m}\in \mathbb{Z}^{n}}\Phi ^{\mathbb{Q}}\left( \mathbf{-}\frac{%
2\pi }{P}\mathbf{m}+i\mathbf{\epsilon },T\right) \exp \left( \frac{2\pi i}{P}%
\left\langle \mathbf{m},\mathbf{y}\right\rangle \right) \right)
\]%
and%
\[
p_{T}^{\mathbb{Q}}\left( \mathbf{y-d}\right) \approx \exp \left(
\left\langle \mathbf{y-d},\mathbf{\epsilon }\right\rangle \right)
\]%
\[
\times \frac{1}{P^{n}}\sum_{\mathbf{m}\in \mathbb{Z}^{n}}\Phi ^{\mathbb{Q}%
}\left( \mathbf{-}\frac{2\pi }{P}\mathbf{m}+i\mathbf{\epsilon },T\right)
\exp \left( \frac{2\pi i}{P}\left\langle \mathbf{m},\mathbf{y-d}%
\right\rangle \right)
\]%
\[
\approx \exp \left( \left\langle \mathbf{y-d},\mathbf{\epsilon }%
\right\rangle \right)
\]%
\[
\times \frac{1}{P^{n}}\sum_{\mathbf{m}\in \Omega _{1/R}^{^{\prime }}}\left(
\Phi ^{\mathbb{Q}}\left( \mathbf{-}\frac{2\pi }{P}\mathbf{m}+i\mathbf{%
\epsilon },T\right) \exp \left( -\frac{2\pi i}{P}\left\langle \mathbf{m},%
\mathbf{d}\right\rangle \right) \right) \exp \left( \frac{2\pi i}{P}%
\left\langle \mathbf{m},\mathbf{y}\right\rangle \right) .
\]%
Since $\ \epsilon _{j}>0$, $2\leq j\leq n,$ $\epsilon
_{1}<-1-\sum_{j=2}^{n}\epsilon _{j}$ then we can apply Theorem 34
 to obtain
\[
V=K\exp \left( -rT\right) \int_{\mathbb{R}^{n}}\left( \exp \left(
y_{1}\right) -\sum_{j=2}^{n}\exp \left( y_{j}\right) -1\right) _{+}p_{T}^{%
\mathbb{Q}}\left( \mathbf{y-d}\right) d\mathbf{y}
\]%
\[
\approx \frac{K\exp \left( -rT\right) }{P^{n}}\sum_{\mathbf{m}\in \Omega
_{1/R}^{^{\prime }}}\left( \Phi ^{\mathbb{Q}}\left( \mathbf{-}\frac{2\pi }{P}%
\mathbf{m}+i\mathbf{\epsilon },T\right) \exp \left( -\frac{2\pi i}{P}%
\left\langle \mathbf{m},\mathbf{d}\right\rangle \right) \right)
\]%
\[
\times \int_{\mathbb{R}^{n}}\left( \exp \left( y_{1}\right)
-\sum_{j=2}^{n}\exp \left( y_{j}\right) -1\right) _{+}\exp \left(
\left\langle \mathbf{y-d},\mathbf{\epsilon }\right\rangle \right) \exp
\left( \frac{2\pi i}{P}\left\langle \mathbf{m},\mathbf{y}\right\rangle
\right) d\mathbf{y}
\]%
\[
=\frac{K\exp \left( -rT-\left\langle \mathbf{d,\epsilon }\right\rangle
\right) }{P^{n}}\sum_{\mathbf{m}\in \Omega _{1/R}^{^{\prime }}}\left( \Phi ^{%
\mathbb{Q}}\left( \mathbf{-}\frac{2\pi }{P}\mathbf{m}+i\mathbf{\epsilon }%
,T\right) \exp \left( -\frac{2\pi i}{P}\left\langle \mathbf{m},\mathbf{d}%
\right\rangle \right) \right)
\]%
\[
\times \int_{\mathbb{R}^{n}}\left( \left( \exp \left( y_{1}\right)
-\sum_{j=2}^{n}\exp \left( y_{j}\right) -1\right) _{+}\exp \left(
\left\langle \mathbf{y},\mathbf{\epsilon }\right\rangle \right) \right) \exp
\left( \frac{2\pi i}{P}\left\langle \mathbf{m},\mathbf{y}\right\rangle
\right) d\mathbf{y}
\]%
\[
=\frac{K\exp \left( -rT-\left\langle \mathbf{d,\epsilon }\right\rangle
\right) }{P^{n}}\sum_{\mathbf{m}\in \Omega _{1/R}^{^{\prime }}}\left( \Phi ^{%
\mathbb{Q}}\left( \mathbf{-}\frac{2\pi }{P}\mathbf{m}+i\mathbf{\epsilon }%
,T\right) \exp \left( -\frac{2\pi i}{P}\left\langle \mathbf{m},\mathbf{d}%
\right\rangle \right) \right)
\]%
\[
\times \frac{\Gamma \left( -\frac{2\pi i}{P}\sum_{s=1}^{n}m_{s}-%
\sum_{s=1}^{n}\epsilon _{s}-1\right) \prod_{s=2}^{n}\Gamma \left(
\frac{2\pi i}{P}m_{s}+\epsilon _{s}\right) }{\Gamma \left( -\frac{2\pi i}{P}%
m_{1}-\epsilon _{1}+1\right) }
\]%
\[
=\widetilde{V}.
\]
$\square$

{\bf Theorem 36}
\begin{em}
\label{pricing1} \bigskip Let in our notations $T>0$, $0<\epsilon _{j}\leq
b_{+,j}$, $2\leq j\leq n$, $b_{-,1}\leq \epsilon
_{1}<-1-\sum_{m=2}^{n}\epsilon _{m}$, $\mathbf{d}:=\left( d_{1},\cdot \cdot
\cdot ,d_{n}\right) ,$ $d_{j}=\ln \left( \frac{S_{0,j}}{K}\right) ,$ $1\leq
j\leq n$, $b=\min \left\{ -b_{-,s},b_{+,s},\text{ }1\leq s\leq n\right\} $,
\[
M\left( P,R\right) :=\left\Vert \left( 2\pi \right) ^{-n}\int_{\mathbb{R}%
^{n}}\exp \left( i\left\langle \cdot \mathbf{-d,x}\right\rangle \right) \Phi
^{\mathbb{Q}}\left( \mathbf{-x}+i\mathbf{\epsilon },T\right) d\mathbf{x}%
\right.
\]%
\[
\left. \mathbf{-}\frac{1}{P^{n}}\sum_{\mathbf{m}\in \Omega _{1/R}^{^{\prime
}}}\left( \Phi ^{\mathbb{Q}}\left( -\frac{2\pi }{P}\mathbf{m}+i\mathbf{%
\epsilon },T\right) \exp \left( -\frac{2\pi i}{P}\left\langle \mathbf{m},%
\mathbf{d}\right\rangle \right) \right) \exp \left( \frac{2\pi i}{P}%
\left\langle \mathbf{m},\mathbf{\cdot }\right\rangle \right) \right\Vert
_{_{L_{\infty }\left( \mathbb{R}^{n}\right) }},
\]%
and $\widetilde{V}$ be the approximant for $V$ from Theorem 35
, then
\[
\delta :=\left\vert V-\widetilde{V}\right\vert
\]%
\[
\ll \frac{K\exp \left( -rT\right) \Gamma \left( -\sum_{s=1}^{n}\epsilon
_{s}-1\right) \prod_{s=2}^{n}\Gamma \left( \epsilon _{s}\right) }{%
\Gamma \left( 1-\epsilon _{1}\right) }
\]%
\[
\times \left( \exp \left( -Pb\right) +\left( mP^{-n}\right) ^{1-\left(
\sum_{s=1}^{n}\nu _{s}^{-1}\right) ^{-1}}\exp \left( -\left( mP^{-n}\right)
^{\left( \sum_{s=1}^{n}\nu _{s}^{-1}\right) ^{-1}}\right) \right)
\]%
\[
+K\exp \left( -rT\right) M\left( P,R\right)
\]%
\[
\times \left\Vert \left( \exp \left( y_{1}\right) -\sum_{j=2}^{n}\exp \left(
y_{j}\right) -1\right) _{+}\exp \left( \left\langle \mathbf{y},\mathbf{%
\epsilon }\right\rangle \right) \right\Vert _{L_{1}\left( \mathbb{R}%
^{n}\setminus \left( \frac{P}{2}-\left\Vert \mathbf{d}\right\Vert _{\infty
}\right) Q_{n}\right) },\text{ }m\rightarrow \infty ,\text{ }P\rightarrow
\infty .
\]
\end{em}

{\bf Proof.}
Let $\widetilde{V}$ be the approximant for $V$. Since $0<\epsilon _{j}\leq
b_{+,j}$, $2\leq j\leq n$, $b_{-,1}\leq \epsilon
_{1}<-1-\sum_{m=2}^{n}\epsilon _{m}$ then we get
\[
\delta =\left\vert V-\widetilde{V}\right\vert
\]%
\[
=K\exp \left( -rT\right) \left\vert \int_{\mathbb{R}^{n}}\left( \left( \exp
\left( y_{1}\right) -\sum_{j=2}^{n}\exp \left( y_{j}\right) -1\right)
_{+}\exp \left( \left\langle \mathbf{y},\mathbf{\epsilon }\right\rangle
\right) \right) \right.
\]%
\[
\times \left( \left( 2\pi \right) ^{-n}\int_{\mathbb{R}^{n}}\exp \left(
i\left\langle \mathbf{y-d,x}\right\rangle \right) \Phi ^{\mathbb{Q}}\left(
\mathbf{-x}+i\mathbf{\epsilon },T\right) d\mathbf{x}\right.
\]%
\[
\left. \left. -\frac{1}{P^{n}}\sum_{\mathbf{m}\in \Omega _{1/R}^{^{\prime
}}}\left( \Phi ^{\mathbb{Q}}\left( -\frac{2\pi }{P}\mathbf{m}+i\mathbf{%
\epsilon },T\right) \exp \left( -\frac{2\pi i}{P}\left\langle \mathbf{m},%
\mathbf{d}\right\rangle \right) \right) \exp \left( \frac{2\pi i}{P}%
\left\langle \mathbf{m},\mathbf{y}\right\rangle \right) \right) d\mathbf{y}%
\right\vert
\]%
\[
:=K\exp \left( -rT\right) \int_{\mathbb{R}^{n}}\left( \exp \left(
y_{1}\right) -\sum_{j=2}^{n}\exp \left( y_{j}\right) -1\right) _{+}\exp
\left( \left\langle \mathbf{y},\mathbf{\epsilon }\right\rangle \right) \mu
\left( \mathbf{y}\right) d\mathbf{y.}
\]%
From the Corollary \ref{collorary-conv} it follows that for chosen $P>0$ and
$m>0$ we have
\[
\left\vert \mu \left( \mathbf{y}\right) \right\vert =\left\vert \left( 2\pi
\right) ^{-n}\int_{\mathbb{R}^{n}}\exp \left( i\left\langle \mathbf{y-d,x}%
\right\rangle \right) \Phi ^{\mathbb{Q}}\left( \mathbf{-x}+i\mathbf{\epsilon
},T\right) d\mathbf{x}\right.
\]%
\[
\left. \mathbf{-}\frac{1}{P^{n}}\sum_{\mathbf{m}\in \Omega _{1/R}^{^{\prime
}}}\left( \Phi ^{\mathbb{Q}}\left( -\frac{2\pi }{P}\mathbf{m}+i\mathbf{%
\epsilon },T\right) \exp \left( -\frac{2\pi i}{P}\left\langle \mathbf{m},%
\mathbf{d}\right\rangle \right) \right) \exp \left( \frac{2\pi i}{P}%
\left\langle \mathbf{m},\mathbf{y}\right\rangle \right) \right\vert
\]%
\[
\ll \exp \left( -Pb\right) +\left( mP^{-n}\right) ^{1-\left(
\sum_{s=1}^{n}\nu _{s}^{-1}\right) ^{-1}}\exp \left( -\left( mP^{-n}\right)
^{\left( \sum_{s=1}^{n}\nu _{s}^{-1}\right) ^{-1}}\right)
\]%
for any $\mathbf{y}\in \frac{P}{2}Q_{n}-\mathbf{d}$. Let us put $\mathbf{m}=%
\mathbf{0}$ in the Theorem 34
. Then
\[
L_{\mathbf{\epsilon }}:=\left\Vert \left( \exp \left( y_{1}\right)
-\sum_{j=2}^{n}\exp \left( y_{j}\right) -1\right) _{+}\exp \left(
\left\langle \mathbf{y},\mathbf{\epsilon }\right\rangle \right) \right\Vert
_{L_{1}\left( \mathbb{R}^{n}\right) }
\]%
\[
=\frac{\Gamma \left( -\sum_{s=1}^{n}\epsilon _{s}-1\right)
\prod_{s=2}^{n}\Gamma \left( \epsilon _{s}\right) }{\Gamma \left(
1-\epsilon _{1}\right) }
\]%
for a chosen $\mathbf{\epsilon =}\left( \epsilon _{1},\cdot \cdot \cdot
,\epsilon _{n}\right) $, $0<\epsilon _{j}\leq b_{+,j}$, $2\leq j\leq n$, $%
b_{-,1}\leq \epsilon _{1}<-1-\sum_{m=2}^{n}\epsilon _{m}$. Observe that $%
\frac{P}{2}Q_{n}-\mathbf{d\supseteq }\left( \frac{P}{2}-\left\Vert \mathbf{d}%
\right\Vert _{\infty }\right) Q_{n}$, where $\left\Vert \mathbf{d}%
\right\Vert _{\infty }:=\max \left\{ \left\vert d_{k}\right\vert ,1\leq
k\leq n\right\} $. Therefore
\[
\int_{\left( \frac{P}{2}-\left\Vert \mathbf{d}\right\Vert _{\infty }\right)
Q_{n}}\left( \exp \left( y_{1}\right) -\sum_{j=2}^{n}\exp \left(
y_{j}\right) -1\right) _{+}\exp \left( \left\langle \mathbf{y},\mathbf{%
\epsilon }\right\rangle \right) \mu \left( \mathbf{y}\right) d\mathbf{y}
\]%
\[
\leq L_{\mathbf{\epsilon }}\left( \exp \left( -Pb\right) +\left(
mP^{-n}\right) ^{1-\left( \sum_{s=1}^{n}\nu _{s}^{-1}\right) ^{-1}}\exp
\left( -\left( mP^{-n}\right) ^{\left( \sum_{s=1}^{n}\nu _{s}^{-1}\right)
^{-1}}\right) \right) .
\]%
Finaly, we have
\[
\int_{\mathbb{R}^{n}\setminus \left( \frac{P}{2}-\left\Vert \mathbf{d}%
\right\Vert _{\infty }\right) Q_{n}}\left( \exp \left( y_{1}\right)
-\sum_{j=2}^{n}\exp \left( y_{j}\right) -1\right) _{+}\exp \left(
\left\langle \mathbf{y},\mathbf{\epsilon }\right\rangle \right) \mu \left(
\mathbf{y}\right) d\mathbf{y}
\]%
\[
\leq M\left( P,R\right) \left\Vert \left( \exp \left( y_{1}\right)
-\sum_{j=2}^{n}\exp \left( y_{j}\right) -1\right) _{+}\exp \left(
\left\langle \mathbf{y},\mathbf{\epsilon }\right\rangle \right) \right\Vert
_{L_{1}\left( \mathbb{R}^{n}\setminus \left( \frac{P}{2}-\left\Vert \mathbf{d%
}\right\Vert _{\infty }\right) Q_{n}\right) }.
\]
$\square$

Assume, for simplicity, $n=2$. Let, as before, $\epsilon
_{1}<-1-\epsilon _{2}$ and $\epsilon _{2}>0$. Since $\left( \exp \left(
y_{1}\right) -\exp \left( y_{2}\right) -1\right) _{+}\geq 0$ if $\exp \left(
y_{1}\right) -\exp \left( y_{2}\right) -1\geq 0$ and $x_{1}\geq 0$ then
\[
\left\Vert \left( \exp \left( y_{1}\right) -\exp \left( y_{2}\right)
-1\right) _{+}\exp \left( \left\langle \mathbf{y},\mathbf{\epsilon }%
\right\rangle \right) \right\Vert _{L_{1}\left( \mathbb{R}^{2}\setminus
\left( \frac{P}{2}-\left\Vert \mathbf{d}\right\Vert _{\infty }\right)
Q_{2}\right) }
\]%
\[
=\int_{L_{1}\left( \mathbb{R}^{2}\setminus \left( \frac{P}{2}-\left\Vert
\mathbf{d}\right\Vert _{\infty }\right) Q_{2}\right) }\left( \exp \left(
y_{1}\right) -\exp \left( y_{2}\right) -1\right) _{+}\exp \left(
\left\langle \mathbf{y},\mathbf{\epsilon }\right\rangle \right) d\mathbf{y}
\]%
\[
:=I_{1}+I_{2},
\]%
where%
\[
I_{1}=\int_{\frac{P}{2}-\max \left\{ d_{1},d_{1}\right\} }^{\infty }\exp
\left( \epsilon _{1}y_{1}\right) \left( \int_{-\infty }^{\ln \left( \exp
\left( x_{1}\right) -1\right) }\left( \exp \left( y_{1}\right) -1-\exp
\left( y_{2}\right) \right) \exp \left( \epsilon _{2}y_{2}\right)
dy_{2}\right) dy_{1}
\]%
and%
\[
I_{2}=\int_{0}^{\frac{P}{2}-\max \left\{ d_{1},d_{1}\right\} }\exp \left(
\epsilon _{1}y_{1}\right) \left( \int_{-\infty }^{-\frac{P}{2}+\max \left\{
d_{1},d_{1}\right\} }\left( \exp \left( y_{1}\right) -1-\exp \left(
y_{2}\right) \right) \exp \left( \epsilon _{2}y_{2}\right) dy_{2}\right)
dy_{1}.
\]%
It is possible to show that
\[
I_{1}\ll \exp \left( -\frac{\epsilon _{2}P}{2}\right)
\]%
and%
\[
I_{2}\ll \exp \left( \frac{\left( \epsilon _{1}+\epsilon _{2}+1\right) P}{2}%
\right)
\]%
as $P\rightarrow \infty $. Hence%
\[
\left\Vert \left( \exp \left( y_{1}\right) -\exp \left( y_{2}\right)
-1\right) _{+}\exp \left( \left\langle \mathbf{y},\mathbf{\epsilon }%
\right\rangle \right) \right\Vert _{L_{1}\left( \mathbb{R}^{2}\setminus
\left( \frac{P}{2}-\left\Vert \mathbf{d}\right\Vert _{\infty }\right)
Q_{2}\right) }
\]%
\[
\ll \exp \left( \frac{P}{2}\max \left\{ -\epsilon _{2},\epsilon
_{1}+\epsilon _{2}+1\right\} \right) ,\text{ }P\rightarrow \infty
\]%
where $\epsilon _{1}+\epsilon _{2}+1<0$, $\epsilon _{2}>0$.

\bigskip


\chapter*{Appendix I: Measure and integral}


\label{Measure and integral}


\section*{$L_{p}$ spaces}


\bigskip For $0<p<\infty $, $l_{p}$ is the space consisting of all sequences
$\mathbf{c}=\left\{ c_{\mathbf{k}},\text{ \ }\mathbf{k}\in \mathbb{Z}%
^{n}\right\} $ satisfying
\[
\sum_{\mathbf{k}\in \mathbb{Z}^{n}}\left\vert c_{\mathbf{k}}\right\vert
^{p}<\infty .
\]%
If $p\geq 1$, then
\[
\left\Vert \mathbf{c}\right\Vert _{p}:=\left( \sum_{\mathbf{k}\in \mathbb{Z}%
^{n}}\left\vert c_{\mathbf{k}}\right\vert ^{p}\right) ^{1/p}
\]%
defines a norm on $l_{p}$. If $p=\infty $, then the norm on $l_{\infty }$ is
defined by%
\[
\left\Vert \mathbf{c}\right\Vert _{\infty }:=\sup_{\mathbf{k}\in \mathbb{Z}%
^{n}}\left\vert c_{\mathbf{k}}\right\vert .
\]%
If $1\leq p\leq \infty $, then $l_{p}$ is a complete normed space with
respect to the norm $\left\Vert \mathbf{c}\right\Vert _{p}$, and therefore
is a Banach space.

Let $1\leq p\leq \infty $ and $\left( \Omega ,\mathcal{F},\upsilon \right) $
be a \textit{measure space} of functions $f:\Omega \rightarrow \mathbb{R}$
such that
\[
\left\Vert f\right\Vert _{p,\upsilon }=\left\Vert f\right\Vert _{p}:=\left\{
\begin{array}{cc}
\left( \int_{\Omega }\left\vert f\right\vert ^{p}d\upsilon \right) ^{1/p}, &
1\leq p<\infty , \\
\mathrm{ess}\sup \left\vert f\right\vert , & p=\infty%
\end{array}%
\right\} <\infty .
\]%
In this case we say that $f\in L_{p}=L_{p}\left( \Omega ,\mathcal{F}%
,\upsilon \right) $. For any $1\leq p\leq \infty $, $L_{p}$ is a Banach
space.

Let $\mathcal{L}$ be the \textit{Lebesgue }$\sigma $\textit{-algebra} and $d%
\mathbf{y}$ be the Lebesgue measure.

{\bf Theorem 37}
\begin{em}
(Young's inequality \cite{stain}, \cite{hlp}) Let $f\in L_{p}\left( \mathbb{R%
}^{n},\mathcal{L},d\mathbf{y}\right) $, $g\in L_{q}\left( \mathbb{R}^{n},%
\mathcal{L},d\mathbf{y}\right) $ and $1/p+1/q=1/r+1$. Then%
\[
\left\Vert h\right\Vert _{r}\leq \left\Vert f\right\Vert _{p}\left\Vert
g\right\Vert _{q}.
\]
\end{em}

{\bf Theorem 38}
\begin{em}
\label{jen} (Jensen inequality) Let $\left( \Omega ,\mathcal{F},\upsilon
\right) $ be a probability space, i.e. $\upsilon \left( \Omega \right) =1$.
Let $f:\Omega \rightarrow \mathbb{R}$ be a $\upsilon $-integrable function
and $g:\mathbb{R}\rightarrow \mathbb{R}$ be a convex function. Then
\[
\int_{\Omega }g\circ fd\upsilon \geq g\left( \int_{\Omega }fd\upsilon
\right) .
\]
\end{em}

\section*{Fubini and Tonelli theorems}


Fubini theorem allows us to compute a double integral using iterated
integrals. As a consequence it gives us sufficient conditions to change the
order of integration. It is one of the central tools of the probability
theory.

{\bf Theorem 39}
\begin{em}
\label{Fubini's theorem} (Fubini theorem) Suppose $\left( A,\mathcal{F}%
_{1},\upsilon _{1}\right) $ and $\left( B,\mathcal{F}_{2},\upsilon
_{2}\right) $ are complete measure spaces. Assume that $f\left( x,y\right) $
is $\upsilon _{1}\times \upsilon _{2}$ measurable on $A\times B$. If
\[
\int_{A\times B}\left\vert f\left( x,y\right) \right\vert d\upsilon
_{1}d\upsilon _{2}<\infty ,
\]%
where the integral is taken with respect to a product measure $\upsilon
_{1}\times \upsilon _{2}$ on $A\times B,$ then
\[
\int_{A\times B}f\left( x,y\right) d\upsilon _{1}d\upsilon
_{2}=\int_{A}\left( \int_{B}f\left( x,y\right) d\upsilon _{2}\right)
d\upsilon _{1}
\]%
\[
=\int_{B}\left( \int_{A}f\left( x,y\right) d\upsilon _{1}\right) d\upsilon
_{2}.
\]
\end{em}

If $\int_{A\times B}\left\vert f\left( x,y\right) \right\vert d\upsilon
_{1}d\upsilon _{2}=\infty ,$ then \ the two iterated integrals from the
right may have different values.

The measure $\upsilon $ is called \textit{$\sigma $--finite} if $\Omega $ is
the countable union of measurable sets with finite measure.

Another important theorem for much of probability theory is the following
statement.

{\bf Theorem 40}
\begin{em}
\label{Tonelli theorem} (Tonelli theorem) Let $\left( A,\mathcal{F}%
_{1},\upsilon _{1}\right) $ and $\left( B,\mathcal{F}_{1},\upsilon
_{2}\right) $ are two $\sigma -$finite measure spaces and $f\left(
x,y\right) $ be a $\upsilon _{1}\times \upsilon _{2}$ measurable function
such that $f\left( x,y\right) \geq 0,$ $\ \forall \left( x,y\right) \in
A\times B$ then
\[
\int_{A\times B}f\left( x,y\right) d\upsilon _{1}d\upsilon
_{2}=\int_{A}\left( \int_{B}f\left( x,y\right) d\upsilon _{2}\right)
d\upsilon _{1}
\]%
\[
=\int_{B}\left( \int_{A}f\left( x,y\right) d\upsilon _{1}\right) d\upsilon
_{2}.
\]
\end{em}

Any probability space has a $\sigma $-finite measure. In this situation
Tonelli theorem simply says that if $f\left( x,y\right) \geq 0,\forall
\left( x,y\right) \in A\times B$ then we can change the order of integration
without a hard condition $\int_{A\times B}\left\vert f\left( x,y\right)
\right\vert d\upsilon _{1}d\upsilon _{2}<\infty $ of Fubini theorem.


\section*{Radon-Nikodym theorem}


Let $(\Omega ,\mathcal{F},\upsilon )$ be a measure space. Assume that $%
\upsilon _{1}$ and $\upsilon _{2}$ are two measures on a \textit{measurable
set} $(\Omega ,\mathcal{F})$ and $\upsilon _{2}(A)=0\Rightarrow \upsilon
_{1}(A)=0$ then we say that $\upsilon _{1}$ is \textit{absolutely continuous}
with respect to $\upsilon _{2}$ (or dominated by $\upsilon _{2}$). In this
case we shall write $\upsilon _{1}\ll \upsilon _{2}$. If $\upsilon _{1}\ll
\upsilon _{2}$ and $\upsilon _{2}\ll \upsilon _{1}$, the measures $\upsilon
_{1}$ and $\upsilon _{2}$ are said to be \textit{equivalent}, $\upsilon
_{1}\asymp \upsilon _{2}$.

{\bf Theorem 41}
\begin{em}
\label{Radon--Nikodim} (Radon-Nikodym) Let $\upsilon _{1}$ and $\upsilon
_{2} $ are two $\sigma $--finite measures on a measure space $(\Omega ,%
\mathcal{F})$ and $\upsilon _{1}\ll \upsilon _{2}$, then there exists a $%
\upsilon _{2}$--measurable function $f$ with the range $R(f)\subset \lbrack
0,\infty )$, denoted by $f=d\upsilon _{1}/d\upsilon _{2}$, such that for any
$\upsilon _{2}$--measurable set $A$ we have
\[
\upsilon _{1}(A)=\int_{A}f\cdot d\upsilon _{2}.
\]
\end{em}

See, e.g. \cite{Shilov--Gurevich} for more information.

\chapter*{Appendix II: Harmonic Analysis}


\label{Harmonic Analysis}


\section*{Plancherel theorem}

To justify an inversion formula we will need Plancherel theorem (see e.g.
\cite{plancherel}). Let in our notation $L_{2}\left( \mathbb{R}^{n}\right)
:=L_{p}\left( \mathbb{R}^{n},\mathcal{L},d\mathbf{y}\right) $.

{\bf Theorem 42}
\begin{em}
\label{Plancherel} (Plancherel) The Fourier transform is a linear continuous
operator from $L_{2}\left( \mathbb{R}^{n}\right) $ onto $L_{2}\left( \mathbb{%
R}^{n}\right) .$ The inverse Fourier transform, $\mathbf{F}^{-1},$ can be
obtained by letting
\[
\left( \mathbf{F}^{-1}g\right) \left( \mathbf{x}\right) =\frac{1}{(2\pi )^{n}%
}\left( \mathbf{F}g\right) \left( -\mathbf{x}\right)
\]%
for any $g\in L_{2}\left( \mathbb{R}^{n}\right) .$
\end{em}

\section*{Riesz-Thorin and Riesz theorems}


The Riesz-Thorin interpolation theorem is an important tool in Harmonic
Analysis and Probability. This theorem bounds norms of linear operators
acting between $L_{p}=L_{p}(\Omega ,\mathcal{F},\upsilon )$ spaces.

{\bf Theorem 43}
\begin{em}
\label{Riesz-Thorin} (Riesz-Thorin, \cite{riesz1}) Let $(\Omega _{1},%
\mathcal{F}_{1},\upsilon _{1})$ and $(\Omega _{2},\mathcal{F}_{2},\upsilon
_{2})$ be $\sigma $-finite measure spaces. Suppose $1\leq
p_{0},p_{1},q_{0},q_{1}\leq \infty $, and let $\boldsymbol{A}$ be a bounded
linear operator $\boldsymbol{A}\in \mathrm{L}\left(
L_{p_{0}},L_{q_{0}}\right) \cap \mathrm{L}\left( L_{p_{1}},L_{q_{1}}\right) $%
. Then
\[
\left\Vert \boldsymbol{A}\left\vert L_{p_{\theta }}\rightarrow L_{q_{\theta
}}\right. \right\Vert \leq \left\Vert \boldsymbol{A}\left\vert
L_{p_{0}}\rightarrow L_{q_{0}}\right. \right\Vert ^{1-\theta }\cdot
\left\Vert \boldsymbol{A}\left\vert L_{p_{1}}\rightarrow L_{q_{1}}\right.
\right\Vert ^{\theta },\forall \theta \in \left[ 0,1\right] ,
\]%
where%
\[
\frac{1}{p_{\theta }}=\frac{1-\theta }{p_{0}}+\frac{\theta }{p_{1}},\text{ \
\ }\frac{1}{q_{\theta }}=\frac{1-\theta }{q_{0}}+\frac{\theta }{q_{1}}.
\]
\end{em}

{\bf Theorem 44}
\begin{em}
\label{riesz theorem} (F. Riesz, \cite{riesz1} v. 2, p. 123) Let $\left(
\Omega ,\mathcal{F},\upsilon \right) $ be a measure space and $\omega _{%
\mathbf{k}}\left( \mathbf{x}\right) $, $\mathbf{k}\in \mathbb{Z}^{n}$ be any
orthonormal and uniformly bounded system over $\Omega $, i.e.%
\[
\int_{\Omega }\omega _{\mathbf{k}}\left( \mathbf{x}\right) \overline{\omega }%
_{\mathbf{m}}\left( \mathbf{x}\right) d\upsilon =\delta _{\mathbf{k,m}%
}:=\left\{
\begin{array}{cc}
1, & \mathbf{k}=\mathbf{m}, \\
0, & \mathbf{k}\neq \mathbf{m}%
\end{array}%
\right.
\]%
and
\[
\sup_{\mathbf{x}\in \Omega }\left\vert \omega _{\mathbf{k}}\left( \mathbf{x}%
\right) \right\vert \leq L,\text{ \ \ }\forall \mathbf{k}\in \mathbb{Z}^{n},
\]%
Let $1\leq p\leq 2$.

\begin{enumerate}
\item If $f\in L_{p}\left( \Omega ,\mathcal{F},\upsilon \right) $, then the
Fourier coefficients%
\[
c_{\mathbf{k}}:=\int_{\Omega }f\left( \mathbf{x}\right) \overline{\omega }_{%
\mathbf{k}}\left( \mathbf{x}\right) d\upsilon
\]%
satisfy the inequality%
\[
\left\Vert \mathbf{c}\right\Vert _{p^{^{\prime }}}\leq L^{2/p-1}\left\Vert
f\right\Vert _{p},
\]%
where $\mathbf{c}=\left\{ c_{\mathbf{k}},\text{ \ \ }\mathbf{k}\in \mathbb{Z}%
^{n}\right\} ,$ $1/p+1/p^{^{\prime }}=1$ and
\[
\left\Vert \mathbf{c}\right\Vert _{q}:=\left( \sum_{\text{\ }\mathbf{k}\in
\mathbb{Z}^{n}}\left\vert c_{\mathbf{k}}\right\vert ^{q}\right) ^{1/q},\text{
\ \ }1\leq q\leq \infty .
\]

\item Given any sequence $\mathbf{c:=}\left\{ c_{\mathbf{k}},\text{ }\mathbf{%
k}\in \mathbb{Z}^{n}\right\} $ with $\left\Vert \mathbf{c}\right\Vert _{p}$
\ finite, there is an $f\in L_{p^{^{\prime }}}\left( \Omega ,\mathcal{F}%
,\upsilon \right) $ satisfying
\[
c_{\mathbf{k}}:=\int_{\Omega }f\left( \mathbf{x}\right) \overline{\omega }_{%
\mathbf{k}}\left( \mathbf{x}\right) d\upsilon
\]%
for all $\mathbf{k}\in \mathbb{N}^{n}$ and
\[
\left\Vert f\right\Vert _{p^{^{\prime }}}\leq L^{2/p-1}\left\Vert \mathbf{c}%
\right\Vert _{p}.
\]
\end{enumerate}
\end{em}


See \cite{riesz1, goldberg,bachman} for more information.



\chapter*{Appendix III: Martingales}


\label{Martingales}

\bigskip

\section*{Martingale methods and pricing}


\label{Martingale}

Observe that every forecast is an average of possible future values. All
possible values that the random variable can assume in an unfolding future
are weighted by the probabilities associated with these values. Hence we
need to compute expected values of random variables $S_{t}$ based on the
information reviled at time $\tau \leq T$. The theory of martingales is
commonly used for these purposes. Martingales (semi-martingales) is an
important class of random sequences with various applications in derivative
pricing. We will need some basic definitions.

{\bf Definition 45}
\begin{em}
A \textit{binary relation} $\preceq$ on a set $A$ is a collection of ordered
pairs $(a,b)$ of elements of A. In other words, it is a subset of the
Cartesian product $A^{2} = A \times A$.
\end{em}

{\bf Definition 46}
\begin{em}
We say that a binary relation $\preceq$ is \textit{antisymmetric} if $a
\preceq b$ and $b \preceq a$ then $a = b$, \textit{transitive} if $a \preceq
b$ and $b \preceq c$ then $a \preceq c$ and \textit{total} if $a \preceq b$
or $b \preceq a$.
\end{em}

{\bf Definition 47}
\begin{em}
A \textit{total order} is a binary relation (denoted by $\preceq $) on some
set $A$ which is transitive, antisymmetric, and total. A set $\mathcal{T}$
paired with a total order $\preceq $ is called a \textit{totally ordered set}
(or a \textit{chain}).
\end{em}

In general, the information used by decision makers will increase as time $t$
passes. It is natural to assume that the decision maker never forgets past
data. Hence, the following definition.

{\bf Definition 48}
\begin{em}
A family of $\sigma $-algebras $\{\mathcal{F}_{t}|\,t\in T\}$, $\mathcal{F}%
_{t}\subset \mathcal{F}$, $\mathcal{F}_{t_{1}}\subset \mathcal{F}_{t_{2}}$
if $t_{1}\preceq t_{2}$ on a given probability space $(\Omega ,\mathcal{F},%
\mathbb{P})$ is called a \textit{current of $\sigma $-algebras} (\textit{a
current of experiments} or \textit{filtration}).
\end{em}

The set $\mathcal{F}_{t}$ can be interpreted as the class of all observed
events in the experiments carried out up to the moment $t$\ inclusively. Fix
an arbitrary totally ordered set $\mathcal{T}$. Let $\{\mathcal{F}%
_{t}|\,t\in \mathcal{T}\}$ be a current of $\sigma $-algebras.

{\bf Definition 49}
\begin{em}
A family of random variables $\{\xi (t),\mathcal{F}_{t},t\in \mathcal{T}\}$
in which the random variables $\xi (t)$ are $\mathcal{F}_{t}$ measurable for
each $t\in \mathcal{T}$ is called a \textit{martingale} if
\[
\mathbb{E}[|\xi (t)|]<\infty ,
\]%
\[
\mathbb{E}[\xi (t)|\,\mathcal{F}_{s}]=\xi (s),\,\,\mathbb{P}-a.s.,\text{ }%
s\preceq t,\,\,s,t\in \mathcal{T}.
\]%
A family $\{\xi (t),\mathcal{F}_{t},t\in \mathcal{T}\}$ is called a \textit{%
submartingale}, if
\[
\mathbb{E}[\xi (t)|\,\mathcal{F}_{s}]\geq \xi (s),\,\mathbb{P}-a.s.,\text{ }%
\,s\preceq t,\,\,s,t\in \mathcal{T},
\]%
and \textit{supermartingale} if
\[
\mathbb{E}[\xi (t)|\,\mathcal{F}_{s}]\leq \xi (s),\,\mathbb{P}-a.s.,\text{ }%
\,s\preceq t,\,\,s,t\in \mathcal{T}.
\]

Super and submartingales are called \textit{semimartingales}.
\end{em}

Remark that the property $\mathbb{E}[\xi (t)|\,\mathcal{F}_{s}]=\xi (s),\,\,%
\mathbb{P}-a.s.,$ $s\preceq t,\,\,s,t\in \mathcal{T}$ means that the best
forecast of unobserved future values is the last observation on $\xi (s)$.
All expectations here are assumed to be taken with respect to the
probability measure $\mathbb{P}$. Observe that a martingale is always
defined with respect to some current of $\sigma $-algebras $\{\mathcal{F}%
_{t}|\,t\in \mathcal{T}\}$ and probability measure $\mathbb{P}$.
Unfortunately, most financial assets are not martingales. For instance, the
price of a bond is expected to increase over time. Also, the stock prices
are expected to increase on average over time. It means that
\[
B_{t}<\mathbb{E}[B_{s}|\,\mathcal{F}_{t}],\,\,t<s<T,
\]%
where $B_{t}$ is the price of a bond maturing at time $t$ and $T$ is a time
horizon. Clearly, it contradicts the condition $B_{t}=\mathbb{E}[B_{s}|\,%
\mathcal{F}_{t}],\,\,t<s$. Similarly, a stock $S_{t}$ will have a positive
expected return. Hence, it does not behave as a martingale. The same
observation is true for the price of European-type options. Although the
majority of financial assets are not martingales, it is still possible to
convert them into martingales.

The majority of known methods of pricing derivatives employ the notion of
arbitrage which reflects market equilibrium. It means that if an arbitrage
portfolio exists, there exist an opportunity of "free lunches". In a real
financial market any arbitrage opportunity will be eliminated by the
activity of brokers who will try to make money using that opportunity and
marked naturally will enter into the state of equilibrium.

Our later discussion shows that no matter what the "true" (or historic)
probabilities are, if there are no arbitrage opportunities, one can
represent the fair market value of a financial instrument using probability
measures constructed under the equilibrium assumption. See \cite{Salih,
Jarrow, Duffie, Hull, Das} for more details. There are two conventional ways
to proceed. The first approach is based on Doob-Meyer's theorem (see, e.g.
\cite{gs3}, p. 25, \cite{Salih}, p. 141).

{\bf Theorem 50}
\begin{em}
(Doob-Meyer decomposition) If $\xi (t)$, $t\geq 0$ is a right-continuous
submartingale with respect to $\mathcal{F}_{t}$, then $\xi \left( t\right) $
admits the decomposition
\[
\xi (t)=M_{t}+A_{t},
\]%
where $M_{t}$ is a right-continuous martingale with respect to probability $%
\mathbb{P}$ and $A_{t}$ is an increasing process measurable with respect to $%
\mathcal{F}_{t}$.
\end{em}

The second approach is based on the idea of changing probability measure to
make $\exp \left( -rt\right) S_{t}$ a martingale. This commonly used in
derivative pricing method is based on Girsanov's theorem and is based on a
proper change of the underlying probability distribution $\mathbb{P}$. More
precisely, if $\exp \left( -rt\right) S_{t}$ is a submartingale, i.e.
\[
\mathbb{E}^{\mathbb{P}}\left[ \exp \left( -rs\right) S_{t+s}|\,\mathcal{F}%
_{t}\right] >S_{t},\,\,\forall s>0,
\]%
where $\mathbb{E}^{\mathbb{P}}\left[ \exp \left( -rs\right) S_{t+s}|\,%
\mathcal{F}_{t}\right] $ is the conditional expectation calculated using a
probability distribution $\mathbb{P}$ then applying Girsanov's theorem we
can find a probability distribution $\mathbb{Q}$ (on the same measure
space), such that
\[
\mathbb{E}^{\mathbb{Q}}\left[ \exp \left( -rs\right) S_{t+s}|\,\mathcal{F}%
_{t}\right] =S_{t},\,\,\forall s>0.
\]%
Hence, $\exp \left( -rs\right) S_{t}$ becomes a martingale. Such probability
distributions $\mathbb{Q}$ are called \textit{equivalent martingale measures}%
.

{\bf Definition 51}
\begin{em}
A \textit{standard Brownian motion} is a random process $X=\{X_{t}|\,t\in
\mathbb{R}_{+}\}$ with \textit{state space} that satisfies the following
properties:

\begin{enumerate}
\item $X_{0}=0$ (with probability $1$).

\item $X$ has stationary increments. That is $\forall s,t\in \lbrack
0,\infty )$, $s<t$, the distribution of $X_{t}-X_{s}$ is the same as the
distribution $X_{t-s}$.

\item $X$ has independent increments, or $\forall t_{1},\cdots t_{n}\in
\lbrack 0,\infty )$ with $t_{1}<\cdots <t_{n}$, the random variables $%
X_{t_{1}},X_{t_{2}}-X_{t_{1}},\cdots ,X_{t_{n}}-X_{t_{n-1}}$ are independent.

\item $X_{t}$ is normally distributed, $\forall t\in \lbrack 0,\infty
)\Rightarrow X_{t}\sim N(0,t).$

\item With probability $1$, $t\mapsto X_{t}$ is continuous on $[0,\infty )$.
\end{enumerate}
\end{em}

{\bf Theorem 52}
\begin{em}
(Girsanov) Consider the probability space $(\Omega ,\mathcal{F},(\mathcal{F}%
_{t})_{0\leq t\leq T},\mathbb{P})$. Assume that $(\Theta _{t})_{0\leq t\leq
T}$ is an adapted to the filtration $(\mathcal{F}_{t})_{0\leq t\leq T}$
process such that $\int_{0}^{T}\Theta _{s}^{2}ds<\infty $ and the process $%
(L_{t})_{0\leq t\leq T}$ is a martingale:
\[
L_{t}:=\exp \left( -\int_{0}^{t}\Theta _{s}dW_{s}-\int_{0}^{t}\Theta
_{s}^{2}ds\right) ,
\]%
where $dW_{t}$ is a standard Brownian motion. Then under the probability $%
\mathbb{P}^{(L)}$ with the density $L_{T}$ with respect to $\mathbb{P}$, the
process $(W_{t}^{\ast })_{0\leq t\leq T}$,
\[
W_{t}^{\ast }:=W_{t}+\int_{0}^{t}\Theta _{s}ds
\]%
is a standard Brownian motion.
\end{em}

\section*{The class of equivalent martingale measures}


\label{EMM}

A market model is called \textit{complete} if the set $EMM$ of all
equivalent martingale measures is a singleton. The necessary and sufficient
conditions for the absence of arbitrage and for the completeness are given
in \cite{Cherny-33}.

{\bf Theorem 53}
\begin{em}
Let $W_{t}$ denote the standard Brovnian motion and $N_{t}$ a standard
Poisson process. Suppose $S_{t}$ is neither increasing nor decreasing. Let $%
\mathcal{F}_{t}:=\sigma \left( S_{u},u\leq t\right) $ be the natural
filtration of $S_{t}$. Then model $S_{t}=S_{0}\exp \left( Z_{t}\right) $ is
complete in the following cases only:

(1) $Z_{t}=\alpha W_{t}+\beta t$, $\left( \alpha ,\beta \right) \in \mathbb{R%
}^{2}\mathbb{\setminus }\left\{ \alpha =0,\beta \neq 0\right\} ;$

(2) $Z_{t}=\alpha W_{\gamma t}+\beta t$, $\left( \alpha ,\beta \right) \in
\mathbb{R}^{2}$, $\gamma >0$ and $\alpha \beta <0$.
\end{em}

We see that the Black-Scholes model is complete in contrast to the
hyperbolic or KoBoL models. It was shown in \cite{44-2015} that the set $EMM$
is so rich that every price in some interval $\left( a,b\right) $ can be
obtained by a particular martingale measure $\mathbb{Q}$. Let $r>0$ be the
constant rate, $\mu $ the drift and $\Pi $ be the L\'{e}vy measure of $Z_{t}$
under $\mathbb{P}$. Let $EMM^{^{\prime }}$ be the subset of all $\mathbb{%
Q\in }EMM$ under which $Z_{t}$ is again L\'{e}vy process. If the system%
\[
\left\{
\begin{array}{c}
\int_{\mathbb{R}}\left( y^{1/2}\left( x\right) -1\right) ^{2}\Pi \left(
dx\right) +\int_{\left\vert x\right\vert >1}\left( \exp \left( x\right)
-1\right) y\left( x\right) \Pi \left( dx\right) <\infty \\
\mu -r+\int_{\mathbb{R}}\left( \left( \exp \left( x\right) -1\right) y\left(
x\right) -\chi _{D}\left( x\right) \right) \Pi \left( dx\right) =0%
\end{array}%
\right.
\]%
has a solution $y:\mathbb{R\rightarrow }\left( 0,\infty \right) $, then $%
EMM\supset EMM^{^{\prime }}\neq \varnothing $.

{\bf Theorem 54}
\begin{em}
(Eberlein-Jacod \cite{44-2015}) Consider the range sets%
\[
I_{e}:=\left\{ \exp \left( -rT\right) \mathbb{E}^{\mathbb{Q}}\left[ H\right]
\left\vert \mathbb{Q\in }EMM\right. \right\} ,
\]%
\[
I_{e}^{^{\prime }}:=\left\{ \exp \left( -rT\right) \mathbb{E}^{\mathbb{Q}}%
\left[ H\right] \left\vert \mathbb{Q\in }EMM^{^{\prime }}\right. \right\} .
\]%
If the L\'{e}vy measure $\Pi $ of $Z_{t}$ under $\mathbb{P}$ satisfies

(1) $\Pi \left( \left( -\infty ,a\right] \right) >0$, $\forall a\in \mathbb{R%
};$

(2) $\Pi $ has no atom and satisfies%
\[
\int_{\left[ -1,0\right) }\left\vert x\right\vert \Pi \left( dx\right)
=\int_{\left[ -1,0\right) }\left\vert x\right\vert \Pi \left( dx\right)
=\infty
\]%
then $EMM$ is not empty, $I_{e}$ is the full interval
\[
\left( \exp \left( -rT\right) H\left( S_{0}\exp \left( rT\right) \right)
,S_{0}\right) ,
\]%
where $H$ is the pay-off function and $I_{e}^{^{\prime }}$ is dense in this
interval.
\end{em}

In order to calculate option prices we need to choose an equivalent
martingale measure in $EMM^{^{\prime }}$. There are two common approaches,
the Esscher transform and the so-called minimal Entropy measure.

Let $Z_{t}$ be a L\'{e}vy process on $(\Omega ,\mathcal{F},(\mathcal{F}%
_{t})_{0\leq t\leq T},\mathbb{P})$. The Esscher transform is any change of $%
\mathbb{P}$ to an equivalent measure $\mathbb{Q}$ with a density process $%
\frac{d\mathbb{Q}}{d\mathbb{P}}\left\vert _{\mathcal{F}_{t}}\right. $ (see
Appendix I, Theorem 41 
 for the definition) of the form
\[
X_{t}=\frac{\exp \left( \theta Z_{t}\right) }{M\left( \theta \right) ^{t}},%
\text{ }\theta \in \mathbb{R},
\]%
where $M\left( \theta \right) $ is the moment generating function of $Z_{t}$.

In general, for any infinitely divisible distribution $\upsilon \left(
dx\right) $ on $\mathbb{R}$ with a finite moment generating function on some
interval $\left( c,d\right) ,$ $c<0<d$ the Esscher transform $\upsilon
_{\theta }\left( dx\right) $ is infinitely divisible for any $\theta \in
\left( c,d\right) $ with L\'{e}vy generating triplet $\left( a_{\theta },\Pi
_{\theta },h_{\theta }\right) $ given by%
\[
a_{\theta }=a,
\]%
\[
\Pi _{\theta }\left( dx\right) =\exp \left( \theta x\right) \Pi \left(
dx\right) ,
\]%
\[
h_{\theta }=h+\theta a+\int_{\mathbb{R}}\left( \exp \left( \theta x\right)
-1\right) \chi _{D}\left( x\right) \Pi \left( dx\right)
\]%
(see \cite{99-2015}). In this case the characteristic exponent should
satisfy (see \cite{bl1}, p. 20),%
\[
\psi ^{\mathbb{Q}}\left( \xi \right) =\psi ^{\mathbb{P}}\left( \xi -i\theta
\right) -\psi ^{\mathbb{P}}\left( -i\theta \right)
\]%
for some $\theta \in \mathbb{R}$. Comparing this with the Theorem 9
we get
\[
r+\psi ^{\mathbb{P}}\left( -i\left( \theta +1\right) \right) -\psi ^{\mathbb{%
P}}\left( -i\theta \right) =0.
\]%
The method of minimal entropy is presented in \cite{89-2015, 56-2015,
95-2015}. See \cite{6-2015, 53-2015, Gerber, 14-2015, 15-2015} for more
information.

\bigskip

\chapter*{Appendix IV: Comparison of numerical methods}


\label{Comparison}

\bigskip

\section*{$m$-Widths}


\label{Widths}


$m$-Widths were introduced by Kolmogorov \cite{kolmogorov} in 1936 to
compare and classify a wide range of numerical methods. Let $X$ be a Banach
space with the norm $\Vert \cdot \Vert $. Kolmogorov's $n$-width $%
d_{n}\left( A,X\right) $ of a symmetric set $A$ in $X$ is defined as
\[
d_{m}\left( A,X\right) =\inf_{L_{m}\subset X}\sup_{A\subset X}\inf_{y\in
L_{m}}\left\Vert x-y\right\Vert ,
\]%
where the last $\inf $ is taken over all subspaces $L_{m}\subset X$ of
dimension $n$. The problem of calculating the $m$-widths usually splits into
two parts: estimating the quantity%
\[
E\left( L_{m},A,X\right) =\sup_{A\subset X}\inf_{y\in L_{m}}\left\Vert
x-y\right\Vert ,
\]%
where $L_{m}$ is a fixed subspace, which gives us a necessary upper bound,
and obtaining a lower estimate of the width $d_{m}\left( A,X\right) $. The
difficulty in finding lower bound for $m$-width is that all $m$-dimensional
subspaces $L_{m}\subset X$ have to be considered. In 1960 Tikhomirov \cite%
{tik2} proved a theorem on the diameter of a ball (see Theorem 61
)
where he first applied an interesting topological method, namely the theorem
of Borsuk-Ulam, on the basis of which he proposed a method of obtaining
lower estimates of widths. \ \ We present here a simple proof of Theorem \ref%
{ball} which is important in our applications.

Let us remind some definitions. Let $X$ be a Banach space with the unit ball
$B$ and $A$ be a compact, centrally symmetric subset of $X$. Let $L_{m+1}$
be an $(m+1)$-dimensional subspace in $X$. Bernstein's $m$-width is defined
as
\[
b_{m}\left( A,X\right) =\sup \left\{ L_{m+1}\subset X\left\vert \sup \left\{
\epsilon >0\left\vert \epsilon B\cap L_{m+1}\subset A\right. \right\}
\right. \right\} .
\]%
The Alexandrov's $m$-width is the value
\[
a_{m}\left( A,X\right) =\inf_{\Sigma _{m}\subset X}\inf_{\sigma
:A\rightarrow \Sigma _{m}}\sup \left\{ \left\Vert x-\sigma \left( x\right)
\right\Vert \left\vert x\in A,\right. \right\} ,
\]%
where the infimum is taken over all $m$-dimensional complexes $\Sigma _{m}$,
lying in $X$ and all continuous mappings $\sigma :A\rightarrow \Sigma _{m}$.
The Urysohn's width $u_{m}\left( A,X\right) $ is the infimum of those $%
\epsilon >0$ for which there exists a covering of $A$ by open sets (in the
sense of topology induced by the norm $\left\Vert \cdot \right\Vert $ in $X$%
) of diameter $<\epsilon $ in $X$ and multiplicity $m+1$ (i.e. such that
each point is covered by $\leq m+1$ sets and some point is covered by
exactly $m+1$ sets). Observe that the width $u_{m}\left( A,X\right) $ was
introduced by Urysohn \cite{uryson} and inspired by the Lebesgue-Brouwer
definition of dimension.

In problems of optimal recovery arise quantities which are known as
cowidths. Let $\left( X,\vartheta \right) $ be a given metric (Banach)
space, $Y$ a certain set (coding set), $A\subset X$, $\Theta $ a family of
mappings $\theta :A\rightarrow Y$, then the respective cowidth can be
defined as
\[
\mathrm{co}^{\Theta }\left( A,X\right) =\inf_{\theta \in \Theta }\sup_{y\in
\theta \left( A\right) }\mathrm{diam}\left\{ \theta ^{-1}\left( y\right)
\cap A\right\} ,
\]%
where%
\[
\theta ^{-1}\left( y\right) =\left\{ x\left\vert x\in X,\text{ }\theta
\left( x\right) =\theta \left( y\right) \right. \right\} .
\]%
In particular, let $Y$ be $\mathbb{R}^{m}$ and $\Theta :A\rightarrow \mathbb{%
R}^{m}$ be a linear application, $\Theta =\mathrm{L}\left( A,\mathbb{R}%
^{m}\right) $, then we get a linear cowidth $\lambda ^{m}\left( A,X\right) $%
. It is easy to check that $\lambda ^{m}=2d^{m}$, where $d^{m}$ is the
Gelfand's $m$-width defined by
\[
d^{m}\left( A,X\right) =\inf \text{ \ }\left\{ L_{-m}\subset X\left\vert
\sup \left\{ \left\Vert x\right\Vert \left\vert x\in A\cap L_{-m}\right.
\right\} \right. \right\} ,
\]%
where $\inf $ is taken over all subspaces $L_{-m}\subset X$ of codimension $%
m $. Letting $Y$ be the set of all $m$-dimensional complexes in $X$ and $%
\Theta =C\left( A,Y\right) $ be the set of all continuous mappings $\theta
:A\rightarrow Y$, then we get Alexandrov's cowidths $a^{m}\left( A,X\right) $%
.

\section*{Functional and operator of best approximation}


\label{b-approx}


\bigskip

Here we present some known facts about functional and operator of best
approximation. Let $X$ be a Banach space with the norm $\Vert \cdot \Vert
_{X}=\Vert \cdot \Vert $. The deviation of $x\in X$ from the non-empty
subset $M\subset X$, i.e.
\begin{equation}
E(x)=E(x,M)=E(x,M,X):=\inf_{y\in M}\Vert x-y\Vert _{X}  \label{bestapprox}
\end{equation}%
is known as the best approximation of $x$ from the set $M$. For a fixed set $%
M\subset X$ the equation (\ref{bestapprox}) defines a functional on $X$, $%
E:X\rightarrow \mathbb{R}_{+}$ which is called the best approximation
functional.

{\bf Proposition 55}
\begin{em}
\label{5.1} Let $M\subset X$ be a linear manifold, then the functional $%
E(\cdot ,M)$ is uniformly continuous, subadditive:
\[
E(x_{1}+x_{2})\leq E(x_{1})+E(x_{2}),{\ }{\ }\forall x_{1},x_{2}\in X,
\]%
positively homogeneous:
\[
E(ax)=|a|E(x),{\ }{\ }\forall a\in \mathbb{R}
\]%
and convex:
\[
E(ax_{1}+(1-a)x_{2})\leq aE(x_{1})+(1-a)E(x_{2}),
\]%
\[
\forall a\in \lbrack 0,1],{\ }{\ }\forall x_{1},x_{2}\in X.
\]
\end{em}

{\bf Proof.}
Let $x_{1}\in X$ and $x_{2}\in X$, then for any $y\in M$
\[
E(x_{1})\leq \Vert x_{1}-y\Vert \leq \Vert x_{1}-x_{2}\Vert +\Vert
x_{2}-y\Vert .
\]%
Taking the infimum on $y\in M$ we find
\[
E(x_{1})\leq \Vert x_{1}-x_{2}\Vert +E(x_{2}),
\]%
or
\[
E(x_{1})-E(x_{2})\leq \Vert x_{1}-x_{2}\Vert .
\]%
Interchanging $x_{1}$ and $x_{2}$ we get $E(x_{2})-E(x_{1})\leq \Vert
x_{1}-x_{2}\Vert $ or
\[
|E(x_{1})-E(x_{2})|\leq \Vert x_{1}-x_{2}\Vert
\]%
which implies the uniform continuity of $E$. To show that $E$ is subadditive
we remark that for any $y_{1}\in X$ and $y_{2}\in X$ we have
\[
E(x_{1}+x_{2})\leq \Vert x_{1}+x_{2}-y_{1}-y_{2}\Vert
\]%
\[
\leq \Vert x_{1}-y_{1}\Vert +\Vert x_{2}-y_{2}\Vert .
\]%
Taking $\inf $ from the right with respect to $y_{1}$ and $y_{2}$ we get $%
E(x_{1}+x_{2})\leq E(x_{1})+E(x_{2})$ which means that $E$ is subadditive.

For any $x\in X$ and $a\in \mathbb{R}\setminus \{0\}$ we have
\[
E(ax)=\inf_{y\in M}\Vert ax-y\Vert =|a|\inf_{y\in M}\Vert x-y/a\Vert
\]%
\[
=|a|\inf_{y\in M}\Vert x-y\Vert =|a|E(x),
\]%
this proves that $E$ is positively homogeneous. Finally, since $E$ is
subadditive and positively homogeneous then it is convex.
$\square$

If the $\inf $ in (\ref{bestapprox}) is attained for $y_{0}\in M$, i.e. $%
E(x)=\Vert x-y_{0}\Vert $, then $y_{0}$ is called an element of best
approximation for $x$ in $M$. The set $M\subset X$ is called an \textit{%
existence set} if for every $x\in X$ there is an element of best
approximation in $X$.

\bigskip

{\bf Proposition 56}
\begin{em}
\label{5.2} Every closed locally compact subset $M\subset X$ is an existence
set. In particular, every finite dimensional subspace of $X$ is an existence
set.
\end{em}

{\bf Proof.}
\bigskip Assume that $x\in X\setminus M$ and $E(x)=c>0$, otherwise the
existence is obvious. By the definition of $\inf $ for every $n\in \mathbb{N}
$ there is such $y_{n}\in M$ that $\Vert x-y_{n}\Vert <E(x)+1/n$ and the
sequence $\{y_{n}\}$ is bounded since%
\[
\Vert y_{n}\Vert =\Vert x-x+y_{n}\Vert \leq \Vert x\Vert +E(x)+1/n
\]%
\[
=\Vert x\Vert +c+1/n.
\]%
Using local compactness of $M$ we may find such a subsequence $\{y_{n_{m}}\}$
that $y_{n_{m}}\rightarrow y_{0}$ as $m\rightarrow \infty $. Remark that $%
y_{0}\in M$ because $M$ is closed. It is clear that
\[
E(x)\leq \Vert x-y_{n_{m}}\Vert <E(x)+1/m_{n},{\ }{\ }n\in \mathbb{N}
\]%
and if we let $m\rightarrow \infty $, we get $\Vert x-y_{0}\Vert =E(x)$,
which means that $y_{0}$ is an element of best approximation.
$\square$

The norm on $X$ is called \textit{strictly convex} if for any $x\in X$ and $%
y\in X$, $\Vert x\Vert =\Vert y\Vert =1$ we have that $\Vert ax+(1-a)y\Vert
<1$ for any $a\in \left( 0,1\right) $. This means that the unit sphere in $X$%
, $\Vert x\Vert =1$ does not contain any segment.

{\bf Proposition 57}
\begin{em}
\bigskip \label{5.3} Let $M$ be a convex subset of a strictly normalized
space $X$. If for some $x\in X$ there is an element of best approximation in
$M$ then this element is unique.
\end{em}

{\bf Proof.}
\bigskip Assume that there are two elements $y_{1}\in M$ and $y_{2}\in M$, $%
y_{1}\neq y_{2}$ of best approximation for $x\in X$,
\[
E(x)=\Vert x-y_{1}\Vert =\Vert x-y_{2}\Vert .
\]%
Since $M$ is convex then for any $a\in \lbrack 0,1]$ the element $%
y_{a}=ay_{1}+(1-a)y_{2}$ is in $M$ and
\[
E(x)\leq \Vert x-y_{a}\Vert =\Vert a(x-y_{1})+(1-a)(x-y_{2})\Vert
\]%
\[
\leq a\Vert x-y_{1}\Vert +(1-a)\Vert x-y_{2}\Vert
\]%
\[
=aE(x)+(1-a)E(x)=E(x).
\]%
This means that the sphere $\{z|{\ }z\in X;{\ }\Vert x-z\Vert =E(x)\}$
contains the segment $y_{a}=ay_{1}+(1-a)y_{2}$, $a\in \lbrack 0,1]$ which is
a contradiction with the strict convexity.
$\square$

The set $M\in X$ with the property that for every $x\in X$ there exists a
unique element of best approximation is called a \textit{Chebyshev set}.

Let $M$ be a Chebyshev set then the operator of \textit{the best
approximation (metric projection)} $P(x)$ is defined by the following
equality
\[
E(x,M)=\Vert x-P(x)\Vert ,{\ }{\ }P(x)\in M.
\]

{\bf Proposition 58}
\begin{em}
\bigskip \label{5.4} If $M$ is a locally compact Chebyshev set in $X$, then
operator $P$ is continuous. If $M$ is a Chebyshev subspace, then $P$ is
homogeneous and, in particular, odd $P(-x)=-P(x)$.\emph{\ }
\end{em}

{\bf Proof.}
\bigskip Let $x_{0}$ be a fixed point in $X$ and $x_{m}\rightarrow x_{0}$,
Observe that
\[
\Vert P(x_{m})-x_{0}\Vert \leq \Vert P(x_{m})-x_{m}\Vert +\Vert
x_{m}-x_{0}\Vert
\]%
\[
=E(x_{m},M)+\Vert x_{m}-x_{0}\Vert
\]%
and the sequence $\{E(x_{m},M)\}$ converge by the Proposition 55
.
Consequently, the sequence $\{P(x_{m})\}$ is bounded. Assume that $%
P(x_{m})\nrightarrow P(x_{0})$. Using local compactness of $M$ we find a
subsequence $P(x_{m_{n}})$ such that $\lim_{n\rightarrow \infty
}P(x_{m_{n}})=z\neq P(x_{0})$. Since $M$ is a Chebyshev set and therefore
closed we have $z\in M$. Taking a limit when $n\rightarrow \infty $ in
\[
\Vert x_{m_{n}}-P(x_{m_{n}})\Vert =E(x_{m_{n}},M)\leq \Vert
x_{m_{n}}-P(x_{0})\Vert
\]%
we get $\Vert x_{0}-z\Vert \leq \Vert x_{0}-P(x_{0})\Vert $, which means
that $z$ is an element of best approximation for $x_{0}$ in $M$. This
contradicts the assumption that $M$ is a Chebyshev set. Hence $%
P(x_{m})\rightarrow P(x_{0})$.

In the case when $M$ is a Chebyshev subspace for any $a\in \mathbb{R}$ we
get
\[
\Vert ax-aP(x)\Vert =|a|\Vert x-P(x)\Vert
\]%
\[
=|a|E(x,M)=\Vert ax-P(ax)\Vert ,
\]%
or $P(ax)=aP(x)$.
$\square$

Let $M=M_{m}$ be an $m$-dimensional Chebyshev subspace of the normed space $%
X $ and $\{x_{1},\cdot \cdot \cdot ,x_{m}\}$ be a basis in $M_{m}$. The
operator of best approximation can be represented as
\begin{equation}
P(x)=\sum_{k=1}^{m}\alpha _{k}(x)x_{k}.  \label{kora}
\end{equation}%
From the Proposition 58
 we get

{\bf Proposition 59}
\begin{em}
\label{5.5} The functionals $\alpha _{k}(x):X\rightarrow M_{m}$, $1\leq
k\leq m$ are homogeneous and continuous.\emph{\ }
\end{em}

{\bf Proof.}
By the Proposition 58
, $P(ax)=aP(x)$, which means that
\[
\sum_{k=1}^{m}\alpha _{k}(ax)x_{k}=\sum_{k=1}^{m}a\alpha _{k}(x)x_{k}.
\]%
The representation (\ref{kora}) is unique, hence for any $a\in \mathbb{R}$
and $x\in X$ we have $a\alpha _{k}(x)=\alpha _{k}(ax)$, $1\leq k\leq m$.
Finally, remark that the convergence in a finite dimensional space $M_{m}$ ($%
\mathrm{dim}M_{m}=m$) is equivalent to componentwise convergence and the
operator $P:{\ }X\rightarrow M_{m}$ is continuous. This implies the
continuity of the functionals $\alpha _{k}$, $1\leq k\leq m$.
$\square$

\bigskip

\section*{Borsuk-Ulam theorem}


\label{Borsuk}

The next statement is an important result and is extensively used in the
calculation of lower bounds for $n$-widths \cite{borsuk}.

{\bf Theorem 60}
\begin{em}
\label{t5.1} \bigskip (Borsuk-Ulam) Let $X$ and $Y$ be finite-dimensional
Banach space over $\mathbb{R}$ or $\mathbb{C}$ with $\mathrm{dim}Y<\mathrm{%
dim}X$ and let $S=S(X)=\{x\in X:\Vert x\Vert =1\}$ be the unit sphere in $X$%
. If $f:S\rightarrow Y$ is a continuous map, then there is a point $x\in S$
such that $f(-x)=f(x)$. In particular, if $f$ is an odd function, then there
is a point $x\in S$ such that $f(x)=0$.
\end{em}

Theorem 60
 was suspected by Ulam and proven by Borsuk and can be
reformulated as following. Let $\Omega $ be a bounded, open, symmetric
neighborhood of $\mathbf{0}$ in $R^{m}$, and $F$ a continuous odd map of the
boundary $\partial \Omega $ into $R^{m-1}$. Then there exists an $x^{\ast
}\in \partial \Omega $ such that $F(x^{\ast })=\mathbf{0}$.

{\bf Theorem 61}
\begin{em}
\label{ball} Let $X_{n+1}$ be any $n+1$ dimensional subspace of a real
normed linear space $X$, and let $B(X_{n+1})$ denote the unit ball of $%
X_{n+1}$. Then
\[
d_{k}(B(X_{m+1}),X)=1,{\ }{\ }k=0,1,2,\cdot \cdot \cdot ,m.
\]
\end{em}

{\bf Proof.}
It is clear that
\[
d_{m}(B(X_{m+1}),X)\leq d_{m-1}(B(X_{m+1}),X)
\]%
\[
\leq \cdot \cdot \cdot \leq d_{0}(B(X_{m+1}),X)=1,
\]%
so it is sufficient to show that $d_{m}(B(X_{n+1}),X)\geq 1$. We show that
for any given $m$-dimensional subspace $L_{m}\in X$ there exists $x\in
\partial B(X_{m+1})$ with zero as a best approximation from $L_{m}$. Let $%
\{x_{1},\cdot \cdot \cdot ,x_{m+1}\}$ and $\{z_{1},\cdot \cdot \cdot
,z_{m}\} $ be bases for $X_{m+1}$ and $L_{m}$ respectively, then for any $%
x\in X_{m+1} $ and $z\in L_{m}$ we have representations
\[
x=\sum_{s=1}^{m+1}a_{s}x_{s},{\ }{\ z}=\sum_{s=1}^{m}b_{s}z_{s}.
\]%
It is sufficient to take $X=\mathrm{lin}\{X_{m+1},L_{m}\}$ in the proof.
Remark that $\mathrm{dim}X=l\leq 2m+1$. Let $\{y_{1},\cdot \cdot \cdot
,y_{l}\}$ be a basis for $X=\mathrm{lin}\{X_{m+1},L_{m}\}$, so any $x\in X$
may be written in the form
\[
x=\sum_{s=1}^{l}c_{s}y_{s}.
\]%
If the norm on $X$ is not strictly convex then it may be replaced by the
norm
\begin{equation}
\Vert x\Vert _{\epsilon }=\Vert x\Vert +\epsilon \left(
\sum_{s=1}^{l}|c_{s}|^{2}\right) ^{1/2}  \label{strict}
\end{equation}%
which is strictly convex. Because $\mathrm{dim}X<2m+1$ we can take the limit
$\epsilon \rightarrow 0$ while maintaining the validity of the theorem. This
means that we can assume that the norm on $X$ is strictly convex which
implies the uniqueness and continuity of the best approximation operator and
also implies its oddness. The domain
\[
\Omega =\left\{ (a_{1},\cdot \cdot \cdot ,a_{n+1}):{\ x}%
=\sum_{s=1}^{m+1}a_{s}x_{s},{\ }\Vert x\Vert <1\right\}
\]%
is a bounded, open, symmetric neighborhood of $\mathbf{0}$ in $\mathbb{R}%
^{m+1}$. For any $a\in \partial \Omega $ let $F(a)=(b_{1},\cdot \cdot \cdot
,b_{m})\in \mathbb{R}^{m}$ denote the vector of coefficients of the best
approximation to
\[
x=\sum_{s=1}^{m+1}a_{s}x_{s}\in \partial B(X_{m+1})
\]%
from $L_{m}$. By the Proposition 59
 the map $F(\cdot ):\partial
\Omega \rightarrow \mathbb{R}^{m}$ is an odd, continuous map of $\partial
\Omega $ into $\mathbb{R}^{m}$. Hence, by the Borsuk-Ulam theorem there
exist an
\[
x^{\ast }=\sum_{s=1}^{m+1}a_{s}^{\ast }x_{s},{\ }{\ }\Vert x\Vert =1,
\]%
for which the zero element is the best approximation from $L_{m}$.
$\square$

\bigskip From the Theorem 61 
 and the definition of Bernstein's $n$%
-widths we get

{\bf Corollary 62}
\begin{em}
\label{bk1} Let $A$ be a compact symmetric set in a Banach space $X$ then,%
\[
d_{m}\left( A,X\right) \geq b_{m}\left( A,X\right) ,\text{ }m=0,1,\cdot
\cdot \cdot .
\]
\end{em}

\section*{Brouwer theorem}


\label{Brower}{}

{\bf Theorem 63}
\begin{em}
(Brouwer) \bigskip For any continuous function $F$ mapping a compact convex
set $B$ into itself there is a point $x\in B$ such that $F\left( x\right) =x$%
.
\end{em}

{\bf Definition 64}
\begin{em}
\bigskip Let $\left( X,\vartheta \right) $ be a \textit{metric space} and $%
F:\left( X,\vartheta \right) \rightarrow \left( X,\vartheta \right) $ be a
continuous map such that $\vartheta \left( x,F\left( x\right) \right) \leq
\epsilon $ for any $x\in X$. In this case we say that $F$ is an $\epsilon $%
\textit{-shift}.
\end{em}

{\bf Corollary 65}
\begin{em}
\label{shift111} Let $X$ be a Banach space with the unit ball $B$, $\dim
X<\infty $. Let $F$ be an $\epsilon $\textit{-shift of }$B$ and $\epsilon
\in \left( 0,1\right) $. Then there is $\delta >0$ such that $\delta
B\subset F\left( B\right) $.
\end{em}

{\bf Proof.}
Pick such $\delta $ that $\epsilon +\delta <1$. If for any $x_{0}\in \delta
B $ we have $x_{0}\in F\left( B\right) $ then $\delta B\subset F\left(
B\right) $ and the statement is proved.

Hence, assume that there is such $x_{0}\in \delta B$ that $x_{0}\notin
F\left( B\right) $. Consider a continuous map $\xi =\Upsilon \left( x\right)
$, $\Upsilon :B\rightarrow \partial B$ which is defined as following. Let $%
l\left( x\right) $, $x\in B$ be a ray from $F\left( x\right) $ which passes
through $x_{0}\in \delta B$ and $\xi :=l\left( x\right) \cap \partial B$.

Assume that $x\in \mathrm{int}B$. Then $x\neq \xi \in \partial B$. Let $x\in
\partial B$. By assumption $x_{0}\notin F\left( B\right) $ and $x_{0}\in
\delta B$ where $\delta <1$. This means that $x_{0}\neq F\left( x\right) $
for any $x\in B$ and $x_{0}\notin \partial B$. By construction, $x_{0}$ is a
convex combination of $F\left( x\right) $ and $\xi $. Hence $x_{0}=\left(
1-\alpha \right) F\left( x\right) +\alpha \xi $, for some $\alpha \in \left(
0,1\right) $. Consequently,%
\[
\left\Vert \xi -F\left( x\right) \right\Vert >\left\Vert \xi
-x_{0}\right\Vert \geq 1-\delta
\]%
and, therefore
\[
\left\Vert x-\xi \right\Vert =\left\Vert x-F\left( x\right) +F\left(
x\right) -\xi \right\Vert
\]%
\[
>1-\delta -\epsilon =1-\left( \delta +\epsilon \right) >0
\]%
since $\delta +\epsilon <1$. This means that $x\neq \xi $. Hence the map $%
\Upsilon $ has not fixed points. This contradicts Brouwer's theorem since
the unit ball $B$ in $X$, $\dim X<\infty $ is compact and convex.
$\square$

Let $\left( A,\vartheta \right) $ be a metric compact space and $%
\left\{ U_{1},\cdot \cdot \cdot ,U_{m}\right\} $ be an \textit{open covering}
of $A$, i.e. $A\subset \cup _{s=1}^{m}U_{s}$. Let $Z$ be a linear metric
space and $\left\{ z_{1},\cdot \cdot \cdot ,z_{m}\right\} $ be a set of
distinct points in $Z$. Let%
\[
\begin{array}{c}
F:A\longrightarrow Z \\
F\left( x\right) =\sum_{s=1}^{m}\lambda _{s}\left( x\right) z_{s},%
\end{array}%
\]%
where%
\[
\lambda _{s}\left( x\right) :=\frac{d_{s}\left( x\right) }{%
\sum_{k=1}^{m}d_{k}\left( x\right) }
\]%
and%
\[
d_{k}\left( x\right) :=\min \left\{ \vartheta \left( x,y\right) \left\vert
y\in A\diagdown U_{k}\right. \right\} .
\]%
Clearly $\lambda _{s}\left( x\right) \geq 0$ since $d_{k}\left( x\right)
\geq 0$. Observe that the functions $\lambda _{s}\left( x\right) $, $1\leq
s\leq m$ are continuous,
\[
\sum_{s=1}^{m}\lambda _{s}\left( x\right) =1
\]%
and $\lambda _{s}\left( x\right) =0$ if $x\notin \overline{U}_{s}$. The set $%
F\left( A\right) $ is called the \textit{nerve} of an open covering $\left\{
U_{1},\cdot \cdot \cdot ,U_{m}\right\} $ generated by the set $\left\{
z_{1},\cdot \cdot \cdot ,z_{m}\right\} $ and is denoted by $\mathcal{N}%
\left( z_{1},\cdot \cdot \cdot ,z_{m}\right) $.

{\bf Proposition 66}
\begin{em}
Let $A$ be a compact in a Banach space $X$. Then for any $\epsilon >0$ there
exists $m=m\left( \epsilon \right) $, a linear manifold $M_{m}$, $\dim
M_{m}=m$ and an $\epsilon $-shift $F:A\rightarrow M_{m}$.
\end{em}

{\bf Proof.}
Since $A$ is a compact then by the Hausdorff theorem for any $\epsilon >0$
there is a finite $\epsilon $-net $\left\{ x_{1},\cdot \cdot \cdot
,x_{m}\right\} $, $m=m\left( \epsilon \right) $ in $A$, i.e. $A$ can be
covered by the union of sets $\epsilon B+x_{s}$, $1\leq s\leq m$, or $A=\cup
_{s=1}^{m}\left( \epsilon B+x_{s}\right) $, where $B$ is the unit open ball
in $X$. Clearly \textrm{aff}$\left\{ x_{1},\cdot \cdot \cdot ,x_{m}\right\} $
is a linear manifold $M_{m}$, $m:=\dim M_{m}\leq n$. The map $F:A\rightarrow
\mathcal{N}\left( x_{1},\cdot \cdot \cdot ,x_{m}\right) $ is a required $%
\epsilon $-shift.
$\square$

{\bf Theorem 66}
\begin{em}
(Tikhomirov \cite{tikhomirov1}, p. 221) \label{ba1} Let $X$ be a Banach
space and $A\subset X$ be a convex symmetric compact. Then%
\[
b_{m}\left( A,X\right) \leq 2a_{m}\left( A,X\right) .
\]
\end{em}

{\bf Poof.}
From the proof of Theorem 61 
 and (\ref{strict}) we may assume that $%
X $ is a finite dimensional Banach space with infinitely smooth and strictly
convex unit ball $B$. Fix an $\left( m+1\right) $-dimensional subspace $%
L_{m+1}$ in $X$. Observe that $L_{m+1}$ is a Chebyshev subspace. Hence the
operator of metric projection $P_{L_{m+1}}:X\rightarrow L_{m+1}$ is well
defined. Assume that
\begin{equation}
2a_{m}\left( A,X\right) <b_{m}\left( A,X\right) -4\epsilon .  \label{jopa1}
\end{equation}%
Let $K_{m}$ be an $m$-dimensional complex and $F:A\rightarrow K_{m}$ be a
continuous map such that
\begin{equation}
\sup \left\{ x\in A\left\vert \left\Vert x-Fx\right\Vert \right. \right\}
\leq a_{m}\left( A,X\right) +\epsilon .  \label{jopa2}
\end{equation}%
Let $z_{1},\cdot \cdot \cdot ,z_{s}$ be the vertices of $K_{m}$ and $\zeta
_{1}:=P_{L_{m+1}}z_{1},\cdot \cdot \cdot ,\zeta _{s}:=P_{L_{m+1}}z_{s}$ are
the elements of the best approximation of $z_{1},\cdot \cdot \cdot ,z_{s}$
in $L_{m+1}$. Since $K_{m}$ is a simplicial complex then any $x\in K_{m}$
can be represented in the form%
\[
x=\sum_{j=1}^{l}\alpha _{s_{j}}z_{s_{j}}.
\]%
Define the maps $P:K_{m}\rightarrow L_{m+1}$,%
\[
Px:=\sum_{j=1}^{l}\alpha _{s_{j}}\zeta _{s_{j}}=\sum_{j=1}^{l}\alpha
_{s_{j}}P_{L_{m+1}}z_{s_{j}}=\sum_{j=1}^{l}P_{L_{m+1}}\alpha
_{s_{j}}z_{s_{j}}
\]%
and $\Psi :=P\circ F$. Observe that $P$ is a simplicial map. Hence $\dim
P\left( K_{m}\right) \leq \dim K_{m}=m$. The diameter of simplexes which
constitute $K_{m}$ can be assumed as small as we pleased. Hence, for any $%
\epsilon >0$ we, by the definition of $P_{L_{m+1}}$ and $P$, may assume that%
\begin{equation}
\max \left\{ y\in K_{m}\left\vert \left\Vert \left( P_{L_{m+1}}-P\right)
y\right\Vert \leq \epsilon \right. \right\} .  \label{jopa3}
\end{equation}%
Therefore, for any%
\[
x\in \left( b_{m}\left( A,X\right) -\epsilon \right) B\cap L_{m+1}\subset A
\]%
we get%
\[
\left\Vert x-\Psi x\right\Vert =\left\Vert x-P\circ Fx\right\Vert
\]%
\[
=\left\Vert x-Fx+Fx-P\circ Fx+P_{L_{m+1}}\circ Fx-P_{L_{m+1}}\circ
Fx\right\Vert
\]%
\[
\leq \left\Vert x-Fx\right\Vert +\left\Vert Fx-P_{L_{m+1}}\circ
Fx\right\Vert +\left\Vert P_{L_{m+1}}\circ Fx-P\circ Fx\right\Vert
\]%
\[
:=J_{1}+J_{2}+J_{3}.
\]%
By assumption (\ref{jopa2}),%
\[
J_{1}\leq a_{m}\left( A,X\right) +\epsilon
\]%
and%
\[
J_{2}=\left\Vert Fx-P_{L_{m+1}}\circ Fx\right\Vert =\inf \left\{ \xi \in
L_{m+1}\left\vert \left\Vert Fx-\xi \right\Vert \right. \right\}
\]%
\[
\leq \left\Vert Fx-x\right\Vert \leq a_{m}\left( A,X\right) +\epsilon .
\]%
From (\ref{jopa3}) it follows that%
\[
J_{3}=\left\Vert P_{L_{m+1}}\circ Fx-P\circ Fx\right\Vert \leq \epsilon .
\]%
Comparing these estimates we get%
\[
\left\Vert x-\Psi x\right\Vert \leq \left( a_{m}\left( A,X\right) +\epsilon
\right) +\left( a_{m}\left( A,X\right) +\epsilon \right) +\epsilon
\]%
\[
\leq 2a_{m}\left( A,X\right) +3\epsilon
\]%
\[
<\left( b_{m}\left( A,X\right) -4\epsilon \right) +3\epsilon =b_{m}\left(
A,X\right) -\epsilon ,
\]%
where we used (\ref{jopa1}). Hence a continuous map $\Psi $ of the ball $%
\left( b_{m}\left( A,X\right) -\epsilon \right) B\cap L_{m+1}$ is an $\left(
b_{m}\left( A,X\right) -\epsilon \right) $-shift. From Corollary \ref%
{shift111} we get%
\[
\dim \left( \Psi \left( b_{m}\left( A,X\right) -\epsilon \right) B\cap
L_{m+1}\right) \geq m+1\text{.}
\]%
But%
\[
\dim \left( \Psi \left( b_{m}\left( A,X\right) -\epsilon \right) B\cap
L_{m+1}\right) \leq \dim K_{m}=m.
\]%
Contradiction proofs that (\ref{jopa1}) is impossible. Consequently, $%
b_{m}\left( A,X\right) \leq 2a_{m}\left( A,X\right) $.
$\square$

\bigskip

\bigskip

\bigskip

\end{document}